\documentclass[a4paper,onecolumn]{article}

\pdfoutput=1
\usepackage[utf8]{inputenc}
\usepackage[english]{babel}
\usepackage[T1]{fontenc}
\usepackage{hyperref}
\usepackage[a4paper, margin=1in]{geometry}
\usepackage{csvsimple}
\usepackage{csquotes}

\usepackage{authblk}

\usepackage[section]{placeins}

\usepackage{amsmath,amssymb,amsthm,bm,mathtools,amsfonts,mathrsfs,bbm,dsfont, physics}
\usepackage[shortlabels]{enumitem}
\usepackage[ruled]{algorithm2e}
\usepackage{multicol}

\usepackage{adjustbox}

\usepackage{subfiles}

\edef\measurepage{\the\dimexpr\pagegoal-\pagetotal-\baselineskip\relax}

\SetKwComment{Comment}{/* }{ */}
\SetKwInOut{Hyperparameters}{Hyperparameters}

\usepackage{tikz}
\usepackage{float}
\usepackage[center]{caption}
\usepackage{subcaption}
\usetikzlibrary{positioning}
\usetikzlibrary{shapes}
\usepackage{changepage}
\usepackage{graphicx}
\usepackage{multirow, bigdelim}
\usepackage{tcolorbox}

\usepackage{makecell}

\usepackage[style=numeric,style=numeric-comp,sorting=none,defernumbers=true,backend=biber]{biblatex}
\usepackage{mathpazo}

\usetikzlibrary{tikzmark,decorations.pathreplacing,calc}

\usepackage[capitalize]{cleveref}

\DeclareRobustCommand{\[}{\begin{equation}}
\DeclareRobustCommand{\]}{\end{equation}}

\newcommand{\ccite}[1]{\textit{``\citefield{#1}{title}, (\citefield{#1}{year})''} \cite{#1}}

\title{\vspace{-20pt} \textbf{Benchmarking a wide range of optimisers for solving the Fermi-Hubbard model using the variational quantum eigensolver}}

\author[1]{Benjamin D.M. Jones}
\author[1]{Lana Mineh}
\author[1,2]{Ashley Montanaro}

\affil[1]{\smaller Phasecraft Ltd.}
\affil[2]{School of Mathematics, University of Bristol.}
\date{\smaller \today}

\begin{document}

\maketitle

\begin{abstract}

We numerically benchmark 30 optimisers on 372 instances of the variational quantum eigensolver for solving the Fermi-Hubbard system with the Hamiltonian variational ansatz. We rank the optimisers with respect to metrics such as final energy achieved and function calls needed to get within a certain tolerance level, and find that the best-performing optimisers are variants of gradient descent such as Momentum and ADAM (using finite difference), SPSA, CMAES, and BayesMGD. We perform gradient analysis, and observe that the step size for finite difference has a very significant impact. We also consider using simultaneous perturbation (inspired by SPSA) as a gradient subroutine: here finite difference can lead to a more precise estimate of the ground state but uses more calls, whereas simultaneous perturbation can converge quicker but may be less precise in the later stages. Finally, we study the quantum natural gradient algorithm: we implement this method for 1-dimensional Fermi-Hubbard systems, and find that whilst it can reach a lower energy with fewer iterations, this improvement is typically lost when taking total function calls into account.
Our method involves performing careful hyperparameter sweeping on 4 instances. We present a variety of analysis and figures, detailed optimiser notes, and discuss future directions.

\end{abstract}

 \begin{figure}[h!]
     \centering
     \includegraphics[width=\textwidth]{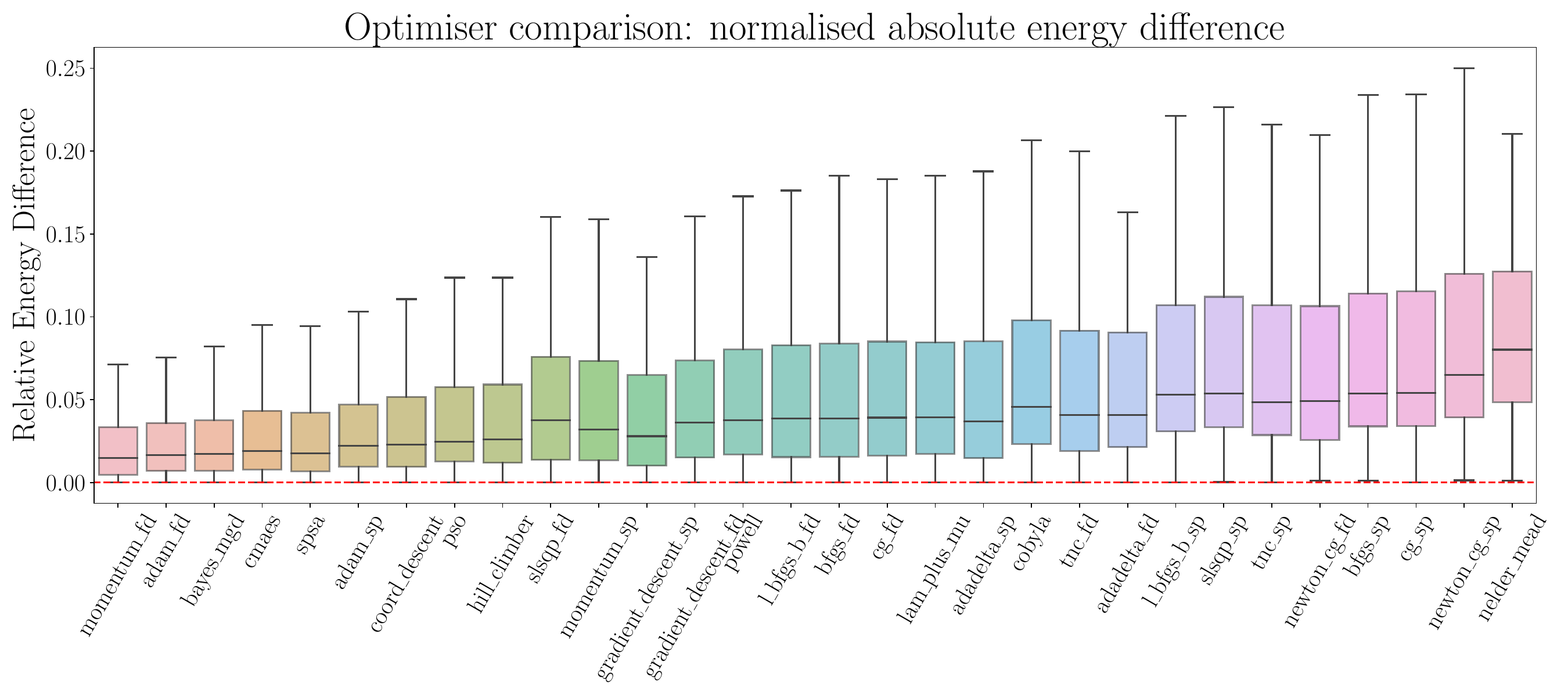}
     \caption{Boxplots of optimisers tested over all 372 VQE for Fermi-Hubbard instances, ordered in terms of performance, i.e. average normalised (divided by the size of the grid $m \times n$) final energy difference with the ground energy. The suffices `fd' and `sp' refer to the finite difference and simultaneous perturbation gradient subroutines respectively. See \cref{fig:boxplot all instances} and \cref{sec:results} for more discussion.}
 \end{figure}

\newpage

\tableofcontents

 \section{Introduction}

The variational quantum eigensolver (VQE) has emerged as a leading approach to simulating ground states of many-body quantum systems using a quantum computer \cite{bharti2022noisy, cerezo2021variational}. The primary appeal of this method is that it can be implemented on current and near-term quantum hardware \cite{stanisic2022observing}, without the need to wait for an error-corrected, universal quantum computer, which could still be many years away \cite{riskreport}. 

Ultimately, VQE provides a cost function to optimise, mapping a list of parameters to a real number. If the cost function is sufficiently expressive, then its minimum argument will correspond to preparation of the ground state of a quantum system of interest on a quantum computer -- a task which is difficult to perform classically. Several studies have demonstrated that VQE could yield an advantage over classical methods in the immediate future \cite{bharti2022noisy, cerezo2021variational, stanisic2022observing, wecker2015progress}.

In practice, a VQE cost function arises from preparing an initial quantum state, applying a parametrised quantum circuit, and undertaking an energy measurement of the final state. Due to the statistical nature of quantum measurements, the resulting cost function is a random variable and hence for a fixed set of parameters, one must run the quantum circuit many times in order to gain a reliable estimate of the expectation value of the final energy. 

Much research has been done on understanding how to design good VQE cost functions (or ansätze)  -- i.e. which initial state to use, the precise form of the parametrised circuit, and how best to perform the energy measurements \cite{wecker2015progress, stanisic2022observing, cade2020strategies, crawford2021efficient}. From the point of view of optimisation however, one can also simply take the cost function as a black box function and study which optimisers may be well suited to finding the minimum of the search space in question.

The field of optimisation algorithms is vast and expansive \cite{kochenderfer2019algorithms, Luke2013Metaheuristics, nocedal1999numerical}. There are many different flavours of optimiser: hill climbing, interpolation and model fitting approaches, evolutionary algorithms, and a wealth of gradient-based approaches, particularly inspired by the field of machine learning \cite{ruder2016overview}. In the context of VQE, there are also several more bespoke optimisation techniques that have been developed, such as coordinate descent (also known as rotosolve or sequential minimal optimization) \cite{nakanishi2020sequential, cade2020strategies, parrish2019jacobi, tseng2001convergence, ostaszewski2021structure}, quantum analytic descent \cite{koczor2022quantum}, quantum natural gradient \cite{stokes2020quantum}, and model gradient descent \cite{sung2020using, stanisic2022observing}.

\paragraph{This work.}

Our goal here is to study a variety of optimisation techniques for solving the Fermi-Hubbard model (i.e. preparing the ground state) using VQE, focusing on a particular circuit ansatz known as the Hamiltonian variational (HV) ansatz \cite{wecker2015progress, cade2020strategies}. After fixing a selection of problem instances, we numerically benchmark several promising algorithms and analyse their performance by considering both final accuracy with the ground energy achieved, and number of cost function calls to get within a certain tolerance of the ground energy. We also consider the quantum natural gradient descent method \cite{stokes2020quantum}, considering it separately to our main numerical study due to ansatz considerations (see \cref{app:qng}).

Although there exist several numerical studies comparing optimisers for VQE (see below), to the best of our knowledge our work is the first to do so specifically in the context of the Hamiltonian variational ansatz for solving Fermi-Hubbard. Our contribution also includes a much larger range of problem instances and optimisers than prior works. It is our hope that this work will serve to inform optimiser choice for near term implementations of VQE on actual quantum hardware. Our main findings can be summarised as follows (see also \cref{sec:discussion}):

\begin{itemize}
\item Momentum or Adam with finite differences are good optimiser choices when considering overall final accuracy with ground energy  -- see \cref{fig:final_errors_boxplot_exclude2x2_ops}.
\item SPSA, CMAES or BayesMGD are good optimiser choices for using a low number of calls -- see \cref{fig:calls_to_within_error}.
\item The simultaneous perturbation gradient subroutine can be beneficial when seeking to minimise number of calls, whereas using finite difference is more effective for final accuracy.
\item We discuss the importance of hyperparameter selection, including choosing an appropriate step size for gradient subroutines (we found that for our purposes 0.4 seemed to be a good choice for finite differences).
\item Any improvements offered by the quantum natural gradient or imaginary time evolution methods seem to be lost when considering the total amount of calls taken (for the systems and implementation considered here).
\end{itemize}

This study is organised as follows. After providing a brief overview of relevant literature and some background on VQE for the Fermi-Hubbard model, we lay out our main method and approach to the bulk of our numerical study in \cref{sec:method}. We give a brief overview of the optimisers considered in \cref{tab:optimiser_summary}, \cref{subsec:optimisers}, but provide more extensive notes in \cref{app:optimisers}. We present our main results and figures in \cref{sec:results}. A brief discussion and conclusion appears in \cref{sec:discussion}. The appendices contain additional figures, some analysis of the quantum natural gradient method, and notes on the optimisers considered in the main text. Data used to generate results for this paper is available at Zenodo \cite{dataset}.

\subsection{Related literature}
\paragraph{VQE Review Articles.}

There exist several review articles on VQE and variational methods \cite{bharti2022noisy, cerezo2021variational, tilly2022variational, fedorov2022vqe, mcardle2019variational}. \cite{bharti2022noisy} and \cite{cerezo2021variational} contain discussions on several optimisation strategies, including gradient calculation, L-BFGS, quantum natural gradient, imaginary time evolution, Hessian based methods, quantum analytic descent, stochastic gradient descent, evolutionary algorithms, reinforcement learning, sequential minimisation, surrogate model based, BOBYQA, Rosalin, Adam and SPSA. We also particularly highlight \cite{tilly2022variational} as it contains detailed discussion and pseudocode, and their Table 11 is a helpful summary of hyperparameters and complexity for several optimisers.

\paragraph{Optimiser comparisons.}

Here we provide a non-exhaustive list of references that compared optimisers in the context of VQE. We point the reader to references contained within these works for additional studies in this direction.

\begin{itemize}[-]

    \item \ccite{lavrijsen2020classical}
    introduce a python software suite, \texttt{Scikit-Quant} \cite{scikit-quant}, containing four optimisers: NOMAD, ImFil, Bobyqa and SNOBfit. They also take BFGS and COBYLA from \texttt{scipy} for baseline comparisons. As problem instances, they used axis rotation and bond stretching and breaking of the ethylene molecule, and the Fermi-Hubbard model on 4 sites. They found that BFGS and COBYLA were best for noise-free cost functions, but the other 4 optimisers generally outperform them, with a ranking depending on factors such as if parameter bounds are available and the quality of initial parameters.

    \item \ccite{sung2020using} contains a helpful introduction section with extra references. They compare Nelder-Mead, BOBYQA, SPSA, SGD using parameter shifts, model gradient descent (MGD) and model policy gradient (MPG), and they introduced the last two as novel optimisers. They use MaxCut, the Sherrington-Kirpatrick model, and the Fermi-Hubbard model as benchmarks, and found the MGD and MPG were the best performers.

    \item In \ccite{bonet2023performance}, they study SLSQP, COBYLA, CMA-ES, and SPSA, and have an interesting section on hyperparameter tuning. They focus on the $H_4$ chain and square and the Hubbard model, and find that SPSA generally performs the best, but CMAES becomes competitive after hyperparameter tuning.

    \item In \ccite{wilson2021optimizing}, the authors compare L-BFGS-B, Nelder-Mead, Bayesian optimisation, evolutionary strategies, and a Long Short Term Memory (LSTM) recurrent neural network model (which they introduce as a meta-learner).

    \item In \ccite{leng2019robust}, they benchmark 6 algorithms: SPSA, RSGF, finite difference stochastic approximation (FDSA), AdamSPSA, AdamRSGF and the Adam variant of FDSA (AdamFDSA). 

    \item \ccite{nannicini2019performance} compared LBFGS, COBYLA, RBFOpt, Powell, and SPSA on several combinatorial optimisation problems,  finding that COBYLA, RBFOpt, LBFGS performed best, although the cost-functions considered were exact.

    \item \ccite{guerreschi2017practical} looked at Nelder-Mead and BFGS (with both finite differences and analytic gradient).

    \item \ccite{romero2018strategies} compares Nelder-Mead, Powell, COBYLA, L-BFGS-B. 

    \item \ccite{wierichs2020avoiding} compares BFGS, ADAM and natural gradient.

    \item \ccite{verdon2019learning} does not directly focus on optimiser choice, yet uses neural networks to suggest good starting parameters, through training on the optimal parameters found in VQE runs.

    \item \ccite{alvertis2024classical}
    conducted a benchmarking study using tensor network methods to evaluate different VQE ansätze for the Fermi-Hubbard model, taking L-BFGS as an optimiser. 

\end{itemize}

\subsection{VQE for Fermi-Hubbard}
\label{subsec:vqe_fh}

The Fermi-Hubbard Hamiltonian is given by
\[
H = t \sum_{\langle j,k \rangle, \sigma} \bigg ( a_{j \sigma}^\dagger a_{k \sigma} + a_{k \sigma}^\dagger a_{j \sigma} \bigg ) + U \sum_{j} n_{j \uparrow} n_{j \downarrow},
\]
where $t$ is the tunneling amplitude (always set to $t=-1$ in this work), $U$ is the Coulomb potential, $n_{j \sigma} = a_{j\sigma}^\dagger a_{j\sigma}$ is the number operator (for spin $\sigma \in \{\uparrow, \downarrow \}$), and $\langle j,k \rangle$ denotes adjacent sites on a rectangular lattice with open boundary conditions.

We use the Jordan-Wigner (JW) mapping from fermions to qubits. For an $m \times n$ grid, this uses $2nm$ qubits, each representing whether the spin up or spin down mode is occupied at a given site. We order the qubits first by spin, and then in a snake like ordering starting at the top row from left to right -- see \cite{cade2020strategies}. The Jordan-Wigner mapping is then as follows (for $j<k$):

\begin{align}
    a_{j}^\dagger a_k + a_k^\dagger a_j &\mapsto \frac{1}{2}(X_j X_k + Y_j Y_k)Z_{j+1} \dots Z_{k-1} \\
    a_{j}^\dagger a_j a_k^\dagger a_k &\mapsto \ketbra{11}{11}_{jk} \\
    &= \frac{1}{4}(\mathbbm{1} - Z_j)(\mathbbm{1} - Z_k).
\end{align}

If the qubits are next to each other in the JW ordering (i.e. $k=j+1$), then for the hopping terms we simply get $ \frac{1}{2}(X_j X_k + Y_j Y_k)$, otherwise there will be an accompanying string of $Z$s between the sites.

The Hamiltonian variational (HV) ansatz \cite{wecker2015progress, cade2020strategies} groups terms in a particular way: $O$ collects all the onsite terms, $H_1$ and $H_2$ respectively represent all the odd and even columns between sites (horizontal hopping terms), and $V_1$ and $V_2$ respectively represent all the odd and even rows between sites (vertical hopping terms). A layer of the HV ansatz is then of the form
\[
e^{i \theta_5 V_2} e^{i \theta_4 H_2}e^{i \theta_3 V_1} e^{i \theta_2 H_1} e^{i \theta_1 O}.
\]

A challenge with implementing this is that for vertical hopping terms that are non-adjacent in the JW ordering, we would have to implement operators such as 

\[
e^{  \frac{ i \theta }{2}(X_j X_k + Y_j Y_k)Z_{j+1} \dots Z_{k-1}}
\]
which as a multi-qubit gate would be difficult to perform on real hardware.

One common way to avoid this string of $Z$ operators is to perform FSWAP gates (corresponding to SWAP for fermionic systems) to make the relevant systems adjacent in the JW picture, so that then only gates of the form $e^{  \frac{ i \theta }{2}(X_j X_k + Y_j Y_k)}$ need to be applied \cite{cade2020strategies, stanisic2022observing}. The FSWAP gate is as follows:

\begin{equation}
    \text{FSWAP} = \begin{pmatrix}
        1 & 0 & 0 & 0 \\
        0 & 0 & 1 & 0 \\
        0 & 1 & 0 & 0 \\
        0 & 0 & 0 & -1 \\
    \end{pmatrix}.
\end{equation}

For example, when considering a $2 \times 2$ system the only non-adjacent hopping terms would be the first column. Ordering the qubits as described above, for a single spin sector this would correspond to the following operator

\begin{align}
 \frac{1}{2}(X_1 X_4 + Y_4 Y_1)Z_{2} Z_3,
\end{align}

which would actually be implemented as

\[
\text{F}_{12} \text{F}_{34} ~ e^{\frac{i \theta}{2} (X_2 X_3 + Y_2 Y_3)}  ~ \text{F}_{12} \text{F}_{34},
\]
where $\text{F}_{jk} $ denotes FSWAP on qubits $j$ and $k$.

Overall, this ansatz results in variational gates of the following form:
\begin{align}
    &e^{\frac{i\theta}{2} (XX + YY)} \\
    &e^{i\theta \ketbra{11}} 
\end{align}
with some fixed gates (FSWAPs) appearing between gates that share the same angle.

\renewcommand{\arraystretch}{1.3}
\begin{table}[H]
    \centering
    \begin{tabular}{|c|c|c|c|c|c|c|}
       \hline
        \multirow{ 2}{*}{Grid size} & \multirow{ 2}{*}{Number of parameters per layer} & \multicolumn{5}{c|}{Number of gates sharing this parameter} \\ \cline{3-7}
        && $O$ & $H_1$ & $H_2$ & $V_1$ & $V_2$  \\ \hline
          $2 \times 1$ & 2 & 4 & 2 & 0 & 0 & 0 \\ 
       $3 \times 1$ & 3 & 6 & 2 & 2 & 0 & 0 \\ 
         $2 \times 2$ & 3 & 8 & 4 & 0 & 4 & 0  \\ 
          $3 \times 2$ & 4 & 12 & 4 & 4 & 6 & 0  \\ 
          $3 \times 3$ & 5 & 18 & 6 & 6 & 6 & 6  \\ \hline
          \multirow{ 2}{*}{$m \times n$} &  \multirow{ 2}{*}{$1 + \min\{2, m-1 \} + \min\{2, n-1 \}$} & \multirow{ 2}{*}{$2mn$} &  \multirow{ 2}{*}{$2n \lceil \frac{m-1}{2} \rceil $} &  \multirow{ 2}{*}{$2n \lfloor \frac{m-1}{2} \rfloor$} &  \multirow{ 2}{*}{$2m \lceil \frac{n-1}{2} \rfloor$} &  \multirow{ 2}{*}{$2m \lfloor \frac{n-1}{2} \rfloor$}  \\ 
         &&&&&& \\\hline
    \end{tabular}
    \caption{Summary of properties of the Hamiltonian variational (HV) ansatz for grids of size $m \times n$, $(m \geq n)$.}
    
\end{table}

 \section{Method}
\label{sec:method}

We can summarise the overall approach to our main numerical study with the following:

\begin{enumerate}[(1)]
    \item Determine the ansätze, problem instances, and target ground energies.
    \item Run optimisers on exact cost functions as a test of the ansatz.
    \item Perform hyperparameter and gradient sweeping:
    \begin{itemize}
        \item We select 4 out of the 372 instances to run sweeping on.
        \item Sweep over step sizes to determine an appropriate finite difference gradient function, and use simultaneous perturbation with SPSA default step size. This establishes two gradient functions to use (which are not included as hyperparameters for gradient-based optimisers).
        \item For each optimiser, we first identify the default hyperparameters (if any). Then we sweep over the 4 instances and take the best hyperparameters.
    \end{itemize}
    \item Run all optimisers on all instances.
    \item Perform data analysis and visualisation.
\end{enumerate}

The data generated during this study (including hyperparameter sweeping, gradient sweeping, and runs on exact and statistical cost functions) is hosted online using Zenodo \cite{dataset}.

\subsection{Instances} \label{subsec:instances}
Recall that the tunneling amplitude is always set to $t=-1$ in this work. We divide our 372 problem instances into four benchmarks as follows. 

An instance is specified by:
\begin{itemize}
    \item $(m,n)$: the grid size.
    \item $U$: Coulomb potential.
    \item The occupation number.
    \item The number of layers in the ansatz.
    \item The number of shots to take in a single energy evaluation.
\end{itemize}

\begin{tcolorbox}
\begin{multicols}{2}
\paragraph{Benchmark 1:} \textit{4 qubits, 12 instances.\\}

\begin{itemize}
\setlength\itemsep{0.6em}
    \item $(m,n) \in \{ (1,2) \}$.
    \item $U \in [2,4,8]$.
    \item Half filling.
    \item 2 and 5 layers.
    \item 1,000 and 10,000 shots.
\end{itemize}

\paragraph{Benchmark 2:} \textit{ 6-12 qubits, 216 instances.}
\begin{itemize}
    \item $(m,n) \in  \{ (1,3), (1,4), (2,2),$ \newline
    $  \phantom{(m,n) \in  \{ } (1,5), (1,6), (2,3)  \}$.
    \item $U \in [2,4,8]$.
    \item Half and quarter filling.
    \item 2, 5 and 8 layers.
    \item 1,000 and 10,000 shots.
\end{itemize}

\columnbreak

\paragraph{Benchmark 3:} \textit{14-16 qubits, 108 instances.}

\begin{itemize}
    \item $(m,n) \in \left \{ (1,7), (1,8), (2,4) \right \}$.
    \item $U \in [2,4,8]$.
    \item Half and quarter filling.
    \item 5, 8 and 10 layers.
    \item 1,000 and 10,000 shots.
\end{itemize}

\paragraph{Benchmark 4:} \textit{18 qubits, 36 instances.}
\begin{itemize}
    \item $(m,n) =(3,3)$.
    \item $U \in [2,4,8]$.
    \item Half and quarter filling.
    \item 5, 8 and 10 layers.
    \item 1,000 and 10,000 shots.
\end{itemize}

\end{multicols}
\end{tcolorbox}

In total this makes $12+216+108+36=372$ instances. For each of these instances, we calculate the ground state energy exactly to use as a benchmark for the VQE performance (clearly these energies do not depend on the number of layers or number of shots). We also isolate four of these instances to be used for hyperparameter sweeping:
\begin{itemize}
    \item $3 \times 1$, $t=1$, $U=4$, quarter-filling, 2 layers,  1,000 shots,
    \item $2 \times 2$, $t=1$, $U=2$, quarter-filling,  5 layers,  10,000 shots,
    \item $5\times1$, $t=1$, $U=8$, half-filling, 8 layers,  10,000 shots,
    \item $3 \times 2$, $t=1$, $U=4$, half-filling, 5 layers,  1,000 shots.
\end{itemize}
We choose these instances as they cover a spread of instance values, whilst also using relatively few qubits to minimise runtime.

\subsection{Initial parameters}
We always take the initial parameters to be $\frac{1}{\texttt{nlayer}}$, where \texttt{nlayer} is the number of layers in the ansatz, as in \cite{cade2020strategies}.

\subsection{Form of cost function} \label{subsec:form}

To determine if the ansätze are sufficiently expressive, we first ran BFGS, L-BFGS-B, Nelder-Mead, Powell and SLSQP from \texttt{scipy} \cite{scipy} on exact cost functions (i.e. exact simulated energy measurements), as for example BFGS was found to perform well on the exact cost function in previous works \cite{cade2020strategies, lavrijsen2020classical} -- see \cref{subsec:exact cost functions} for results.

For our main numerical study, we take cost functions with statistical noise calculated from sampling a number of shots (either 1,000 or 10,000) and taking the average. This is done by directly simulating the measurements (as opposed to for example adding a noise profile on top of the exact cost function). In this work we do not consider models incorporating physical noise in the circuit.

\subsection{Gradients}

Several of the optimisers we consider require access to a gradient function (such as gradient descent, Momentum, ADAM, SLSQP, BFGS). There are multiple choices of gradient functions here, such as finite differences, parameter shift rules \cite{wierichs2022general}, exact simulated gradient, or stochastic approximations to the gradient. In this work, we use finite differences, and also consider using the simultaneous perturbation gradient that appears in SPSA.

\paragraph{Finite differences.\\}

The finite difference formula for computing the $k$\textsuperscript{th} element of the gradient vector, using step size $\epsilon$, is given by

\[
\nabla f(\theta)_k = \frac{f(\theta + \epsilon e_k) - f(\theta - \epsilon e_k) }{2\epsilon},
\]
where $e_k$ are vectors with $1$ at entry $k$ and $0$ elsewhere. This takes $2\nu$ function evaluations to calculate, for $\nu$ parameters.

As there is statistical uncertainty in the cost function (due to taking a finite number of samples), a trade-off exists when selecting the step size. If the step size is too small, the statistical noise brings error to the gradient approximation. Similarly if the step size is too large, we will incur an error due to the finite difference approximation. See \cite{mari2021estimating} for an interesting discussion and calculation of this trade-off.

\begin{figure}
    \centering
    \includegraphics[width=\textwidth]{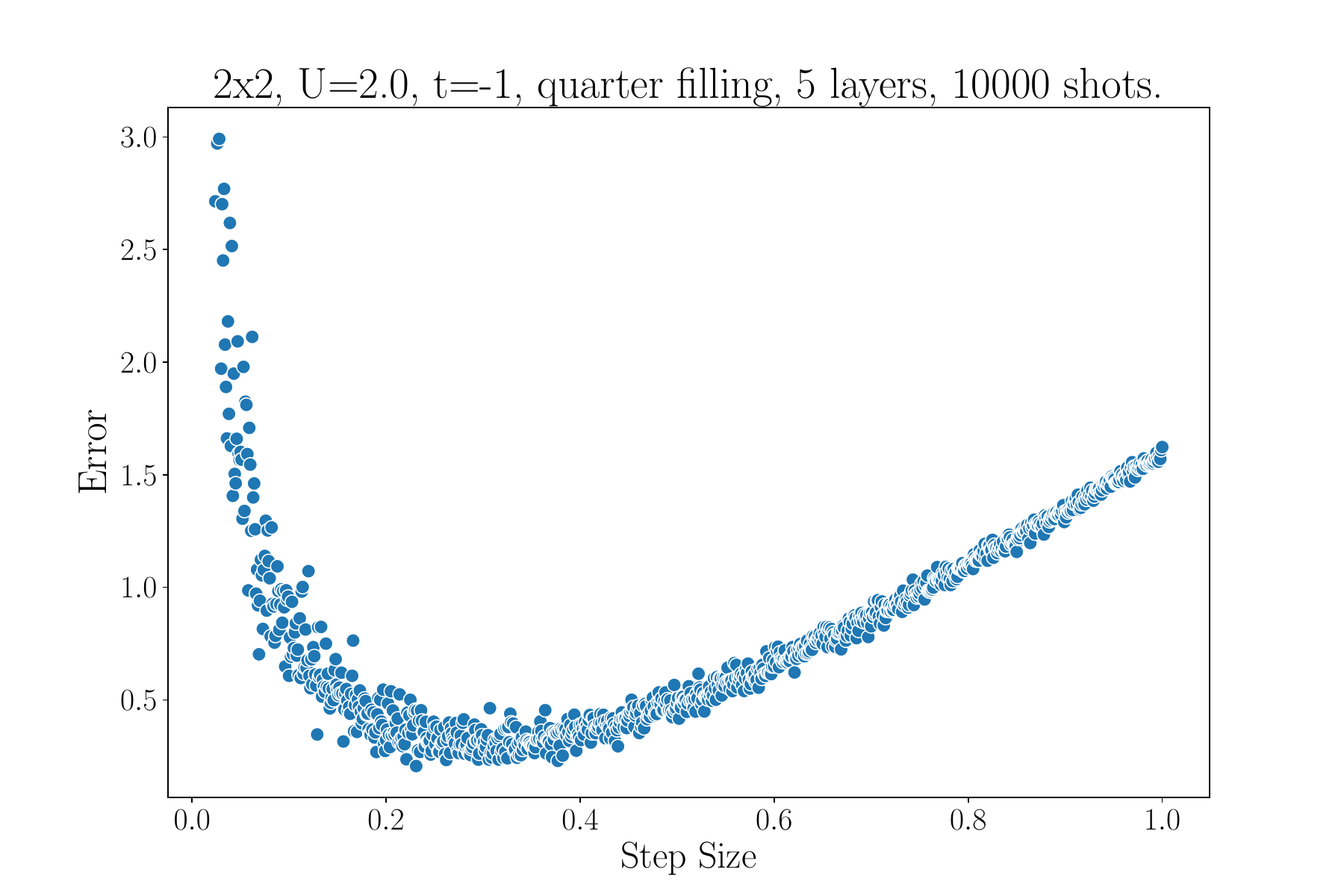}
    \caption{Plot of finite difference step size with absolute error (norm of difference) with the exact gradient. We see a trade-off between the error introduced by finite difference, and the error appearing from the statistical noise.}
    \label{fig:gradient_tradeoff}
\end{figure}

\begin{figure}
    \centering
    \includegraphics[width=\textwidth]{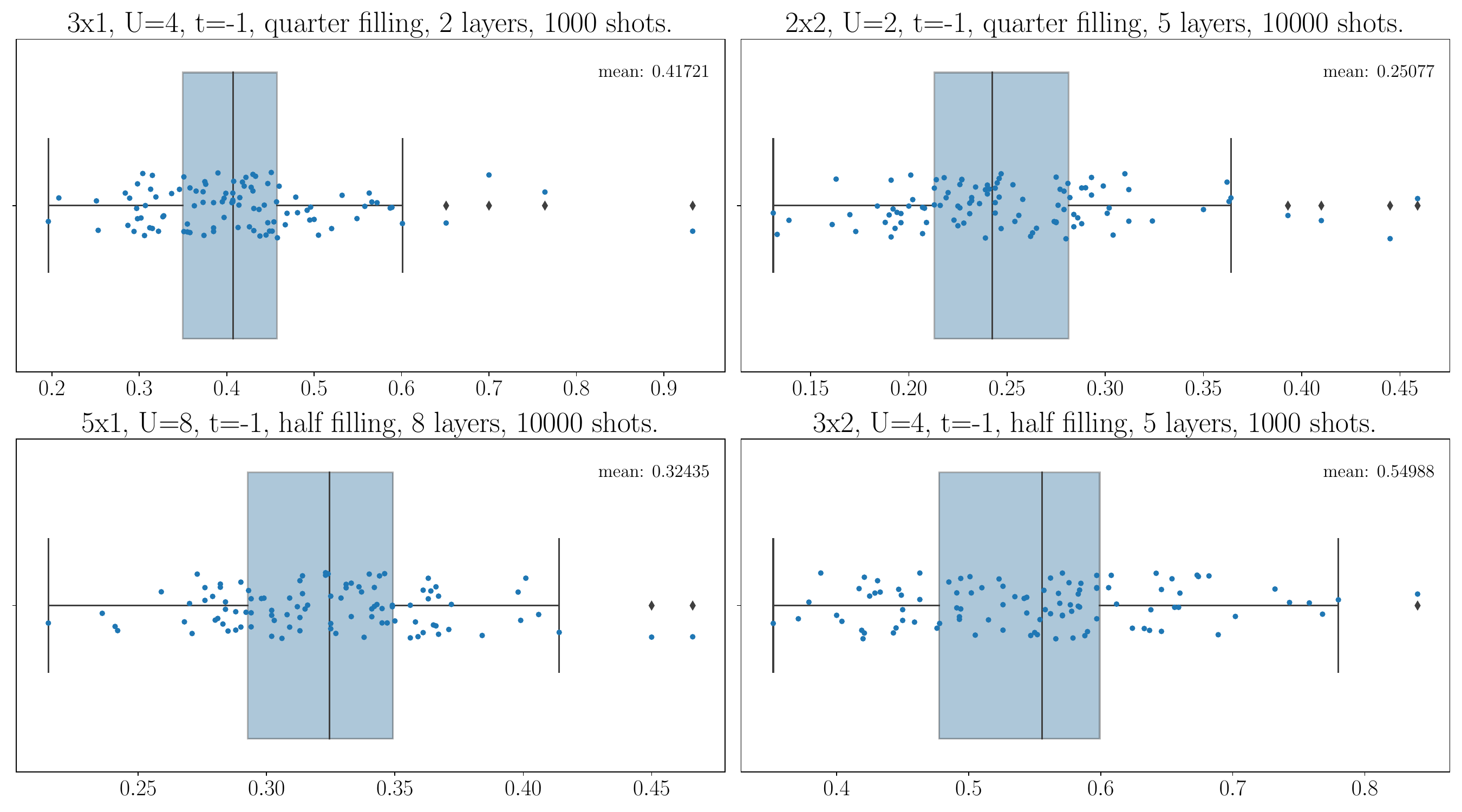}
    \caption{Gradient sweeping: boxplots of best step size found for 100 random points on four instances that minimises the error with the exact gradient. The x axis denotes the best step size found for each of the 100 random points. From this we use a step size of 0.4 in numerical simulations, as the average of the four means to 1 decimal place.}
    \label{fig:gradient_over_instances}
\end{figure}

Over the four hyperparameter sweeping instances, we sampled 100 random points (i.e. a random set of parameters), and swept over 999 step sizes from 0.001 to 0.999. Based on this, we choose a step size of 0.4, see \cref{fig:gradient_tradeoff} for a plot of this trade-off for a particular instance, and \cref{fig:gradient_over_instances} for a comparison of the optimal step sizes for the 4 sweeping instances considered. We arrive at the value of 0.4 by taking the average of the best step size over these 4 instances to one decimal place: $\frac{1}{4}(0.41721 + 0.25077 + 0.32435 + 0.54988) \approx 0.4$. Note that choosing an arbitrary value (such as 0.01) could result in all gradient-based optimisers performing poorly.

\paragraph{Simultaneous Perturbation. \\}

Let $\Delta$ be a random vector with entries uniformly in $\{\pm 1 \}$. The gradient arising in SPSA can be written as
\[
\nabla f(\theta)_k = \frac{f(\theta + \epsilon \Delta) - f(\theta - \epsilon \Delta) }{2 \epsilon \Delta_k}.
\]

This takes two function evaluations to calculate, independently of the number of parameters. Inspired by SPSA, we employ simultaneous perturbation as a gradient subroutine. We set the step size here to be 0.15, as in e.g. \cite{cade2020strategies}.

\subsection{Hyperparameter sweeping}

Many of the optimisers we consider have hyperparameters, such as the learning rate $\eta$ in a simple gradient descent algorithm:
\[
\theta_{t+1} = \theta_t - \eta \nabla E(\theta_t).
\]

For each optimiser with hyperparameters, we conduct hyperparameter sweeping using the \texttt{Optuna} package \cite{optuna_2019}. We use the following search spaces depending on the type of hyperparameter we are considering, for example:
\begin{itemize}
    \item Categorical: choose randomly from the options.
    \item Positive float: choose uniformly randomly from two orders of magnitude above and below the default value, on a log scale.
    \item Float between 0 and 1: uniformly random between 0 and 1.
\end{itemize}

We then define a function that takes the hyperparameters as input and outputs the best value found after running for only 100 calls of the cost function. We only allow a short number of calls to prioritise hyperparameters that `get off to a good start', and to avoid choosing hyperparameters arbitrarily if they all reach the ground state (i.e. based on numerical error). We pass this function to \texttt{Optuna} and allow it 1,000 evaluations to try and find the best hyperparameters within the defined search space. We then take the best values for each of the four instances and the default hyperparameters to run on all instances, giving at most 5 sets of hyperparameters to run on each instance per optimiser. In \cref{app:optimisers} we include plots for each optimiser showing the proportion of times that each set of hyperparameters performed the best, and we find that the sensitivity to hyperparameter choice is highly optimiser dependent.

\subsection{Termination criteria and data recorded}

We allow each run of a single optimiser on a given instance to use 5000 calls of the cost function (note each call includes the number of shots, so we actually run the quantum circuit at most $5000 \times \texttt{nshots} \times \texttt{nmeas}$ times (where $\texttt{nmeas}$ is the number of measurements required to get a full single energy evaluation), or for no more than 1 hour, whichever happens sooner. We record every call of the cost function by wrapping it and saving to a \texttt{CSV} file. Also note that we record both the exact value and the value from the statistical cost function, but the optimisers only have access to the statistical cost function. At each cost function call we record (see \cref{tab:data_collected}):

\begin{itemize}
    \item The noisy cost function value.
    \item The exact energy.
    \item The standard error.
    \item The parameters up to 6 decimal places.
    \item The number of calls made.
    \item The total number of measurements made (i.e. taking into account number of shots and having to measure the different terms separately).
    \item The time taken.
\end{itemize}

\begin{table}[H]
    \centering
    \begin{tabular}{|l|l|l|l|l|l|}
    \hline
        iter & value & params & exact value & nmeas & time \\ \hline
        1 & -1.725 & [0.5, 0.5, 0.5, 0.5, 0.5, 0.5] & -1.766045 & 3000 & 0.0 \\ \hline
        2 & -1.531 & [0.35, 0.65, 0.65, 0.35, 0.65, 0.65] & -1.612306 & 6000 & 0.000909\\ \hline
        3 & -1.409 & [0.65, 0.35, 0.35, 0.65, 0.35, 0.35] & -1.515276 & 9000 & 0.001617 \\ \hline
        4 & -1.616 & [0.586272, 0.693444, 0.693444, 0.586272, 0.693444, 0.413728] & -1.600992 & 12000 & 0.002409 \\ \hline
    \end{tabular}
    \caption{Example table of data collected.}
    \label{tab:data_collected}
\end{table}

In total this generated approximately $37$GB of data, which is hosted using Zenodo \cite{dataset}.

\subsection{Optimisers}
\label{subsec:optimisers}

\renewcommand{\arraystretch}{1.5}
\setlength{\tabcolsep}{4pt}
\begin{table}[H]
    \centering
    \begin{tabular}{|c||c|c|c|c|c|} \hline
    Optimiser &  Black-box & \makecell{Gradient \\ based} & \makecell{Number of \\ hyperparameters} & Implementation & References  \\[2ex] \hline \hline
    \hyperref[subsec:hill_climber]{Hill Climber} & Yes & No & 2 & In house. & \cite{Luke2013Metaheuristics} \\
    \hyperref[subsec:spsa]{SPSA} & Yes & Yes & 5 & In house. & \cite{spall1998overview, spall98implementation} \\
    \hyperref[subsec:bayes_mgd]{BayesMGD} & Uses uncertainty & No & 7 & In house. & \cite{stanisic2022observing, sung2020using}   \\
    \hyperref[subsec:coord_descent]{Coordinate Descent} & No & No & 1 & In house. & \cite{cade2020strategies, nakanishi2020sequential, tseng2001convergence, ostaszewski2021structure, parrish2019jacobi}   \\
    \hyperref[subsec:lam_mu]{$\lambda + \mu$} & Yes & No & 11 &  \cite{deap} & \cite{Luke2013Metaheuristics}\\
    \hyperref[subsec:pso]{PSO} & Yes & No & 6 & \cite{deap} & \cite{Luke2013Metaheuristics}\\
    \hyperref[subsec:cmaes]{CMAES} & Yes & No & 1 & \cite{deap} & \cite{deap, Luke2013Metaheuristics, hansen2001completely, hansen2016cma}\\
    \hyperref[subsec:grad_descent]{Gradient Descent} & Yes & Yes & 1 & In house. & \cite{ruder2016overview}  \\
    \hyperref[subsec:Momentum]{Momentum} & Yes & Yes & 3 & In house. & \cite{ruder2016overview}   \\
    \hyperref[subsec:adam]{Adam} & Yes & Yes & 4 &  In house. & \cite{ruder2016overview} \\
    \hyperref[subsec:adadelta]{AdaDelta} & Yes & Yes & 1 & In house. & \cite{ruder2016overview} \\
    \hyperref[subsec:nelder_mead]{Nelder-Mead}& Yes & No & 2 &  \cite{scipy} & \cite{gao2012implementing, nelder1965simplex} \\
    \hyperref[subsec:cobyla]{COBYLA} & Yes & No & 0 & \cite{scipy} & \cite{powell1994direct}  \\
    \hyperref[subsec:powell]{Powell} & Yes & No & 0 & \cite{scipy} &  \cite{ powell2007view, kochenderfer2019algorithms}  \\
    \hyperref[subsec:bfgs]{BFGS} & Yes & Yes & 0 & \cite{scipy} & \cite{kochenderfer2019algorithms} \\
    \hyperref[subsec:l_bfgs_b]{L-BFGS-B} & Yes & Yes & 0 & \cite{scipy} & \cite{kochenderfer2019algorithms}\\\
    \hyperref[subsec:tnc]{TNC} & Yes & Yes & 0 & \cite{scipy} & \cite{nocedal1999numerical, kochenderfer2019algorithms}\\
    \hyperref[subsec:slsqp]{SLSQP} & Yes & Yes & 0 & \cite{scipy} & \cite{nocedal1999numerical} \\
    \hyperref[subsec:cg]{CG} & Yes & Yes & 0 & \cite{scipy} & \cite{nocedal1999numerical, kochenderfer2019algorithms}\\
    \hyperref[subsec:newton_cg]{Newton CG} & Yes & Yes & 0 & \cite{scipy} & \cite{nocedal1999numerical, kochenderfer2019algorithms} \\
\hline
    \hyperref[app:qng]{Natural Gradient}& No & Yes & 1 & In house. & \cite{stokes2020quantum} \\\hline
\end{tabular}
    \caption{Overview of optimisers considered in this study, see \cref{app:optimisers} for more in-depth notes. We distinguish Natural Gradient from the others as due to ansatz considerations we consider it separately in \cref{subsec:results:qng} and \cref{app:qng}. We also highlight that there is some arbitrariness in the number of hyperparameters taken for a given optimiser, and that for the gradient-based optimisers we use both finite differences and simultaneous perturbation as gradient subroutines.}
    \label{tab:optimiser_summary}
\end{table}

 \section{Results}

\label{sec:results}

\subsection{Exact cost functions}
\label{subsec:exact cost functions}

\begin{figure}[htbp!]
    \centering
    \includegraphics[width=\textwidth]{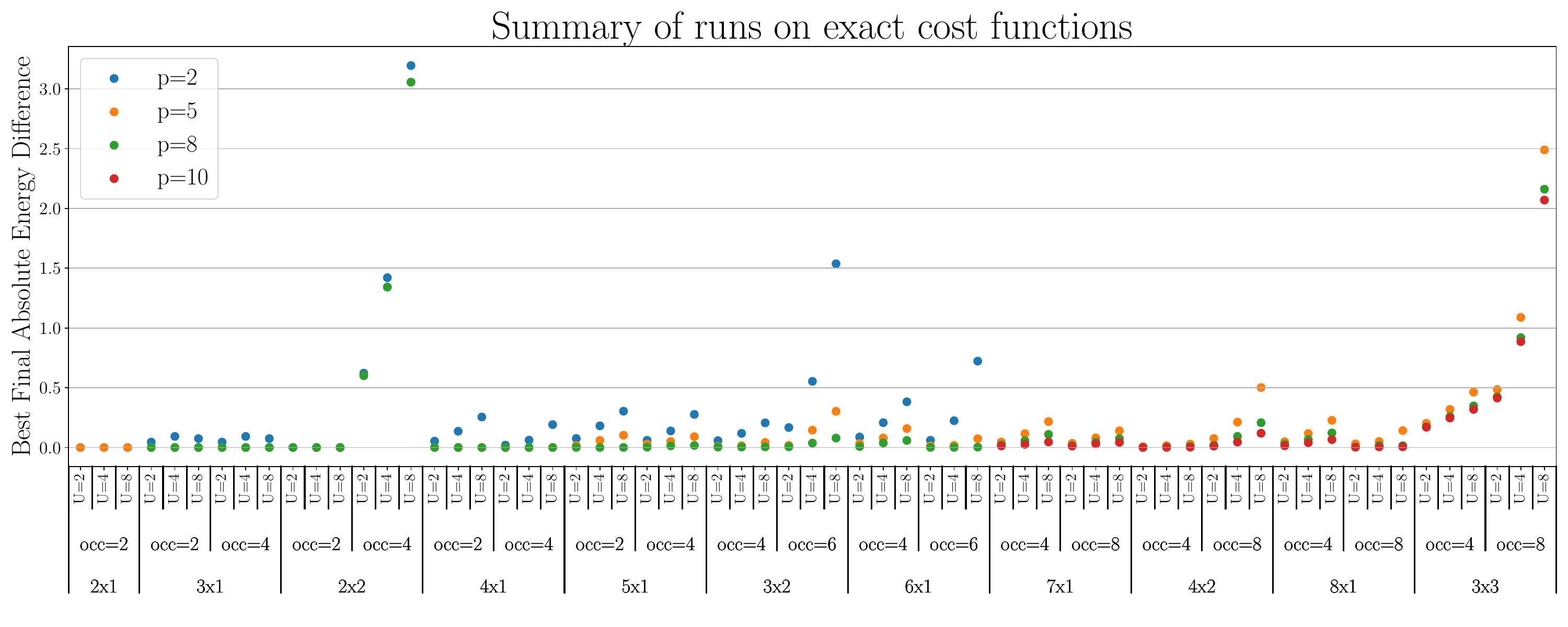}
    \caption{We ran BFGS, L-BFGS-B, Nelder-Mead, Powell and SLSQP (from \texttt{scipy} \cite{scipy}) on exact cost functions for the instances considered, to determine the expressivity of the ansätze in the noiseless case. We plot the closest final energy to the ground state. Here ``occ'' refers to the occupation number, i.e. the sum of spin up and spin down particles (which we set to be the same) --  these values correspond to quarter and half filling.}
    \label{fig:exact_runs}
\end{figure}

We first show results for running VQE on the exact cost functions, that is, for the instances considered above but instead of sampling a number of shots statistically, using a cost function that returns the exact energy.  We ran BFGS, L-BFGS-B, Nelder-Mead, Powell, and SLSQP from the \texttt{scipy} \cite{scipy} package (in \cite{cade2020strategies} it was found that BFGS performed well on the exact cost function), and then plotted the best for each run. This serves as a test of the ansatz, and we see in \cref{fig:exact_runs} that on the whole, for sufficient depth the optimisers were able to reach the ground state. There are certain anomalies, for example $2 \times 2$ half filling, which are as a result of a degenerate ground state (see \cite{cade2020strategies} for a brief discussion on this).

\subsection{Statistical cost functions}

\begin{figure}[h!]

\centering
     \begin{subfigure}[b]{0.49\textwidth}
         \centering
\includegraphics[width=\textwidth]{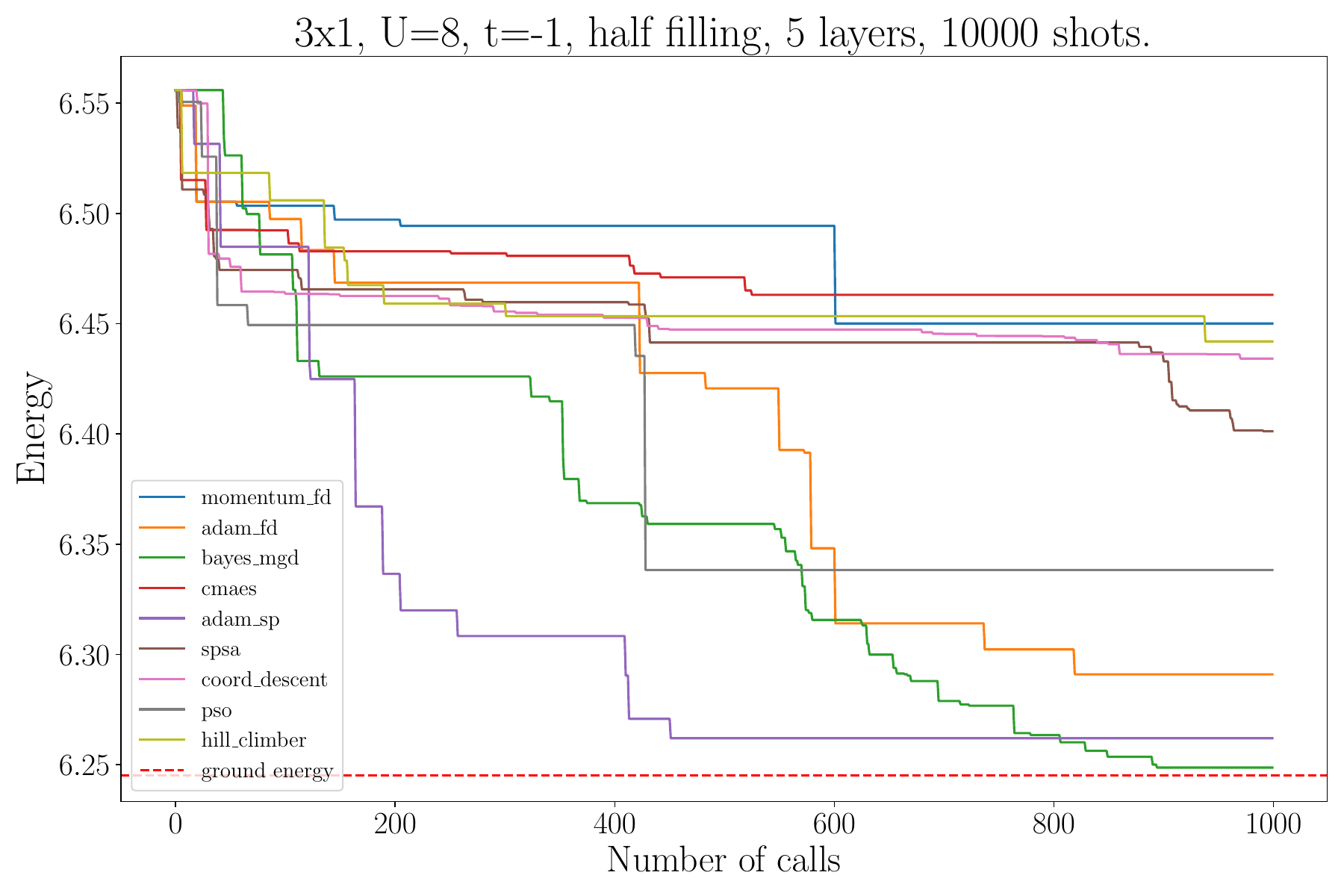}
         \caption{}
         \label{subfig:runs1}
     \end{subfigure}\hfill
     \begin{subfigure}[b]{0.49\textwidth}
         \centering
\includegraphics[width=\textwidth]{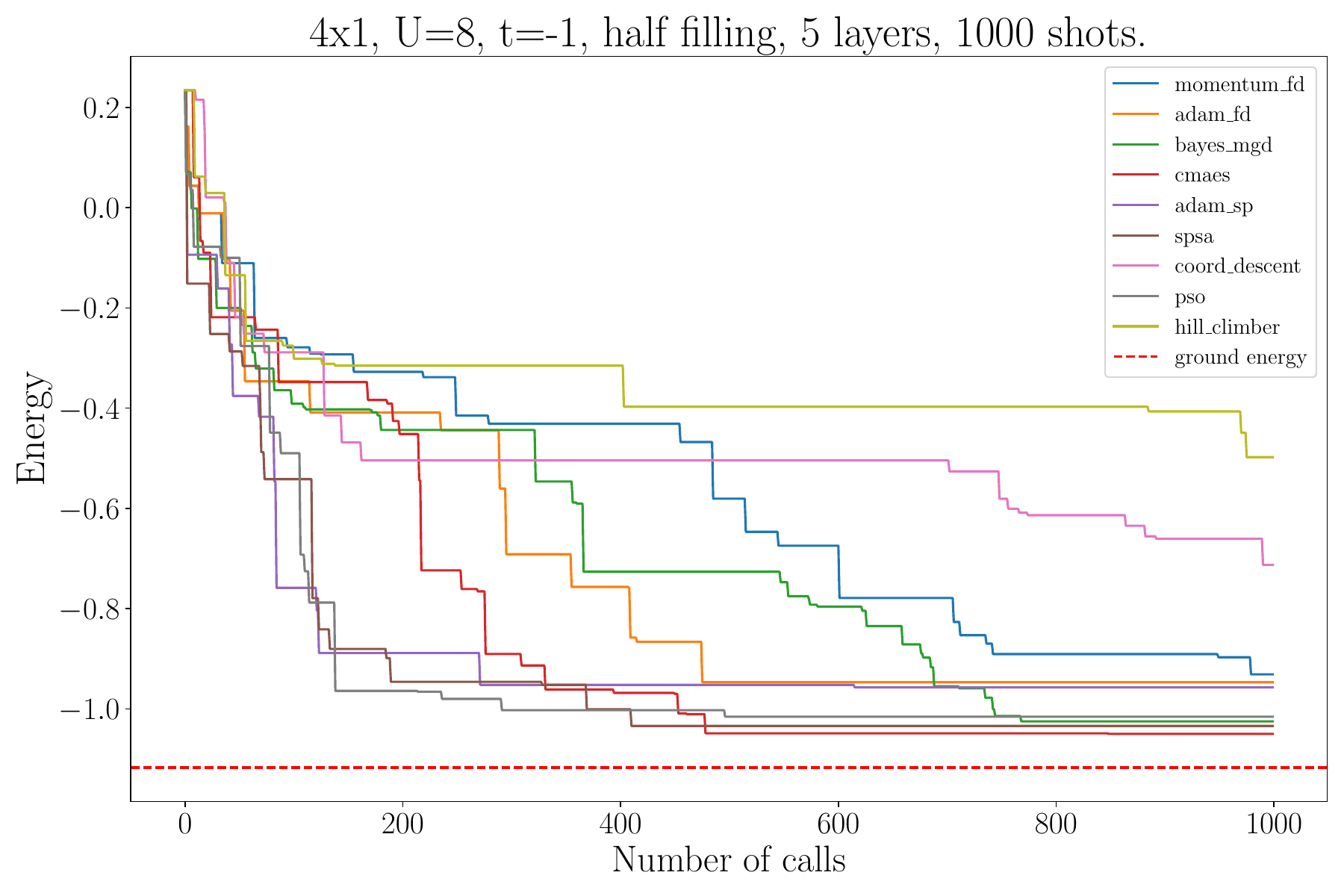}
        \caption{}
        \label{subfig:runs2}
     \end{subfigure}
     \caption{Plots of individual runs, displaying the exact energy throughout a run (although the optimisers only have access to a statistical cost function).}

     \label{fig:individual_runs}
   
     \end{figure}

Here we display results when the cost function takes statistical noise into account, some plots of individual runs are shown n \cref{fig:individual_runs}. Note that although the optimisers themselves only have access to statistical cost functions, in \cref{subfig:runs1} and \cref{subfig:runs2} we plot the best \text{exact} energy found so far, (so that the graphs are always monotonically decreasing, and can never drop below the ground energy).

As we collected data over 372 instances, it is infeasible to study all of the plots of individual runs to extract insights. In order to gain an overall picture of optimiser performance across the spread of instances, there are two natural metrics to focus on:
\begin{enumerate}[(1)]
    \item The difference in final energy with the ground energy.
    \item The number of calls required to get within some tolerance (e.g. 0.01) of the ground energy.
\end{enumerate}

\subsubsection{Comparing accuracy}
\begin{figure}[htbp!]
    \centering
    \includegraphics[width=0.9\textwidth]{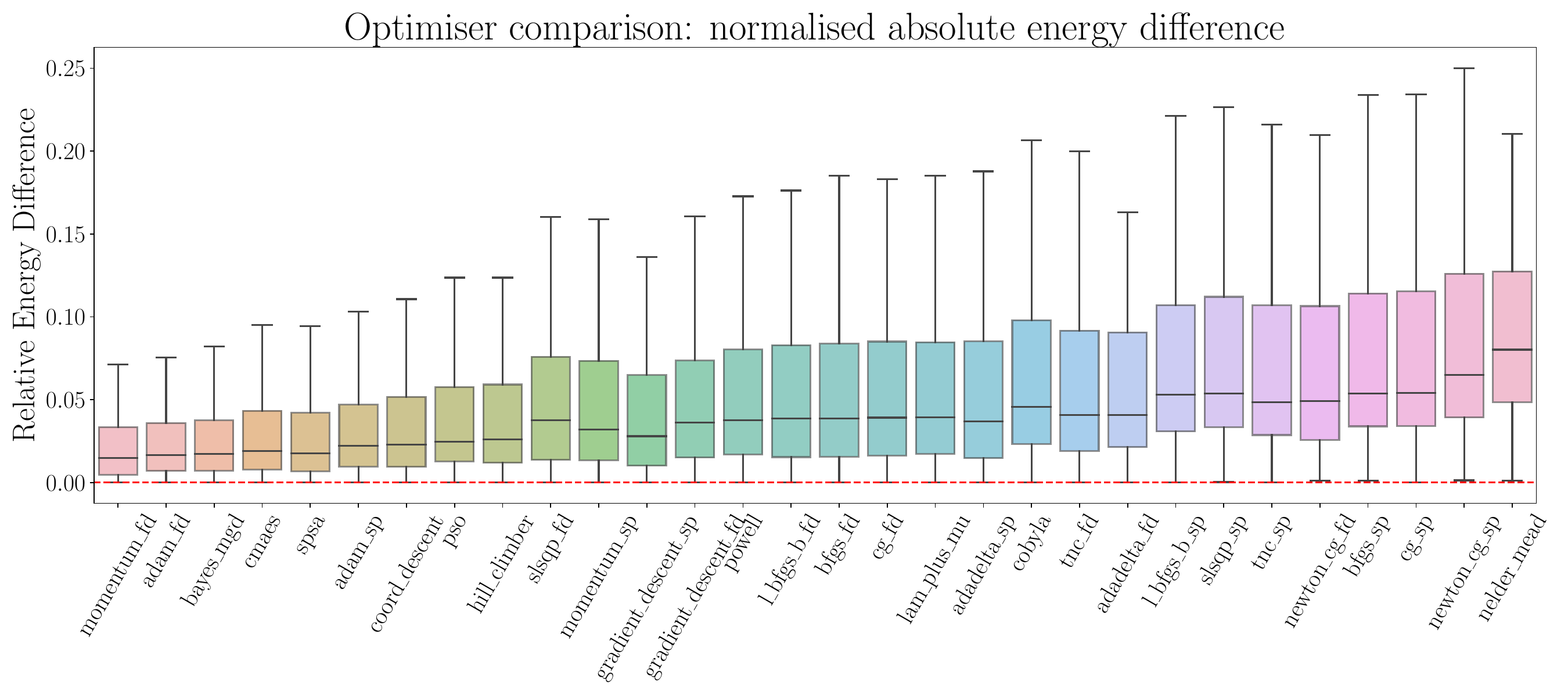}
    \caption{This plot summarises data collected over all optimisers and all instances (372 in total). For each optimiser, we take the best final energy (over different runs, e.g. for multiple hyperparameter settings), calculate the difference with the ground energy and normalise it (by dividing by $mn$ for an $m \times n$ grid). For each optimiser an underlying boxplot is shown (with outliers removed), ordered by their mean. The suffixes `fd' and `sp' respectively refer to using finite difference (with step size 0.4) and simultaneous perturbation (with step size 0.15) as gradient functions.}
    \label{fig:boxplot all instances}
\end{figure}

\begin{figure}[htbp!]
    \centering
    \includegraphics[width=0.8\textwidth]{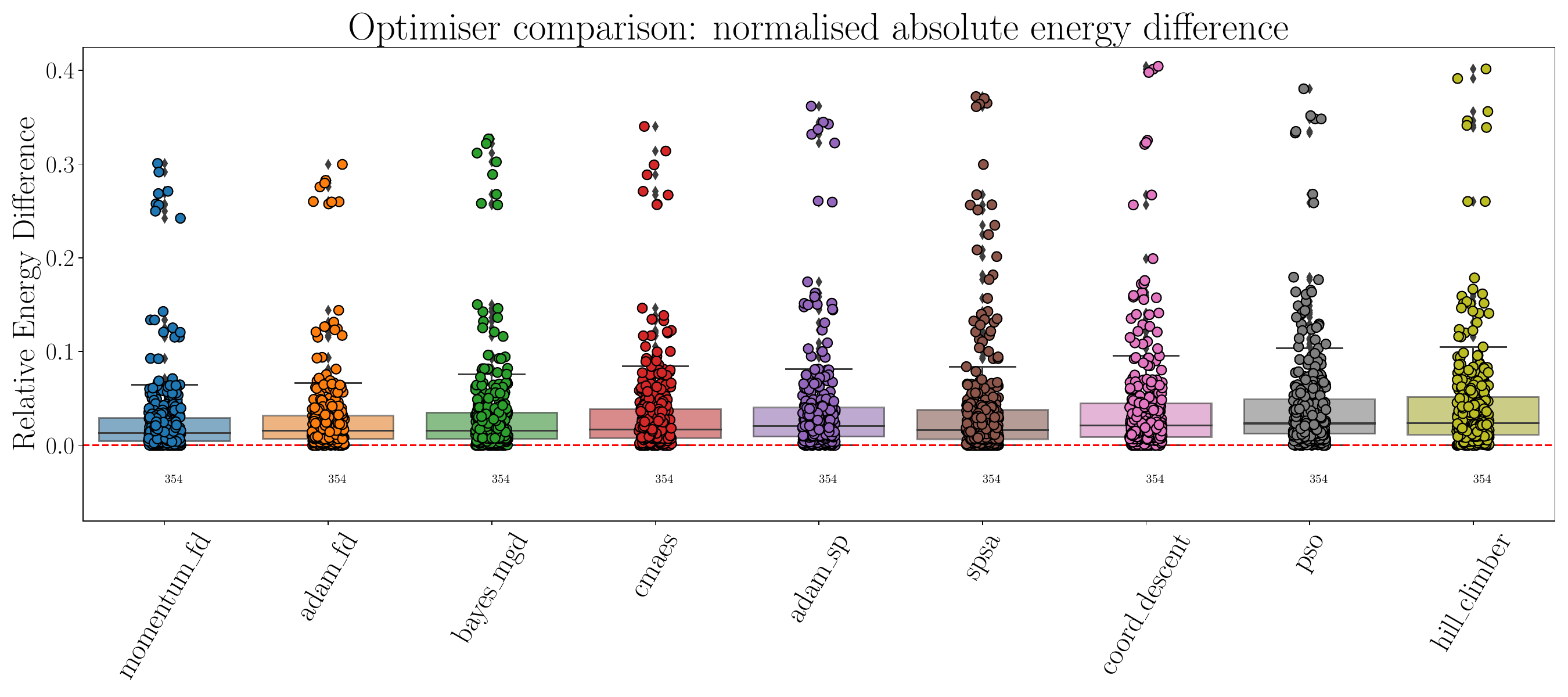}
    \caption{This plot is as in \cref{fig:boxplot all instances}, but only displaying the best optimisers, removing the $2 \times 2$ half-filling anomalous case (leaving 354 instances to run on), and overlaying the individual points over the boxplots. The number below each optimiser corresponds to the number of data points (instances).}
    \label{fig:final_errors_boxplot_exclude2x2_ops}
\end{figure}

To tackle the first of these, in \cref{fig:boxplot all instances} we use a boxplot to display the best energies achieved for each optimiser over all 372 instances studied (see \cref{subsec:instances}). The optimisers are ordered by performance, and we see that by this metric the top performers were Momentum and Adam with finite difference as gradient subroutine (with step size 0.4), BayesMGD, CMAES and SPSA. Similarly, \cref{fig:final_errors_boxplot_exclude2x2_ops} shows the average final errors for a selection of the better performing optimisers, whilst also excluding the anomalous $2 \times 2$ half-filling case (see the comment at the beginning of this section). Note that these figures use data that is split evenly between taking 1,000 and 10,000 shots.

\begin{figure}[htbp!]
\centering

\begin{subfigure}{0.5\textwidth}
    \centering
    \includegraphics[width=\textwidth]{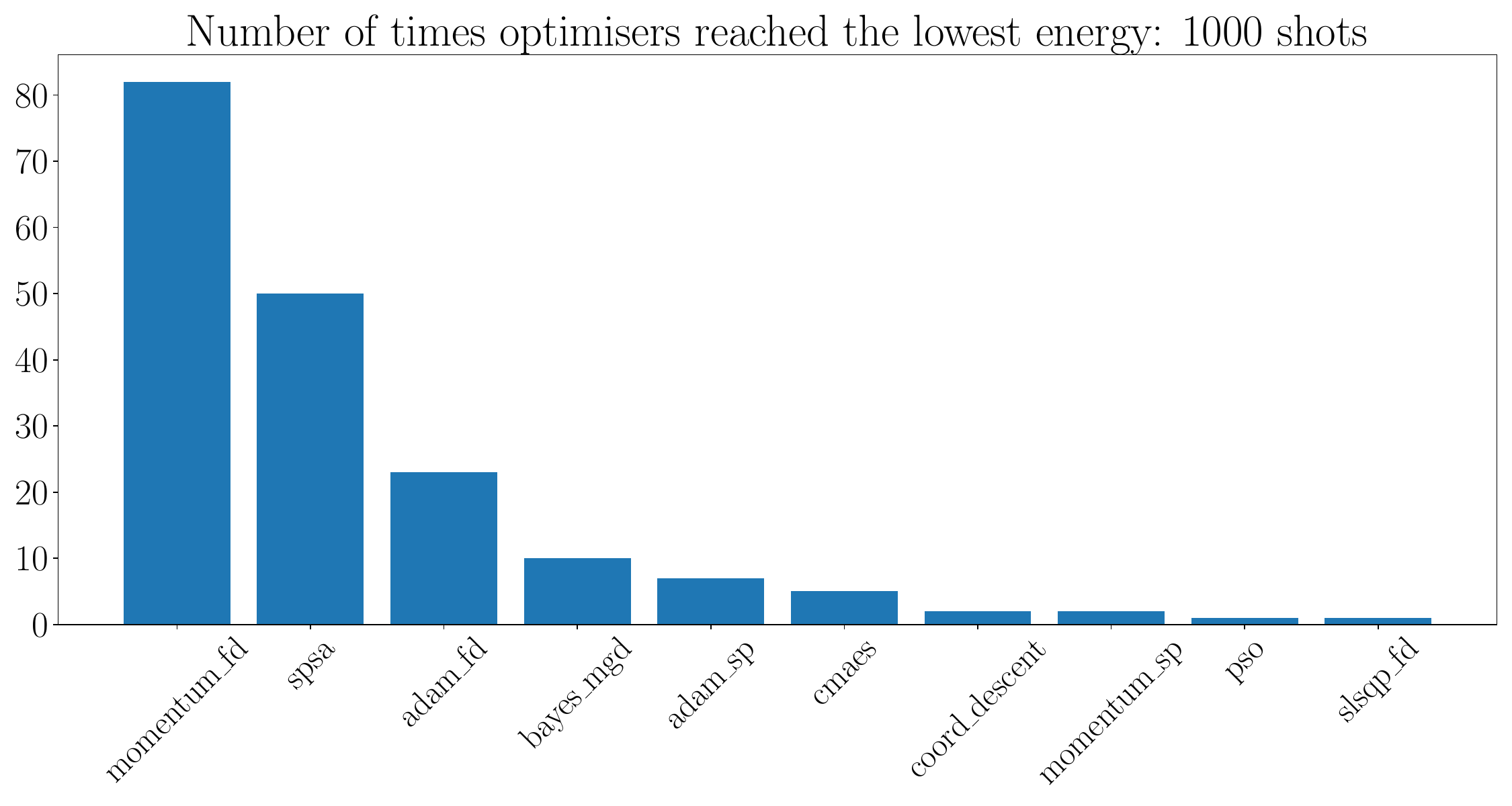}
\end{subfigure}~
\begin{subfigure}{0.5\textwidth}
    \centering
    \includegraphics[width=\textwidth]{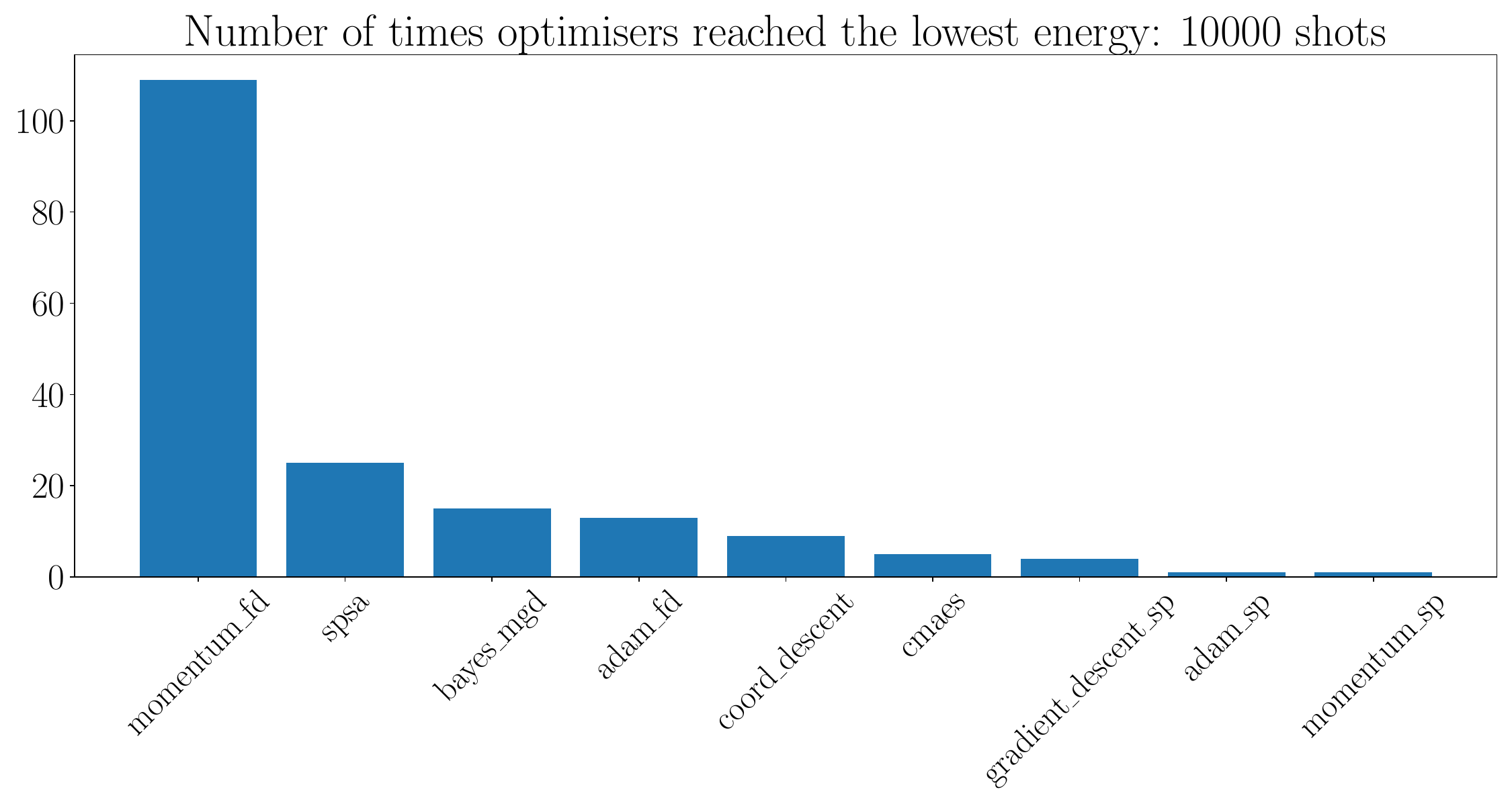}
\end{subfigure}

\begin{subfigure}{0.5\textwidth}
    \centering
    \includegraphics[width=\textwidth]{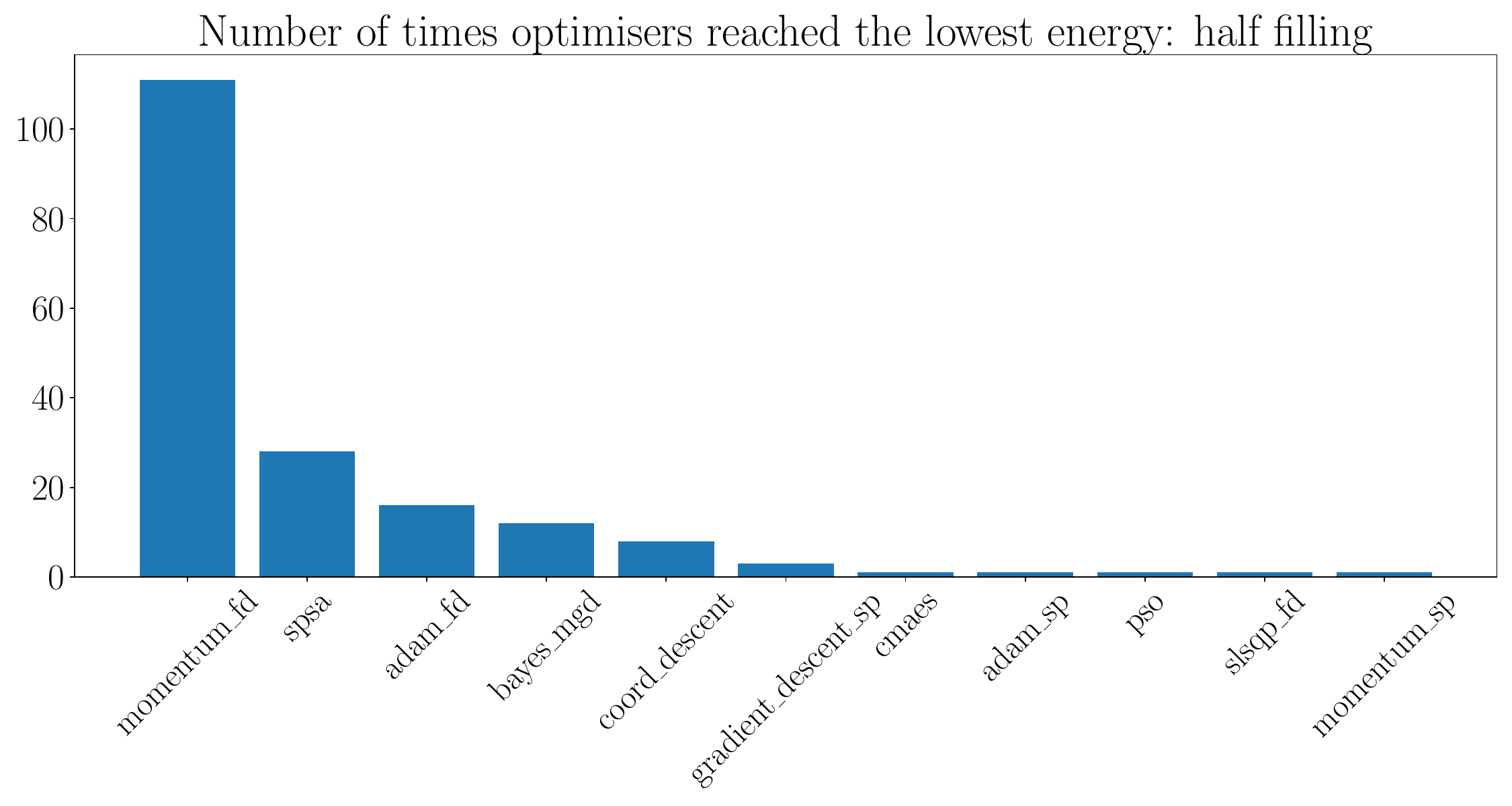}
\end{subfigure}~
\begin{subfigure}{0.5\textwidth}
    \centering
    \includegraphics[width=\textwidth]{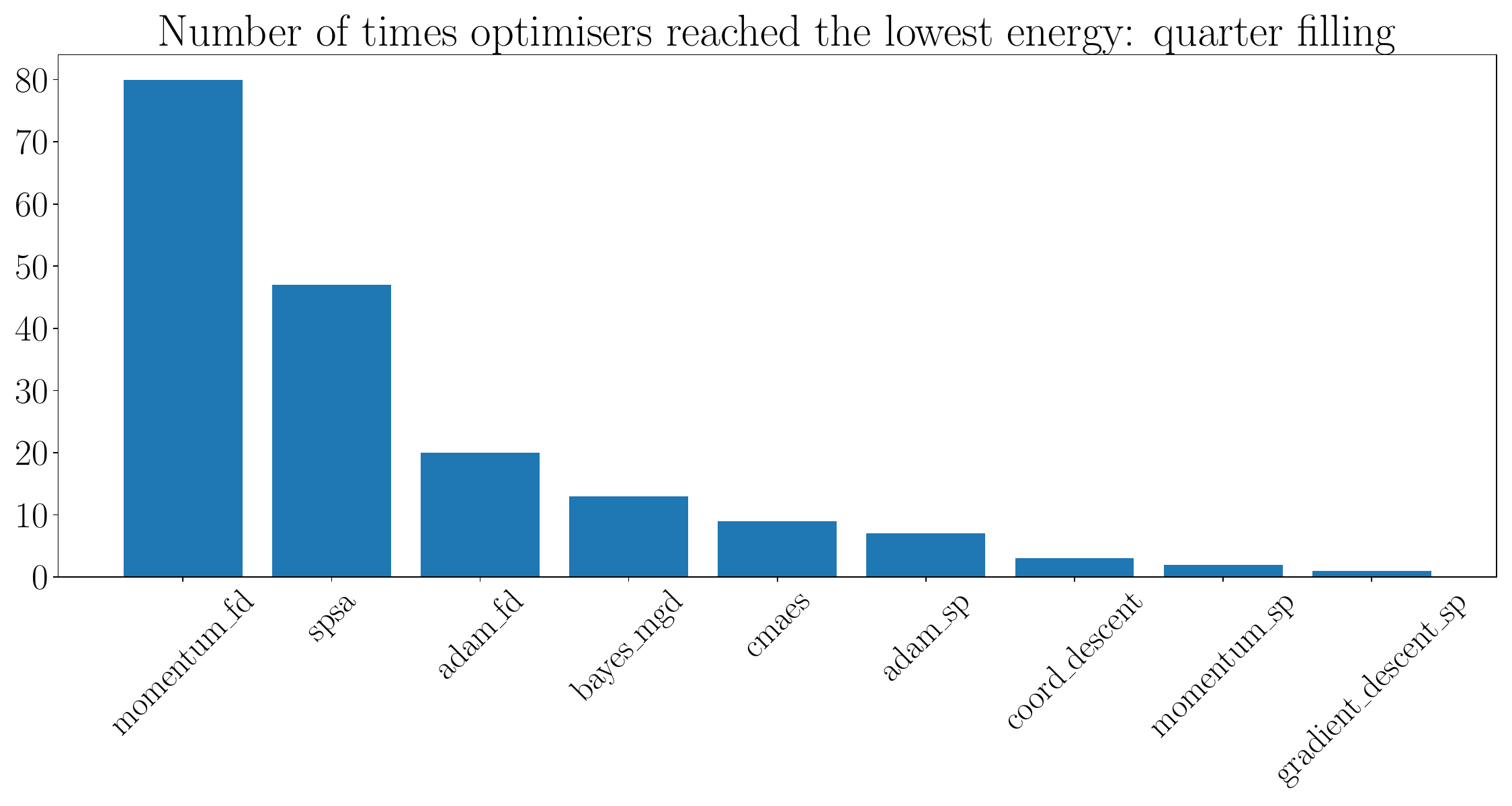}
\end{subfigure}

\caption{These barplots display the number of times each optimiser outperformed all the others, for different types of instances. We first rounded the final values to 5 decimal places to avoid numerical error. We see that Momentum with finite difference as a gradient found the lowest energy the most times, particularly in the 10,000 shot or half filling regimes.}
\label{fig:winners}

\end{figure}

The barplots in \cref{fig:winners} displays on how many instances each optimiser outperformed all of the others. We first rounded the final values to 5 decimal places to counteract any numerical imprecision. It is clear that Momentum with finite differences performed the best on the majority of instances, with this superiority especially apparent when using a higher number of shots and for half filling.

\subsubsection{Comparing number of calls}

\begin{figure}[htbp!]
\centering

\begin{subfigure}{\textwidth}
    \centering
    \includegraphics[width=\textwidth]{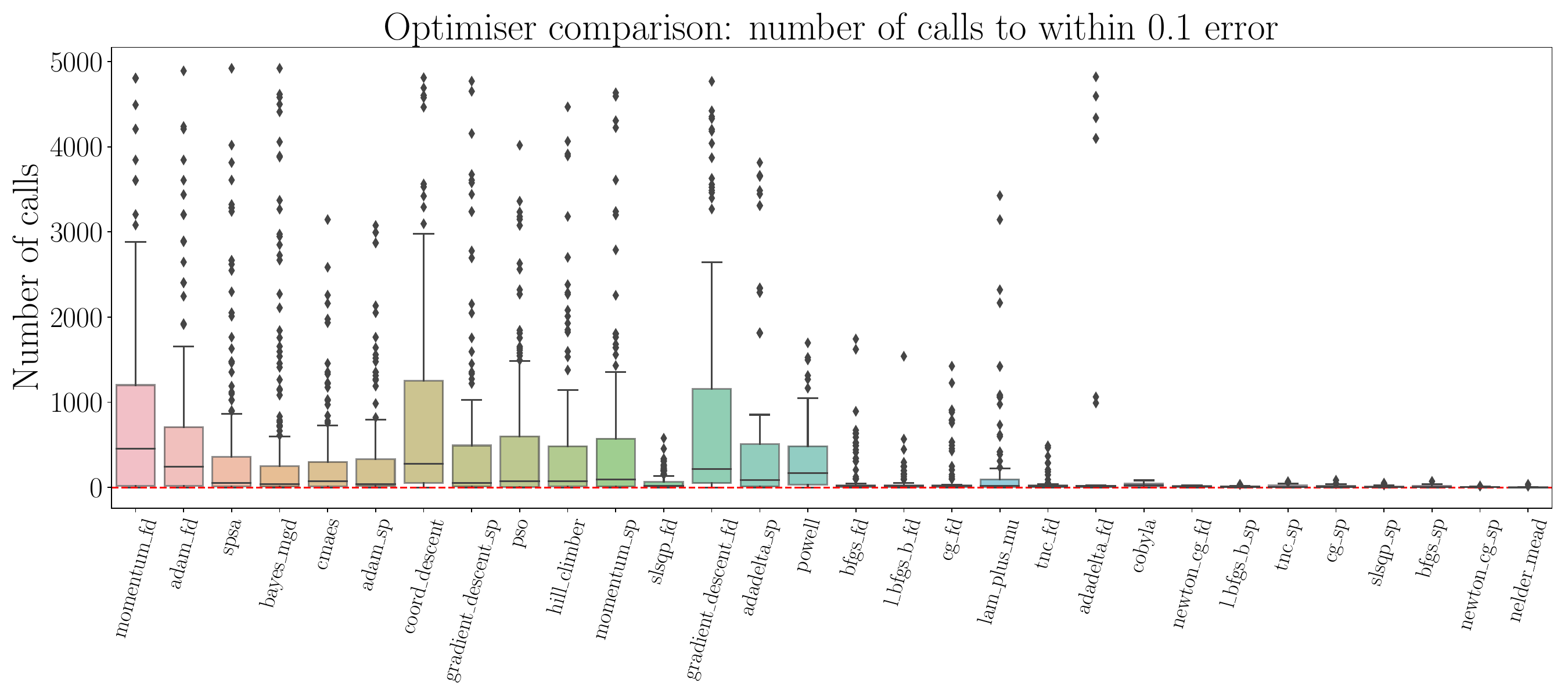}
\end{subfigure}

\begin{subfigure}{\textwidth}
    \centering
    \includegraphics[width=\textwidth]{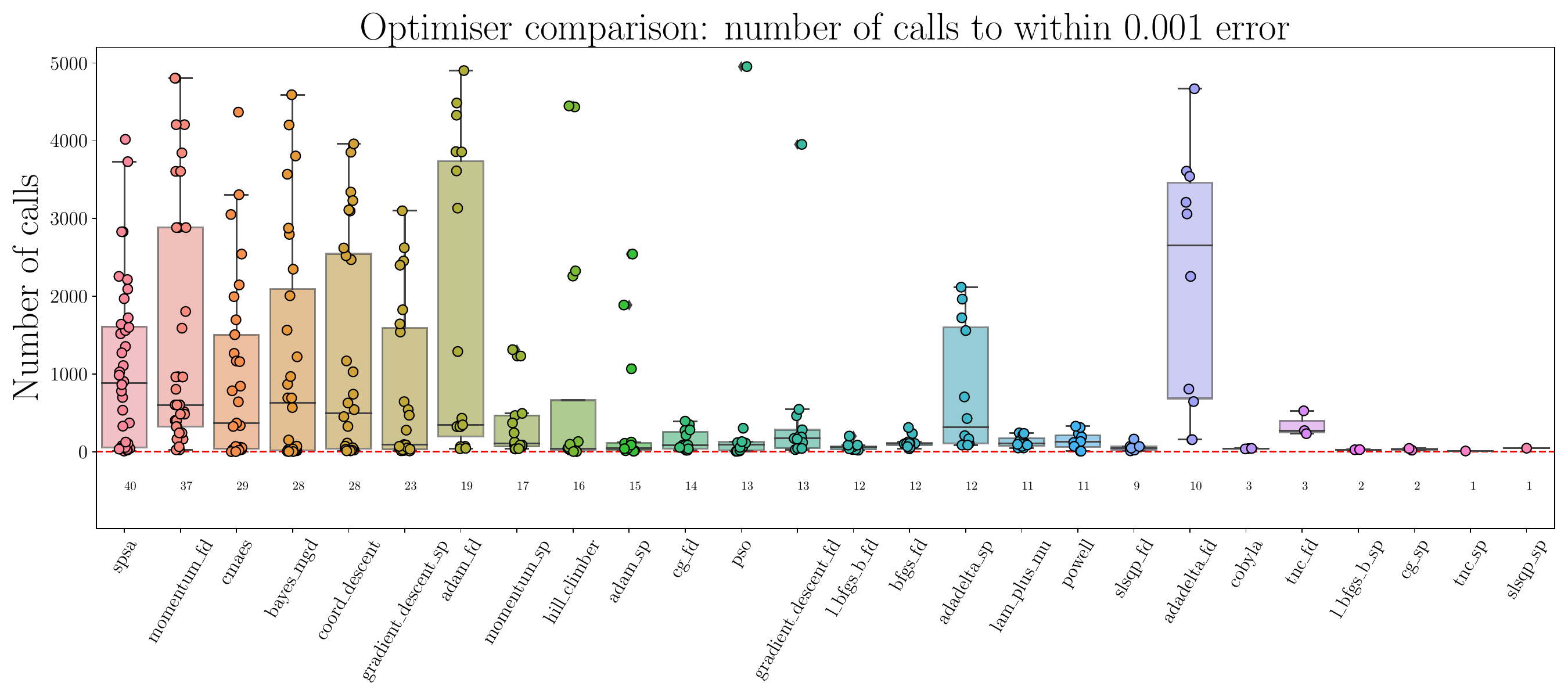}
\end{subfigure}

\begin{subfigure}{\textwidth}
    \centering
    \includegraphics[width=\textwidth]{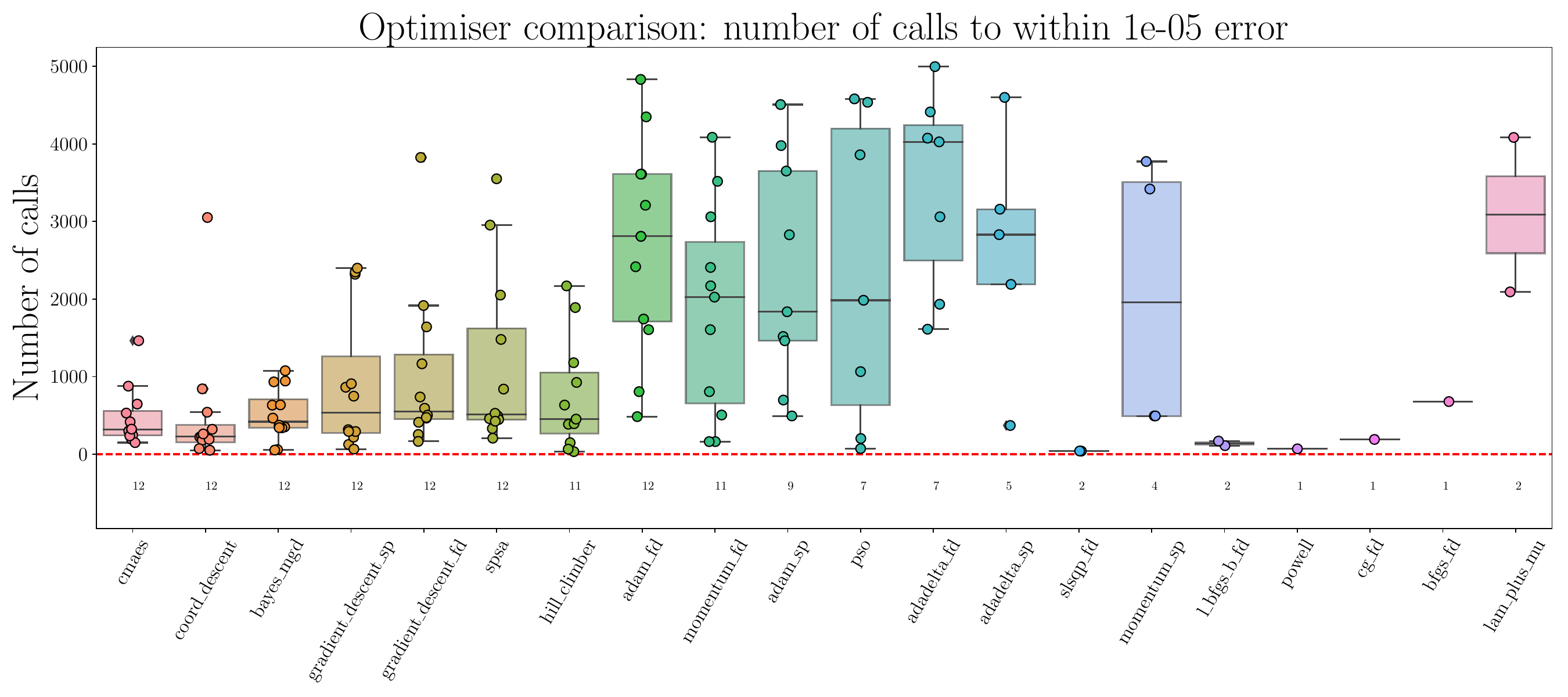}
\end{subfigure}

\caption{Comparison of number of calls required to reach a certain error with the ground state energy. The number under each optimiser refers to the number of instances (out of 372) that the optimiser was able to reach this tolerance, and the optimisers are ordered in terms of performance (specifically, by the number of times they were able to get within the given error, followed by the minimum average number of calls).}
\label{fig:calls_to_within_error}

\end{figure}

We also consider the second metric mentioned above, namely the number of calls required to reach a certain tolerance of error, in \cref{fig:calls_to_within_error}. Under this metric, we see some variety in the best-performing optimiser. For example, we see that finite difference methods can use significantly more calls to reach a certain tolerance than other methods such as SPSA or CMAES.

\begin{figure}[htbp!]
    \centering
    \includegraphics[width=0.8\textwidth]{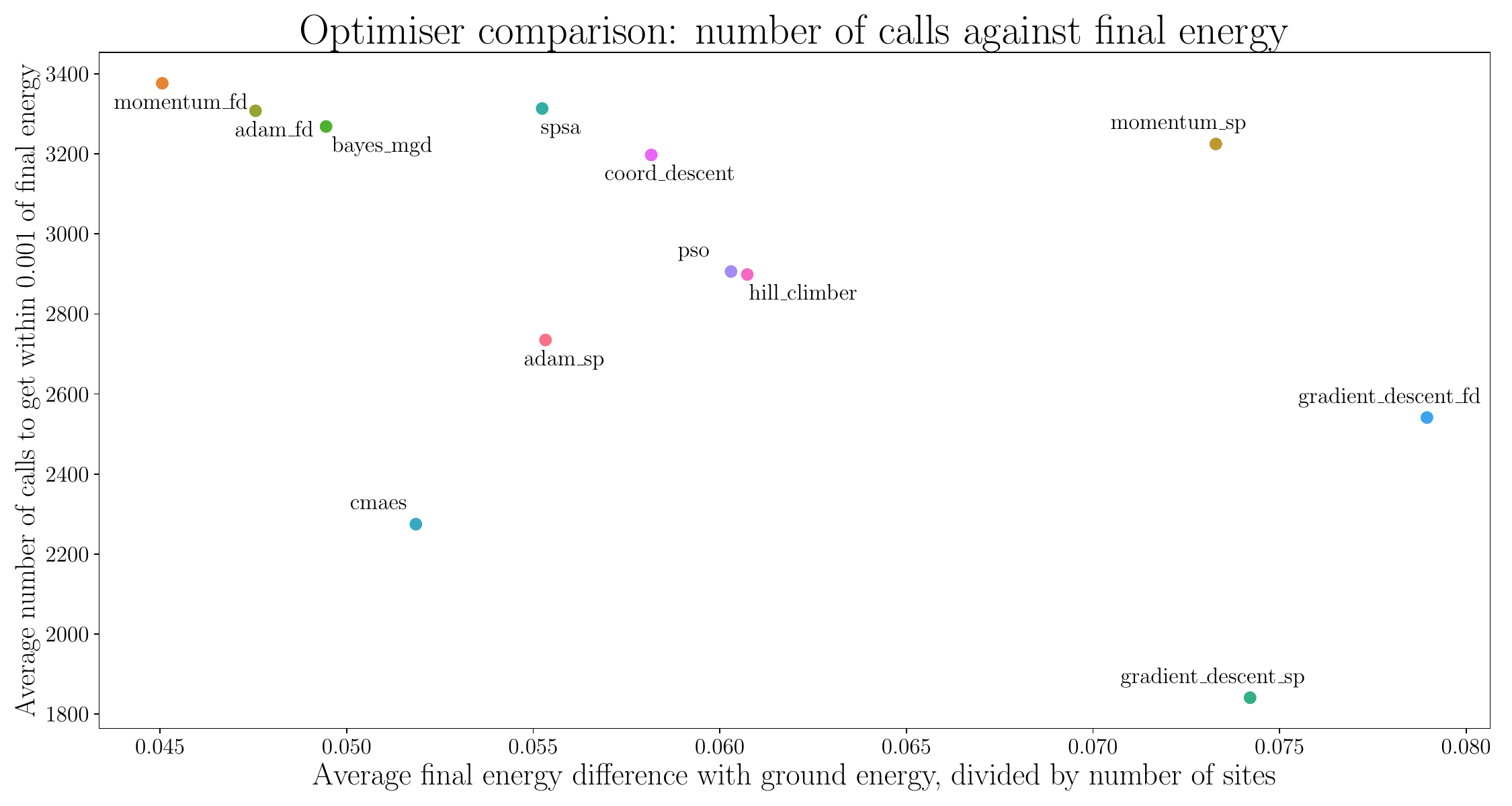}
    \caption{For each instance, we consider on average how many calls it took each optimiser to get within 0.001 of their final energy over the run (not the ground energy), and also the average final energy obtained. For example, we see that Momentum and Adam with finite differences achieves a good final energy, yet they also can take a lot of calls to get there. CMAES seems to offer a good balance between these two perspectives.}
    \label{fig:combined_energy_calls}
\end{figure}

In \cref{fig:combined_energy_calls} we combine the two main metrics discussed above of final energy obtained and number of calls to get within a fixed tolerance of the ground energy. Specifically, we plot the average number of calls to get within 0.001 of the final energy that the optimiser obtained on that run. For example, we can see that the average final energy obtained is similar for SPSA and Adam with simultaneous perturbation as a gradient subroutine, but the latter appears to use fewer calls in general. We also see that CMAES may offer a balanced compromise between these two metrics.

\subsubsection{Comparing gradients}

\begin{figure}[h!]
\centering
     \begin{subfigure}[b]{\textwidth}
         \centering
         \includegraphics[width=0.7\textwidth]{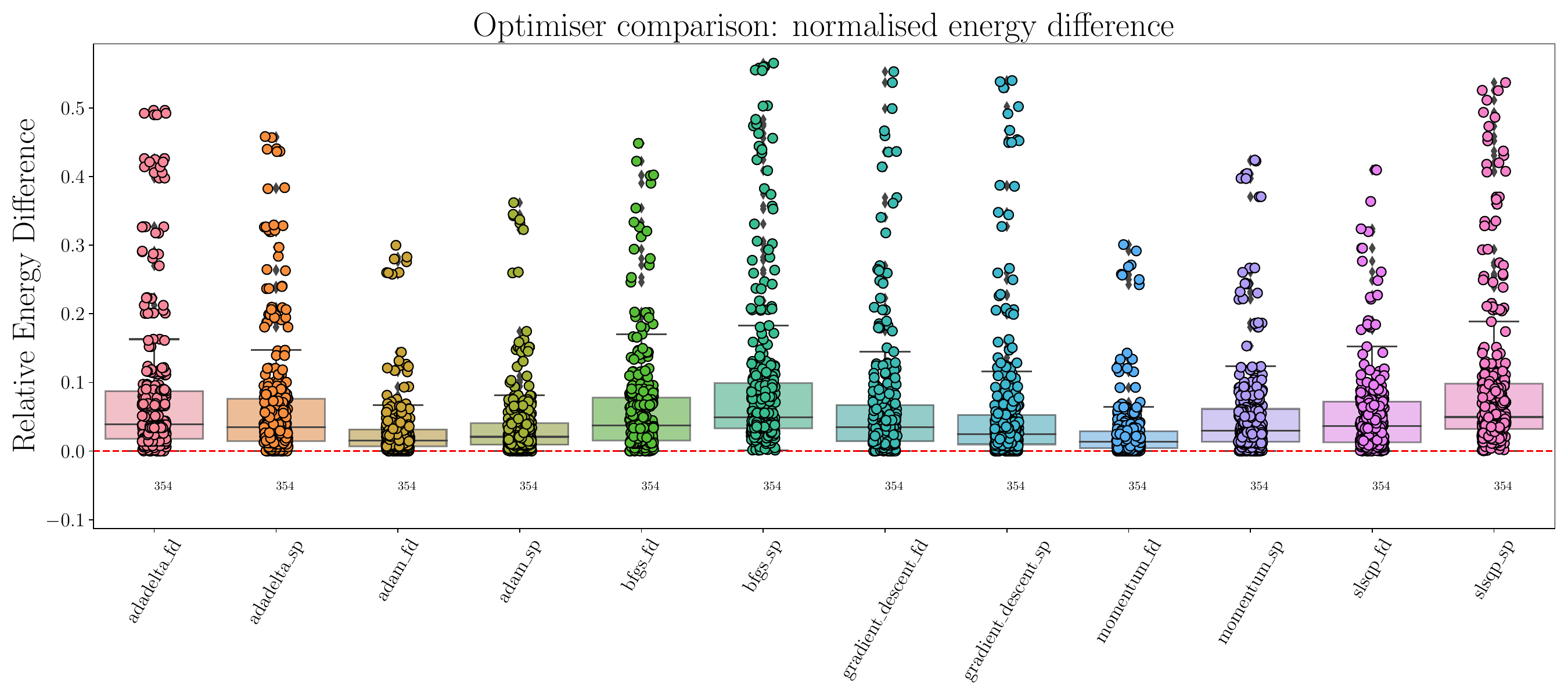}
     \end{subfigure}
     
     \begin{subfigure}[b]{\textwidth}
         \centering
         \includegraphics[width=0.65\textwidth]{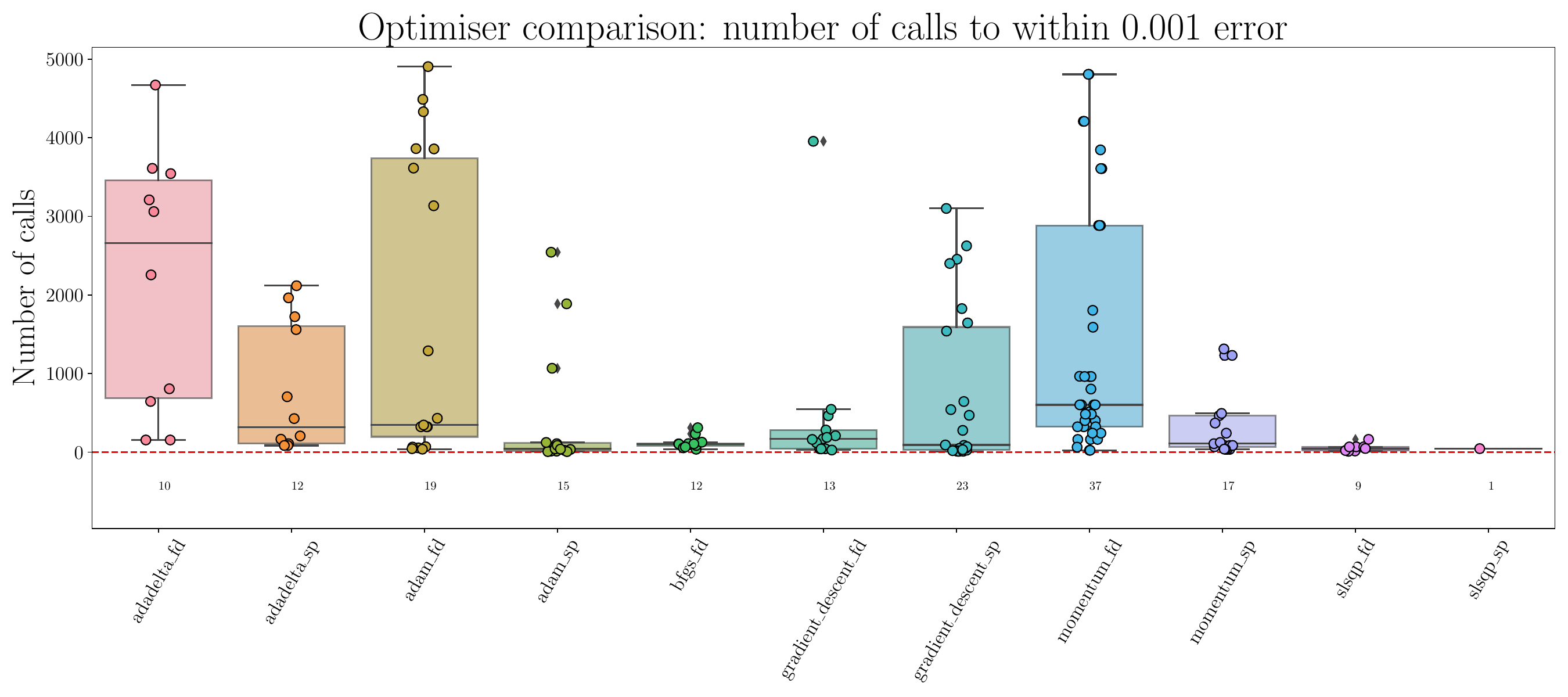}
     \end{subfigure}

     \caption{Comparison of gradient-based optimisers, and of finite difference (FD) and simultaneous perturbation (SP) as gradient subroutines, ordered alphabetically. We observe that in general the FD approach leads to a lower overall energy, whereas the SP approach uses fewer calls.}
     \label{fig:gradient_based}

 \end{figure}

\noindent\begin{minipage}{\linewidth}
    \hspace{12pt} Finally, in \cref{fig:gradient_based} we plot some of the gradient-based optimisers to directly compare the use of finite difference and simultaneous perturbation as gradient subroutines. We see that as expected, FD uses more calls than SP to get within 0.001 error of the ground state (as SP uses 2 evaluations per gradient evaluation, as opposed to 2 times the number of parameters for FD), but in general FD seems to obtain a lower overall error, for example more of the runs using FD gradient are able to get within 0.001 error. 
\end{minipage}
\newpage

 \subsection{Quantum natural gradient descent}
\label{subsec:results:qng}

In this subsection we consider the quantum natural gradient (QNG)\cite{stokes2020quantum} and closely related imaginary time evolution (ITE) \cite{mcardle2019variational} approaches, deferring some background and details regarding our implementation to \cref{app:qng} (in particular we restrict to $1$-dimensional systems due to ansatz considerations). The idea behind QNG is to perform gradient descent on a more appropriate coordinate space, taking account for geometrical properties of quantum states.  One of the main contributions of \cite{stokes2020quantum} is to argue that the relevant geometry is given by the Fubini-Study metric tensor, or quantum Fisher information (QFI). 

\begin{figure}[htbp!]
     \centering
     \begin{subfigure}[b]{\textwidth}
         \centering
         \includegraphics[height = 0.16\textheight]{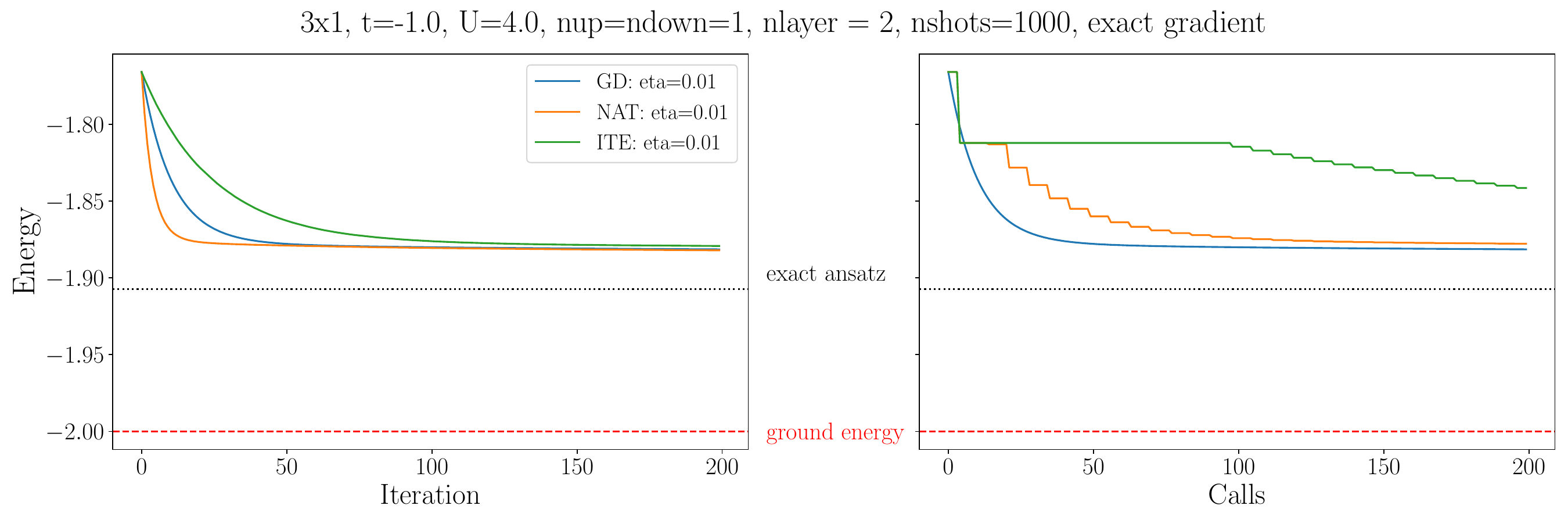}
     \end{subfigure}

      \begin{subfigure}[b]{\textwidth}
     \centering
     \includegraphics[height = 0.16\textheight]{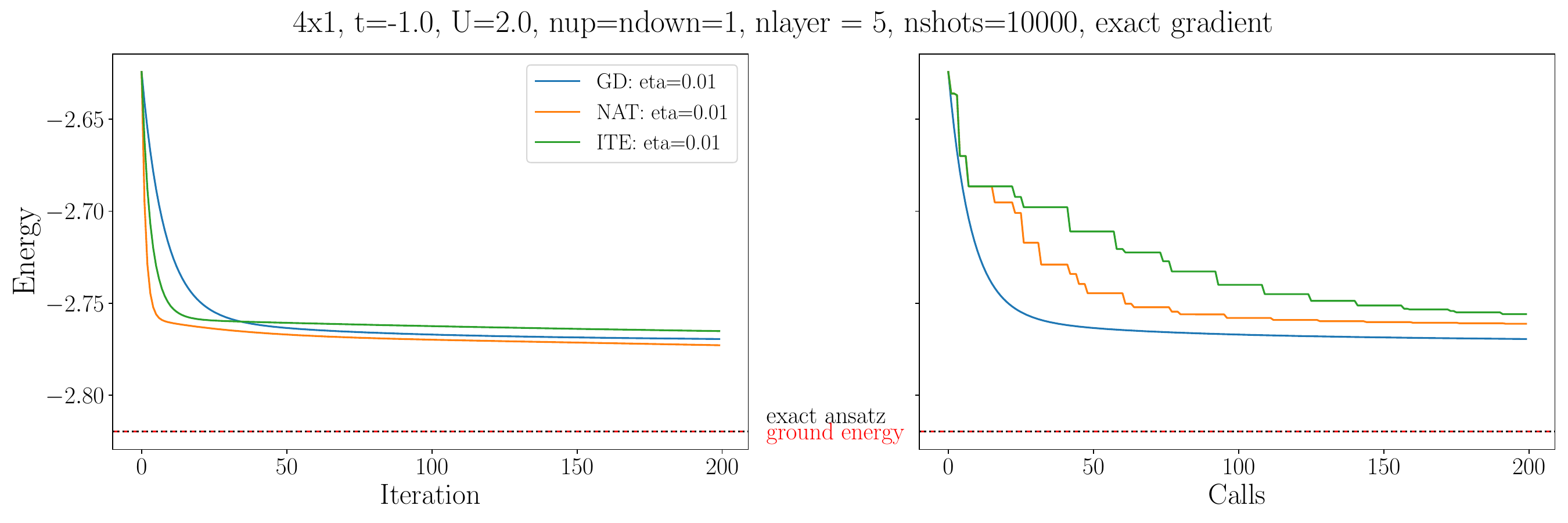}

     \end{subfigure}

           \begin{subfigure}[b]{\textwidth}
     \centering
     \includegraphics[height = 0.16\textheight]{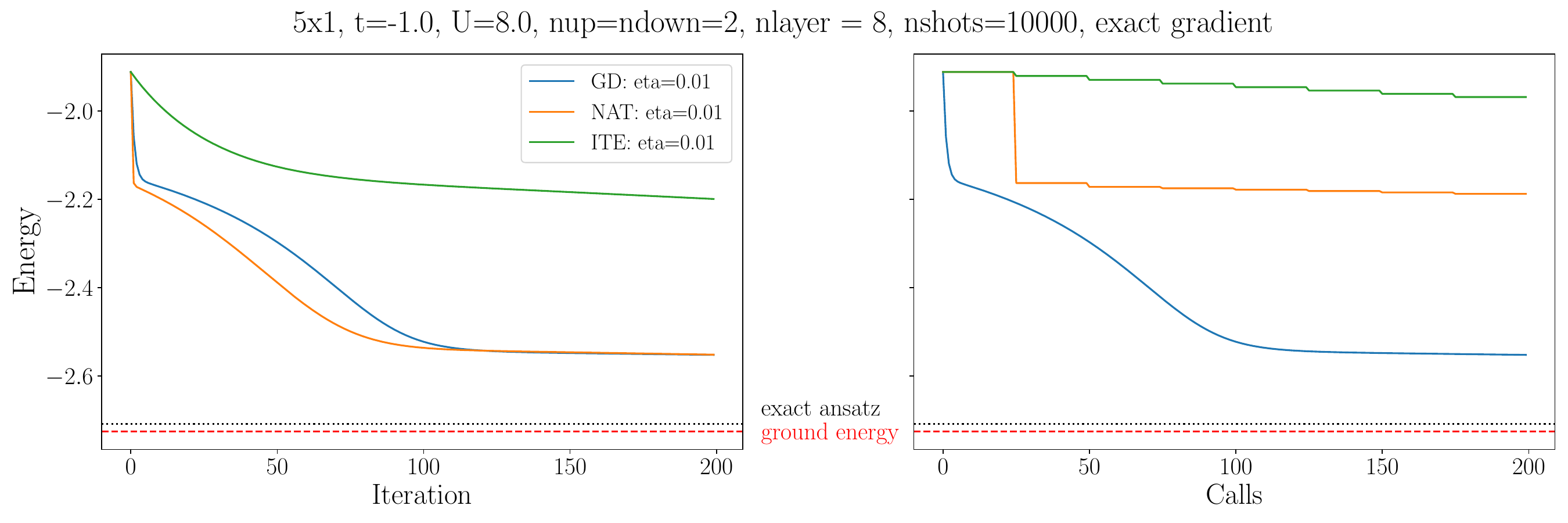}

     \end{subfigure}

               \begin{subfigure}[b]{\textwidth}
     \centering
     \includegraphics[height = 0.16\textheight]{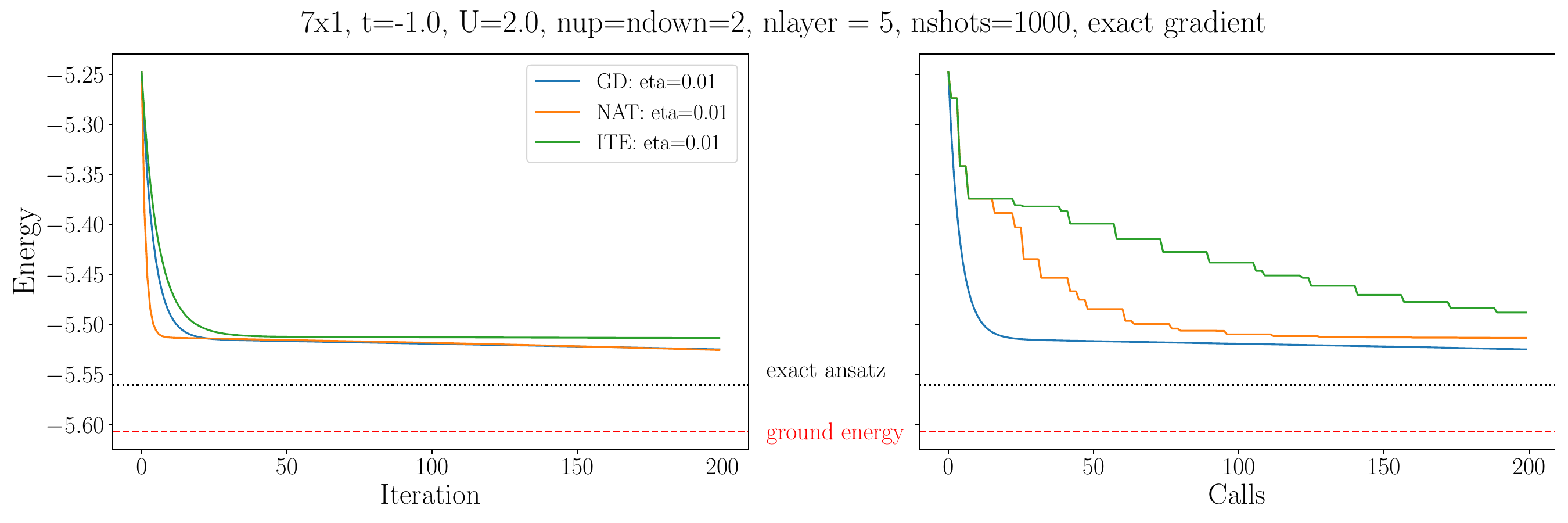}

     \end{subfigure}

     \caption{Comparison of vanilla gradient descent (GD), natural gradient descent (NAT) and imaginary time evolution (ITE) when given access to the exact gradient. The figures on the left correspond to plotting over iterations, whereas on the right they are over total calls. Here GD uses 1 call per iteration, whereas NAT and ITE both use $\nu$ + 1 calls per iteration ($\nu$ being the number of parameters, plus an extra call to evaluate at the current parameters). We see in general that although the natural gradient may be a more effective gradient, it loses this advantage when considering the total number of calls. In the figures `exact ansatz' refers to the best energy achieved when optimising the exact cost function (no statistical noise), and `eta' refers to the learning rate. }

     \label{fig:QNG_exact_gradient}
         
\end{figure}

\begin{figure}[htbp!]
     \centering
     \begin{subfigure}[b]{\textwidth}
         \centering
\includegraphics[height = 0.16\textheight]{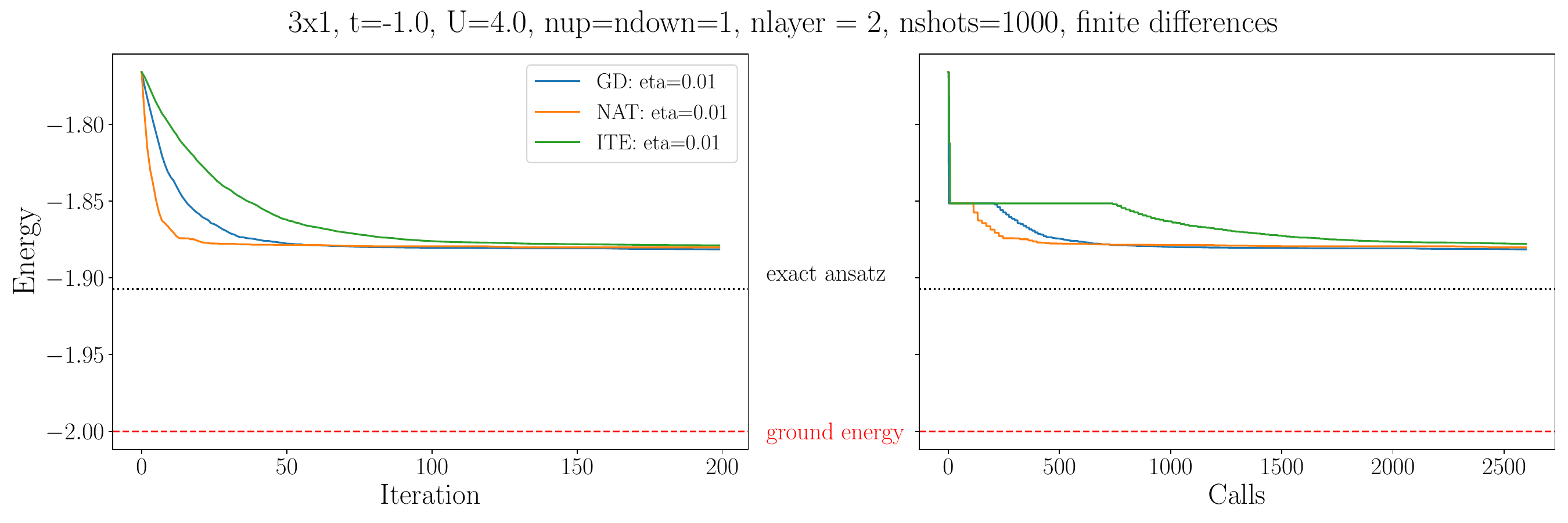}
     \end{subfigure}

      \begin{subfigure}[b]{\textwidth}
     \centering
     \includegraphics[height = 0.16\textheight]{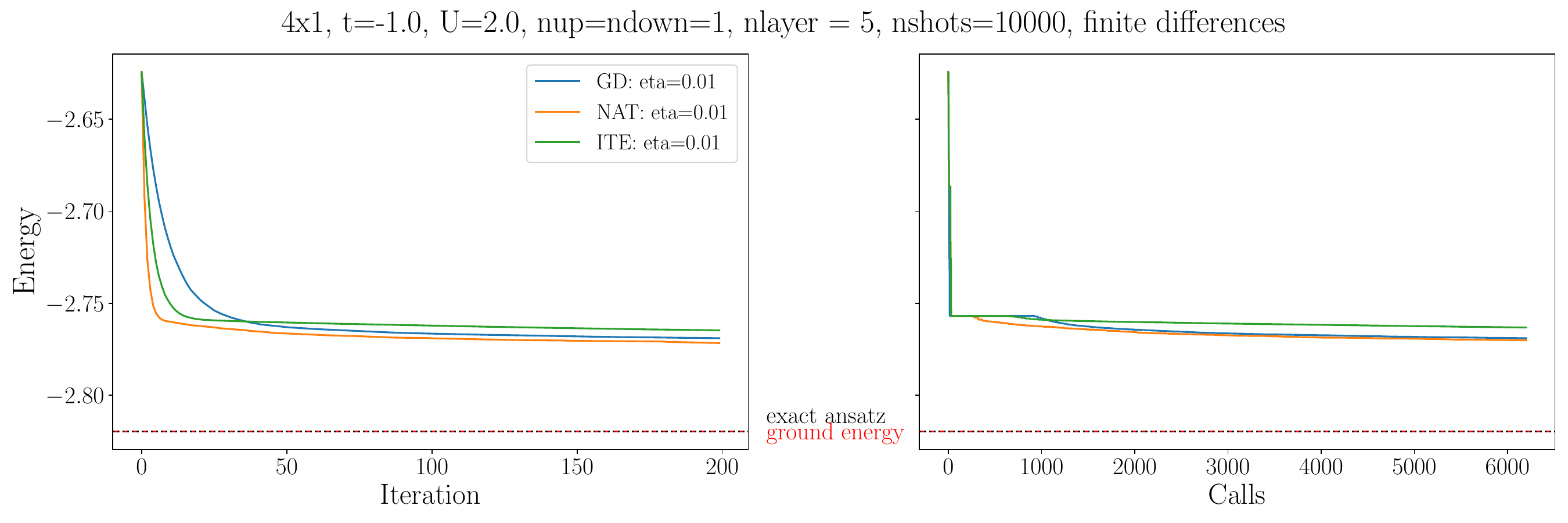}

     \end{subfigure}

           \begin{subfigure}[b]{\textwidth}
     \centering
     \includegraphics[height = 0.16\textheight]{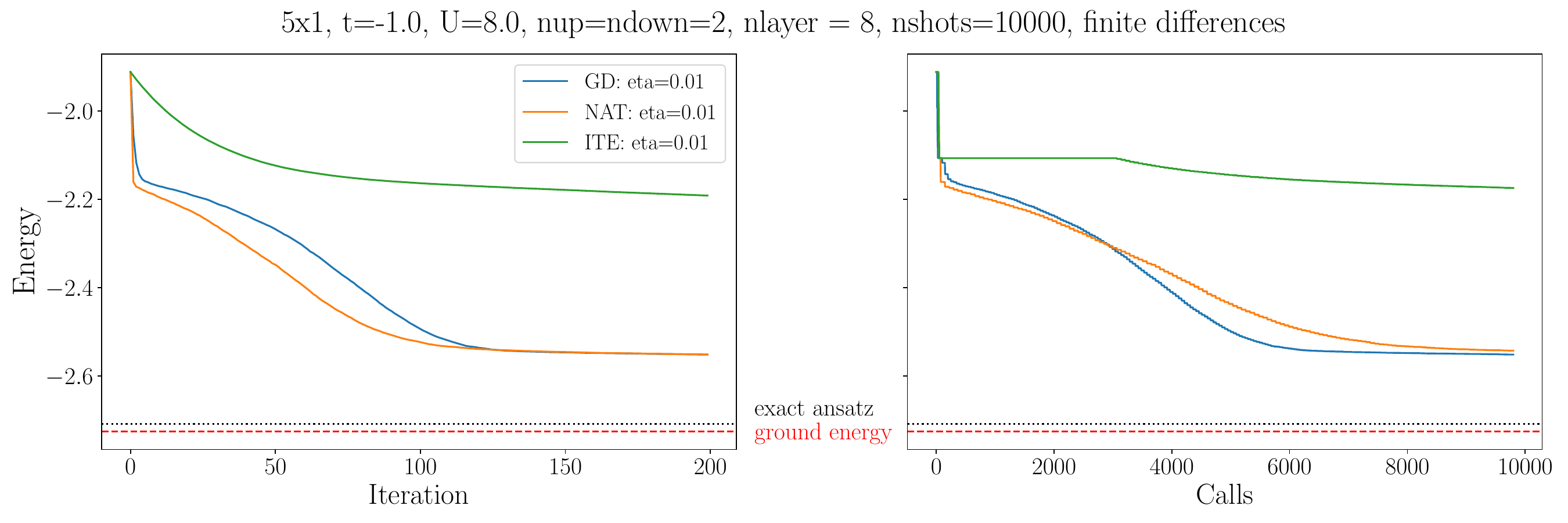}

     \end{subfigure}

                \begin{subfigure}[b]{\textwidth}
     \centering
     \includegraphics[height = 0.16\textheight]{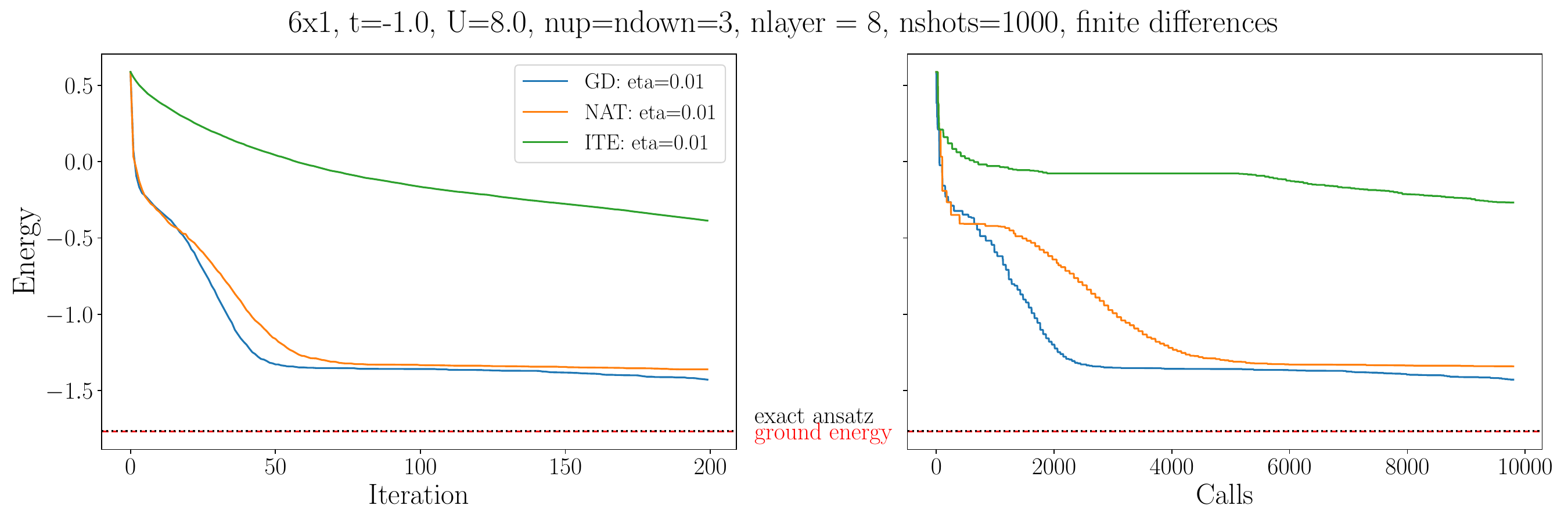}

     \end{subfigure}

               \begin{subfigure}[b]{\textwidth}
     \centering
     \includegraphics[height = 0.16\textheight]{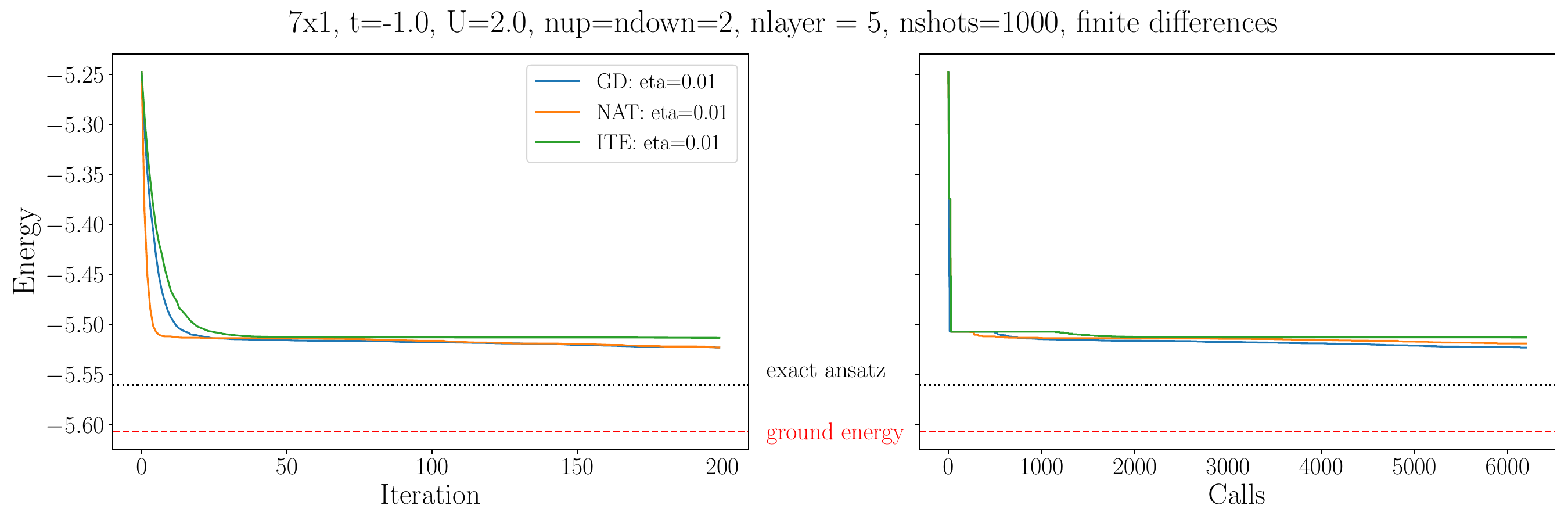}

     \end{subfigure}

  \caption{Comparison of vanilla gradient descent (GD), natural gradient descent (NAT) and imaginary time evolution (ITE) when using finite differences to calculate the gradient (this uses $2\nu$ evaluations, for $\nu$ parameters). The figures on the left correspond to plotting over iterations, whereas on the right they are over total calls. Here GD uses $2\nu + 1$ calls per iteration, whereas NAT and ITE both use $3\nu$ + 1 calls per iteration ($\nu$ being the number of parameters, plus an extra call to evaluate at the current parameters). We see here that the natural gradient can retain its advantage even when considered over total calls taken, as the relative overhead increase from $2\nu$ to $3\nu$ calls is not so significant. In the figures `exact ansatz' refers to the best energy achieved when optimising the exact cost function (no statistical noise), and `eta' refers to the learning rate. }
    \label{fig:QNG_finite_difference}
         
\end{figure}

\begin{figure}[htbp!]
     \centering
     \begin{subfigure}[b]{\textwidth}
         \centering
         \includegraphics[height = 0.16\textheight]{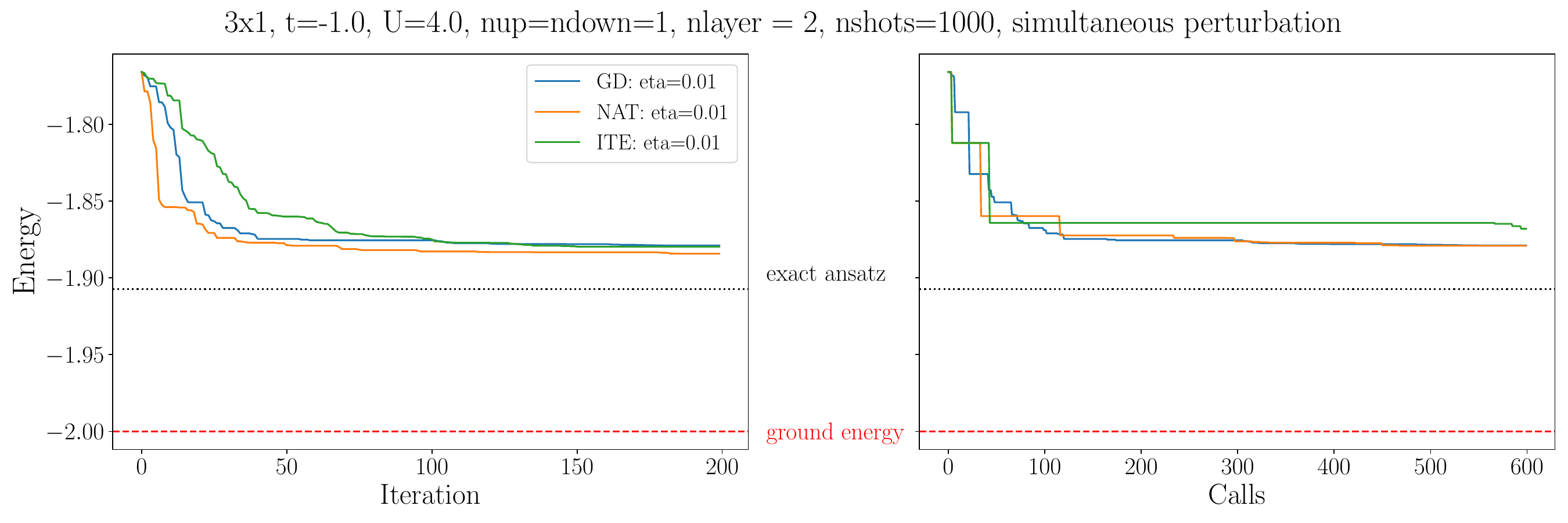}

     \end{subfigure}

      \begin{subfigure}[b]{\textwidth}
     \centering
     \includegraphics[height = 0.16\textheight]{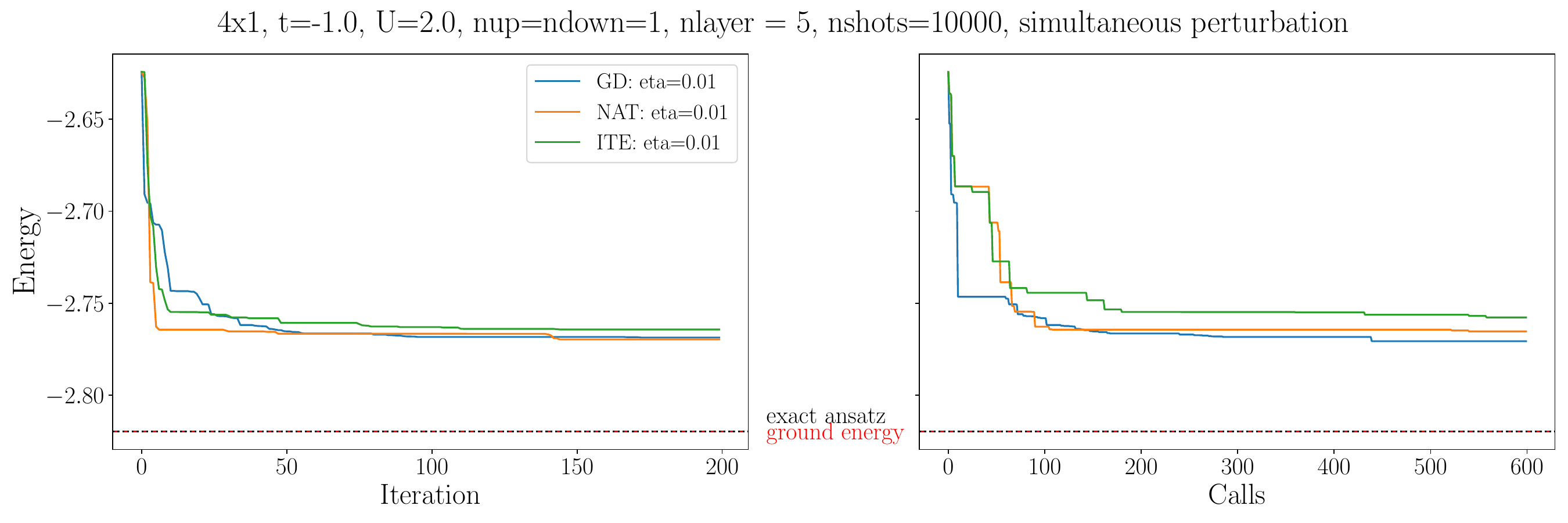}

     \end{subfigure}

           \begin{subfigure}[b]{\textwidth}
     \centering
     \includegraphics[height = 0.16\textheight]{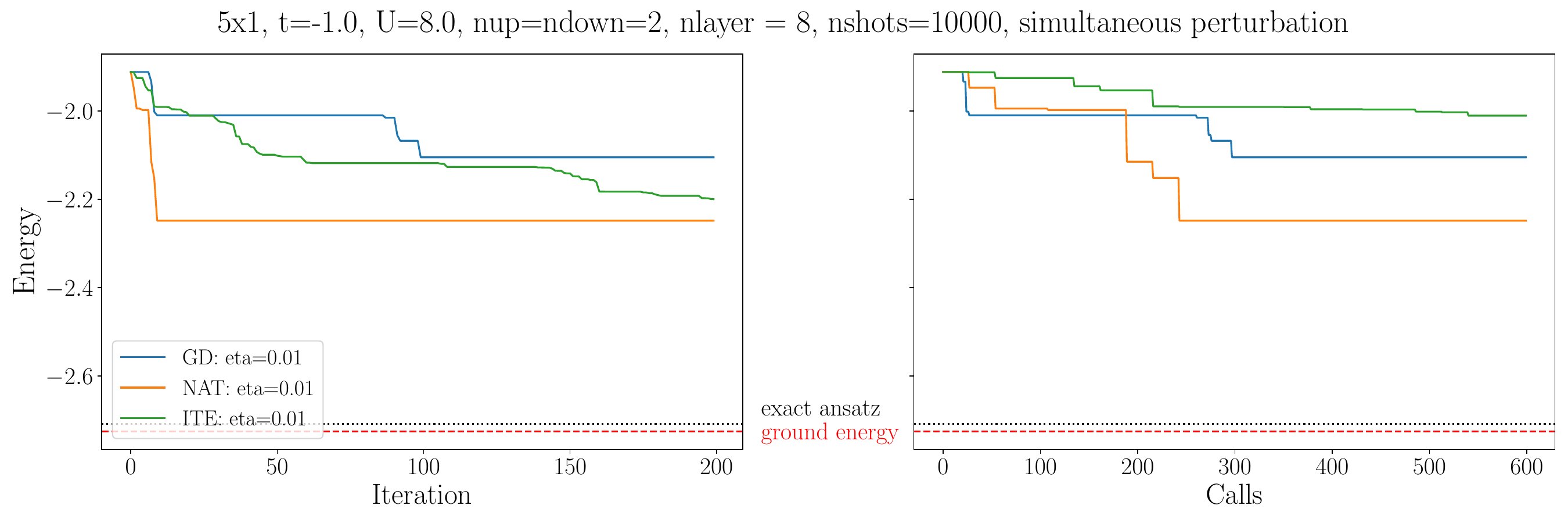}

     \end{subfigure}

                \begin{subfigure}[b]{\textwidth}
     \centering
     \includegraphics[height = 0.16\textheight]{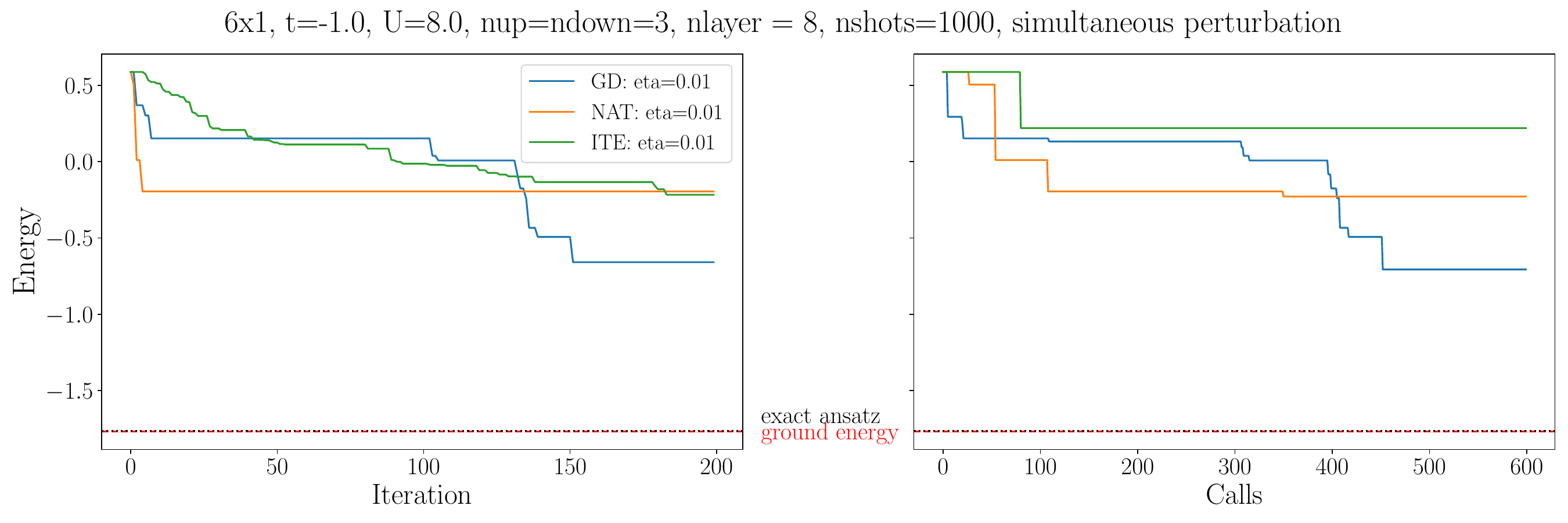}

     \end{subfigure}

               \begin{subfigure}[b]{\textwidth}
     \centering
     \includegraphics[height = 0.16\textheight]{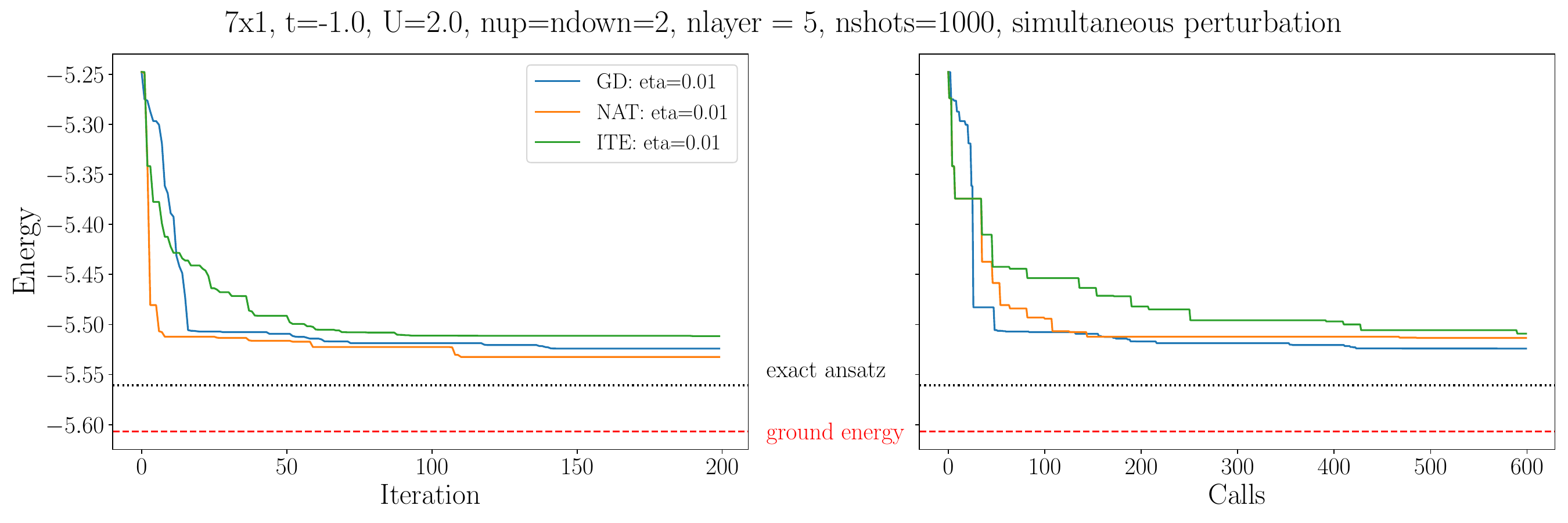}

     \end{subfigure}

      \caption{Comparison of vanilla gradient descent (GD), natural gradient descent (NAT) and imaginary time evolution (ITE) when using simultaneous perturbation to calculate the gradient (this uses $2$ evaluations). The figures on the left correspond to plotting over iterations, whereas on the right they are over total calls. Here GD uses $3$ calls per iteration, whereas NAT and ITE both use $\nu$ + 3 calls per iteration ($\nu$ being the number of parameters, plus an extra call to evaluate at the current parameters). We see here that in general any advantage gained by NAT or ITE seems to be lost when considering the total number of calls used. It is also interesting that ITE seems to perform significantly better given simultaneous perturbation as a gradient subroutine, compared to the exact gradient or finite differences. In the figures `exact ansatz' refers to the best energy achieved when optimising the exact cost function (no statistical noise), and `eta' refers to the learning rate. }
           \label{fig:QNG_simultaneous_perturbation}
      
\end{figure}

In \cref{fig:QNG_exact_gradient}, \cref{fig:QNG_finite_difference} and \cref{fig:QNG_simultaneous_perturbation} we respectively show results when using the exact gradient, finite differences, and simultaneous perturbation as the gradient function. We plot standard gradient descent, natural gradient descent, and imaginary time evolution, all with the same learning rate $\eta=0.01$. We show plots over each iteration for the three methods (where an iteration is defined as a single gradient descent step), as well as over the total number of calls. 

The natural gradient seems to perform better than the naive gradient as a subroutine (see the plot over iterations in \cref{fig:QNG_exact_gradient}), but this advantage seems to be lost when considering the extra calls it takes. As finite differences takes $2\nu$ (for $\nu$ parameters) calls to calculate a single gradient, it could be that the extra $\nu$ calls to also evaluate the diagonal fisher information matrix for the natural gradient could be worthwhile.
However it seems that even this regime, from \cref{fig:QNG_finite_difference} we see that the natural gradient does not seem to perform well enough to warrant the overall increase in calls from $2\nu$ to $3\nu$.

Possible future directions would include extending to non-linear FH systems, exploring other ansätze, attempting to estimate the covariance more directly as opposed to approximating them numerically, and attempting to calculate non-diagonal entries in the QFI matrix (see \cref{app:qng}).

 \section{Discussion}
\label{sec:discussion}

Clearly the choice of optimiser can make a big difference, as can be seen from the contrasting performance of optimisers. Our results show that careful consideration and selection is required to ensure the effectiveness of the variational quantum eigensolver.

\paragraph{Which are the best optimisers?\\}

Firstly, we can clearly see from the above plots that the best-performing optimisers include: Momentum with finite difference gradient, Adam with finite difference gradient, Adam with simultaneous perturbation gradient, SPSA, CMAES, BayesMGD, and Coordinate Descent.

We can also provide some concrete suggestions:
\begin{itemize}[-]
\item \textit{Best for final accuracy with ground energy: } Momentum or Adam with finite differences -- see \cref{fig:final_errors_boxplot_exclude2x2_ops}
\item \textit{Best for using a low number of calls: } SPSA, CMAES or BayesMGD -- see \cref{fig:calls_to_within_error}.
\item \textit{Best offering a balance between these two: } CMAES  -- see \cref{fig:combined_energy_calls}.
\end{itemize}

\paragraph{Finite differences vs simultaneous perturbation\\}

We can also see that using finite differences as a gradient subroutine would be preferable when prioritising final energy obtained, whereas simultaneous perturbation could be more appropriate when seeking to use a small number of calls. One could also consider initially using simultaneous perturbation to get close to the ground energy before switching to finite differences to fine tune.

We found that the step size used in finite differences plays a big role. It can be appealing to simply set an arbitrary step size (e.g. 0.01), however we have shown in \cref{fig:gradient_tradeoff} and \cref{fig:gradient_over_instances} that it is important to give this some thought. For instances comparable to those studied here, we suggest a step size of around 0.4.

\paragraph{Impact of hyperparameter selection\\}

Choosing good hyperparameters is also an important task. In \cref{app:optimisers} we give detailed summaries of the hyperparameters taken for each optimiser, and find that the sensitivity of an optimiser to its hyperparameters can vary wildly. For example, CMAES seems to be relatively insensitive to the initial standard distribution given (see \cref{fig:cmaes_hparams}), whereas the performance of SPSA seemed to depend quite heavily on the hyperparameters chosen (\cref{fig:spsa_hparams}), and BayesMGD seemed to perform well on all hyperparameters except the default ones (\cref{fig:bayes_hparams}).

\paragraph{Future ideas \\}
There are many possible continuations of a work such as this. We have focused on the Fermi-Hubbard model and the Hamiltonian variational ansatz. A natural extension would be to consider different models, for example the Transverse Field Ising model. One would expect that optimiser performance would be correlated across different underlying physical problems, although it could be the case that certain optimisers are better suited for the search space of specific physical systems. Similarly, there are many other choices for VQE ansatz. The primary appeal of the HV ansatz is that it uses a low number of parameters, and for systems beyond $3\times 3$ the number of parameters is independent of the system size. One could also consider ansätze such as the number preserving (NP) ansatz \cite{cade2020strategies}, or the most expressive ansatz possible by freeing up all available parameters. It would be interesting to study which optimisers are well suited to more expressive ansatz and which are able to get the most out of a limited number of parameters. There is also much arbitrariness in what one considers a hyperparameter, and one could always promote more variables to hyperparameters and perform more in depth sweeping.

Another interesting direction would be to consider multi-stage approaches. For example in \cite{cade2020strategies}, they considered increasing the number of shots as the optimisation progresses, and one could consider something similar for a specific optimisation routine. Examples could include:
\begin{itemize}[-]
    \item Starting with an ansatz with a low number of parameters, and then freeing up more parameters in the later stages.
    \item Changing the optimiser midway through.
    \item Switching the gradient function midway through: for example starting with simultaneous perturbation, and then switching to finite differences for greater accuracy in the later stages.
\end{itemize}

Finally, one could consider novel optimisation techniques. The bulk of this numerical study has focused on black-box optimisers, however there exist several more bespoke VQE optimisation techniques (one example being coordinate descent). In \cref{subsec:results:qng} we studied an application of the quantum natural gradient for 1-dimensional Fermi-Hubbard systems. It would be valuable to further develop theoretical tools to study VQE specific optimisers, which may be uniquely placed to take advantage of the specific problem structure of variational quantum algorithms.

Another approach that would be interesting to study in more depth is quantum analytic descent \cite{koczor2022quantum}. The algorithm proceeds by constructing a classical model of the cost function around a reference point, and then doing gradient descent (classically) on this model to find the minimum. It is primarily intended for high precision regimes, and sophisticated techniques are used to update the model whilst using a low and variable number of shots. It would be interesting to extend the algorithm from \cite{koczor2022quantum} to the HV ansatz and our setting here (as the original paper does not account for fixed gates between gates sharing the same parameter, and uses a variable number of shots), and also potentially consider connections to the coordinate descent algorithm \cite{cade2020strategies}, and parameter shift rules \cite{wierichs2022general}, as they all involve constructing trigonometric models of the cost function.

\section*{Acknowledgments}

We would like to thank the Phasecraft team for their support with this work, and Bálint Koczor for helpful discussions. This work was supported by Innovate UK (grant no. 44167) and has received funding from the European Research Council (ERC) under the European Union’s Horizon 2020 research and innovation programme (grant no. 817581).

\addcontentsline{toc}{section}{References}

\printbibliography
 
\appendix

\addtocontents{toc}{\protect\setcounter{tocdepth}{1}}

\section{Additional figures}

\begin{figure}[ht!]
    \centering
\includegraphics[width=\textwidth]{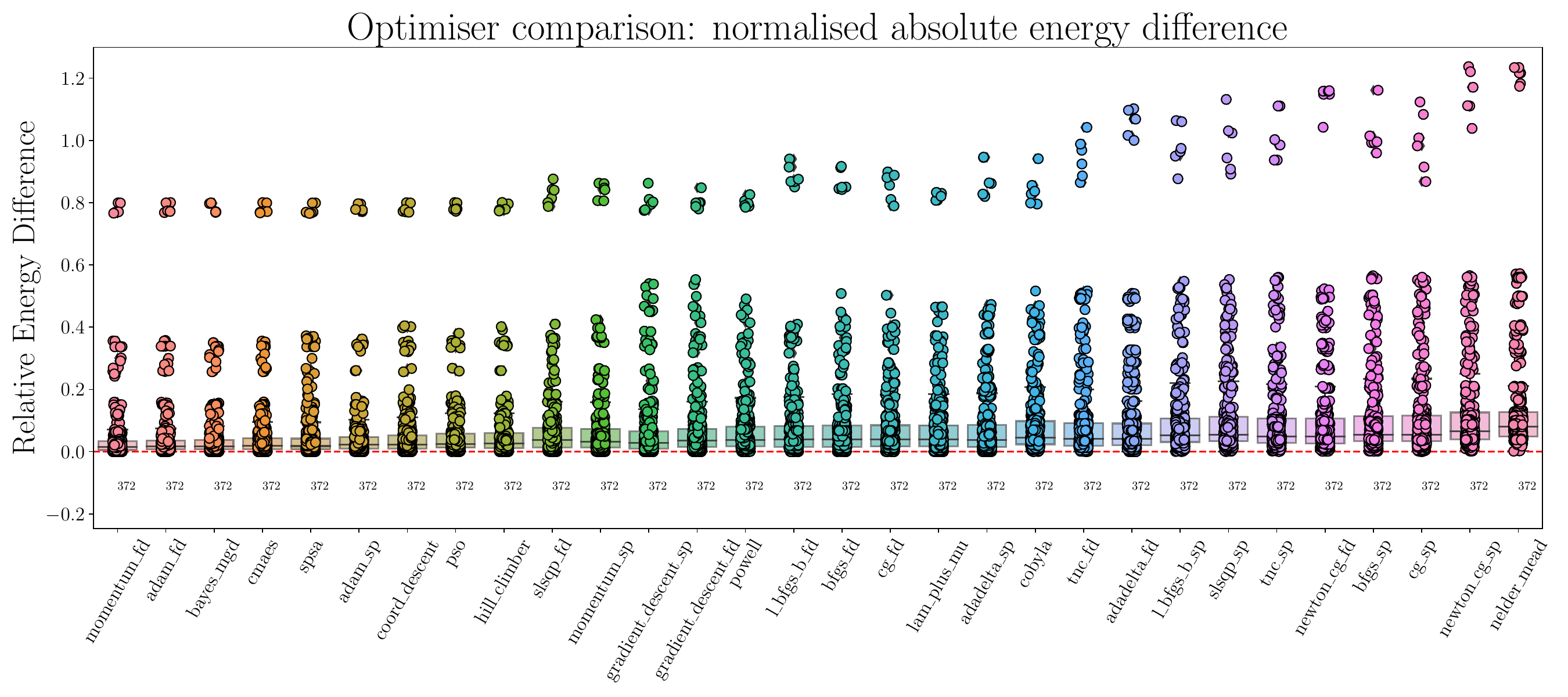}
    \caption{This plot summarises data collected over all optimisers and all instances (372 in total). For
each optimiser, we take the best final energy (over different runs, e.g. for multiple hyperparameter
settings), calculate the difference with the ground state energy and divide by the size of the grid ($mn$
for an $m \times n$ grid) to get a normalised measure. For each optimiser both the individual points and an
underlying boxplot are shown. The optimisers are ordered from left to right
by their mean of these values. The suffixes ‘fd’ and ‘sp’ respectively refer to using finite difference
(with step size 0.4) and simultaneous perturbation (with step size 0.15) as gradient functions.
\label{fig:final_errors_boxplot}}
\end{figure}

In \cref{fig:final_errors_boxplot} we present boxplots as in \cref{fig:boxplot all instances}, but overlaid with all data points.

\begin{figure}[h!]
    \centering
    \includegraphics[width=\textwidth]{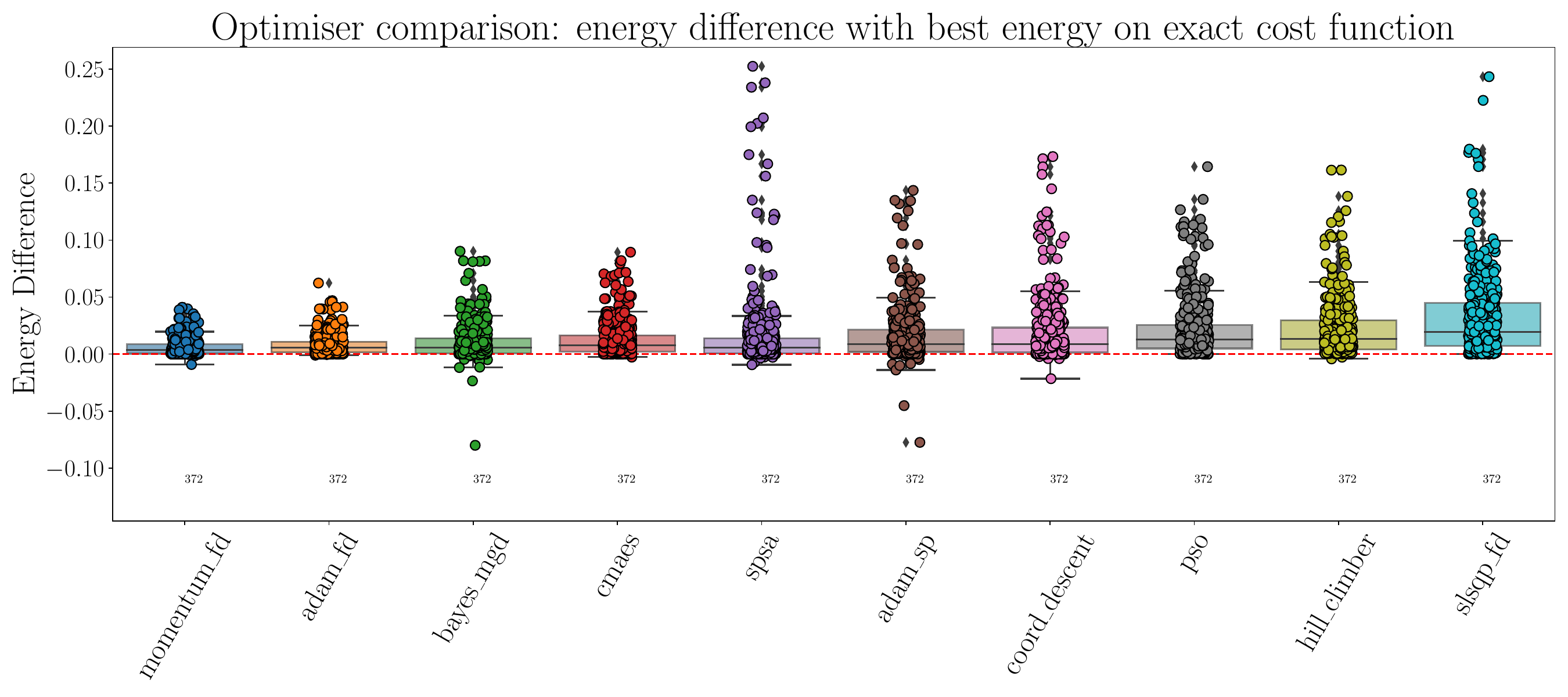}
    \caption{Comparison of the final energy difference achieved by optimisers and the best optimiser performance on statistical cost functions (see \cref{fig:exact_runs}), over all 372 instances. We see that occasionally certain optimisers were able to outperform the \texttt{scipy} optimisers that ran on the exact cost functions. }
    \label{fig:final_errors_boxplot_exact}
\end{figure}

We also compare the final energies achieved with that of the best energy achieved on the exact cost functions in \cref{fig:final_errors_boxplot_exact}. One could expect that the best energy value found on the exact cost functions would provide a lower bound on the best value achieved with statistical cost functions, as adding statistical uncertainty makes the optimisation problem much more difficult. However we see that some optimisers were occasionally able to outperform the optimisers used on the exact cost functions (which were all \texttt{scipy} optimisers, see \cref{subsec:form}). Perhaps such optimisers could be good candidates to run on exact cost functions in future.


\begin{figure}[htbp!]
    \centering
    \includegraphics[width=0.85\textwidth]{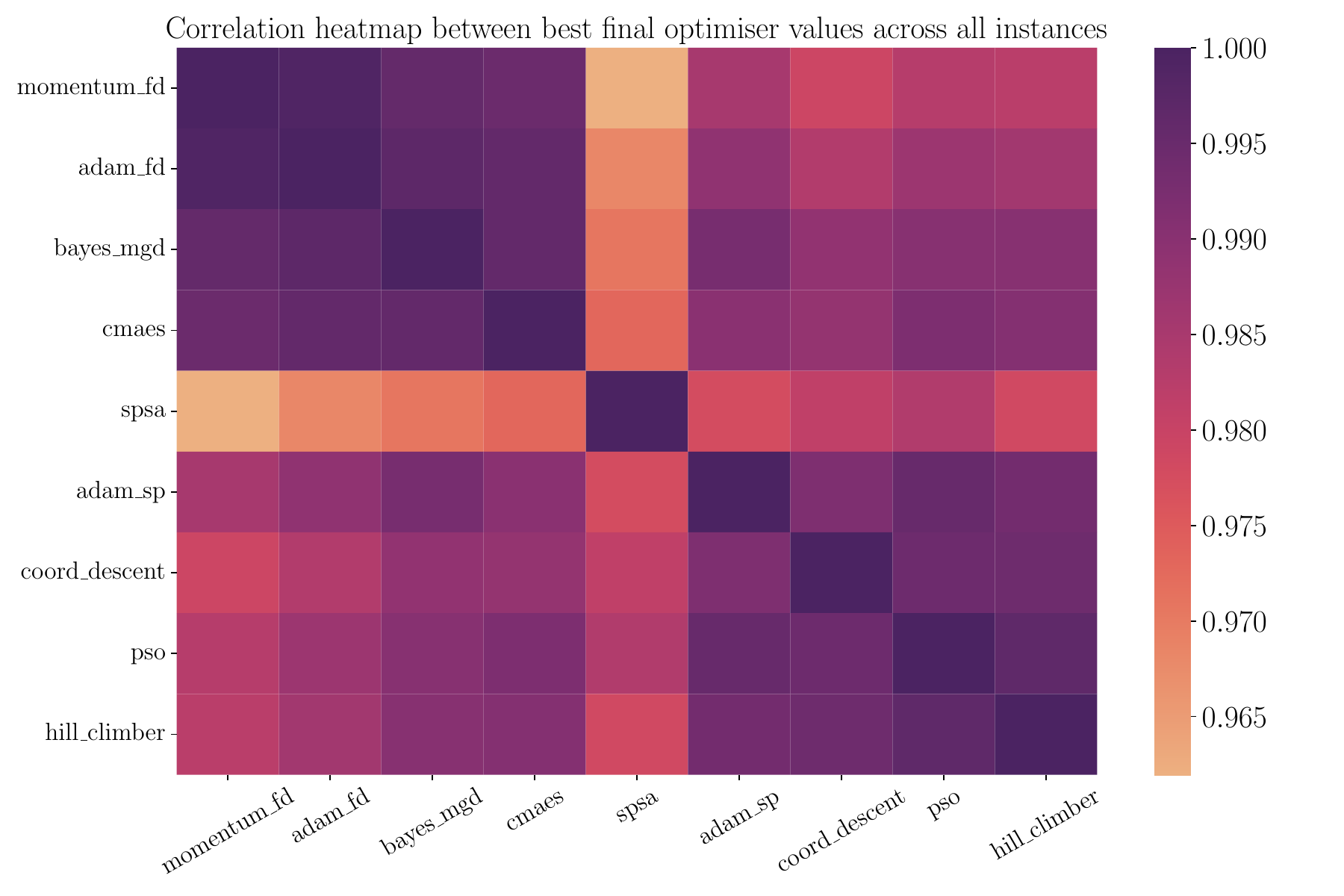}
    \caption{Here we display a heatmap for some of the better performing optimisers. We see strong correlation between optimiser performance across instances, as is to be expected.}
    \label{fig:heatmap_ops}
\end{figure}

\begin{figure}[htbp!]
    \centering
    \includegraphics[width=0.9\textwidth]{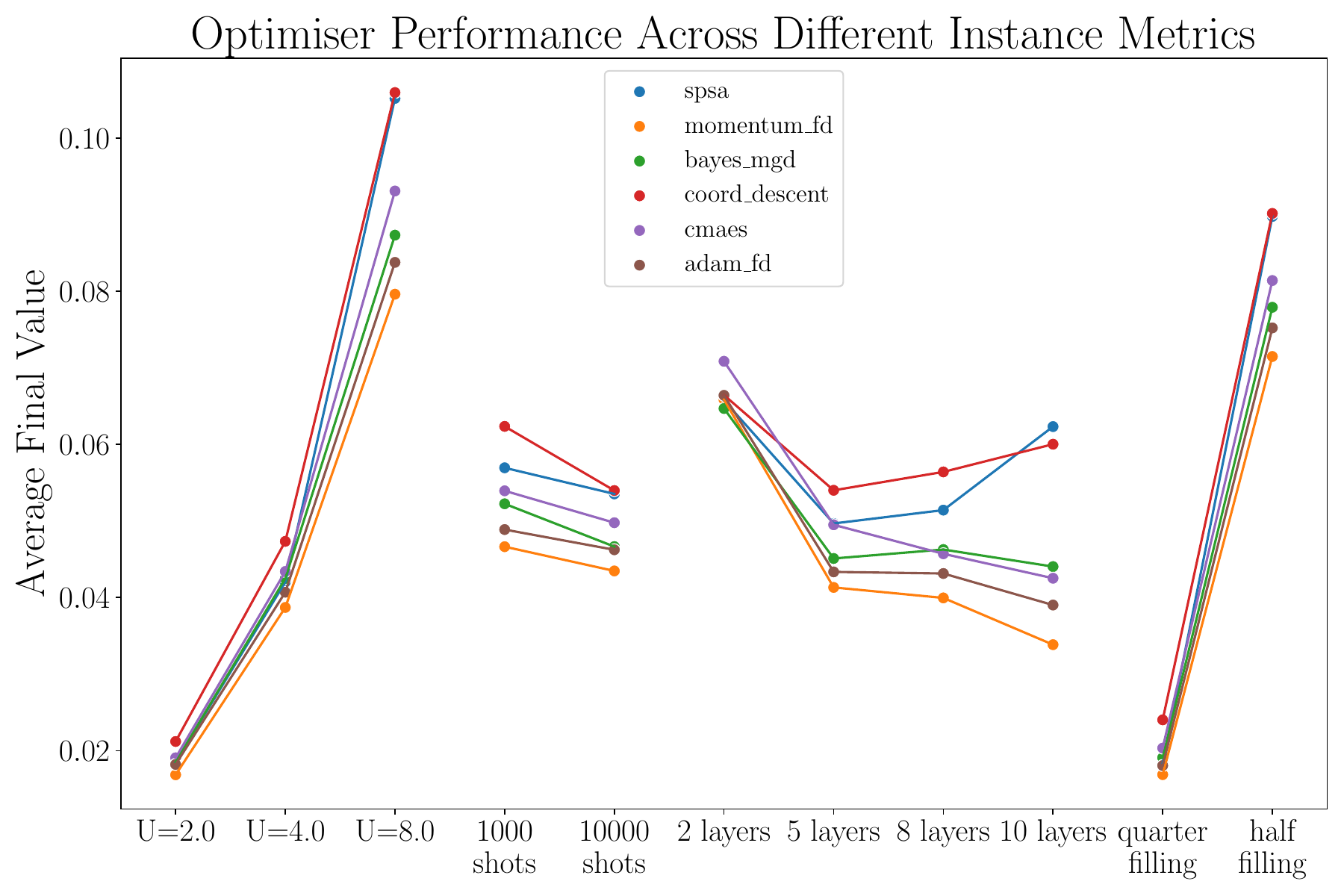}
    \caption{Here we average each optimisers normalised (i.e. divided by the grid size $mn$) final energy difference with the ground energy over all instances whilst keeping an instance metric fixed. For example, we see that increasing the Coulomb potential $U$ in general increases the normalised distance away from the ground energy that the optimisers were able to achieve. We also note that SPSA and coordinate descent have more difficulty than other optimisers when increasing the number of layers.}
    \label{fig:line_comparison}
\end{figure}

One can also ask the following question: do optimisers always perform better on the same instance? That is, how strong is the correlation between the performance of the optimiser and the instance considered? In general one would expect this to be strongly correlated, and we show a corresponding heatmap for some of the best optimisers in \cref{fig:heatmap_ops} which agrees with this intuition. We can see that SPSA has slightly worse correlation with respect to the other optimisers (still a positive correlation of more than 0.96). To explore this, we plot in \cref{fig:line_comparison} the average performance of some of the best optimisers when considering different instance metrics. We can pull out many natural conclusions: that decreasing the Coulomb potential $U$, increasing the number of shots, and considering quarter filling instead of half-filling all result in an overall better final (normalised by the grid size) average error. However we also notice that SPSA and coordinate descent seem to struggle as the number of layers increases, in contrast to the other optimisers. This may go some way in explaining the slightly lower correlation between SPSA and the other optimisers in \cref{fig:heatmap_ops}. 

 \section{Quantum natural gradient descent}
\label{app:qng}

The quantum natural gradient \cite{stokes2020quantum, yamamoto2019natural} has been introduced as a quantum generalisation of the natural gradient \cite{amari1998natural, martens2020new}, and can be motivated by seeking to perform gradient descent on the most natural geometry of the search space in question. One can argue that standard gradient descent on VQE-type circuits is implicitly using the $l_2$ norm in parameter space (essentially copied from \cite{stokes2020quantum}):

\begin{align}
\theta_{t+1} &= \theta_t - \eta \nabla f(\theta_t)  \\
&=  \underset{x \in \mathbbm{R}^d}{\text{arg} \min} \left [ \langle x - \theta_t, \nabla f(\theta_t) \rangle + \frac{1}{2 \eta} \norm{x - \theta_t}_2^2 \right ]
\end{align}
which can be seen from differentiating the minimisation argument with respect to $x$ and setting it to zero. While other works have argued that the $l_1$ norm may be more appropriate \cite{harrow2021low}, the main contribution of \cite{stokes2020quantum} is to argue that the relevant geometry is given by the Fubini-Study metric tensor, or quantum Fisher information (QFI). For quantum states parametrised by $\ket{\psi(\bm{\theta})} = \ket{\psi(\theta_1, \dots, \theta_\nu) }$, the QFI  is
\[
\mathcal{F}_{ij} = \text{Re} \left [\braket{\partial_i \psi}{\partial_j \psi} - \braket{\partial_i \psi}{\psi}\braket{\psi}{\partial_j \psi} \right ]
\]
and $\ket{\partial_j \psi} := \frac{\partial}{\partial \theta_j} \ket{\psi(\bm{\theta})}$. The update rule is then 
\[
\theta_{t+1} = \theta_t - \eta \mathcal{F}^{-1}(\theta_t) \nabla E(\theta_t)
\]

Here we consider the quantum natural gradient \cite{stokes2020quantum} and closely related imaginary time \cite{mcardle2019variational} approaches for the HV ansatz. The challenge with this ansatz in this context is that we have gates sharing the same parameter, with fixed gates (FSWAPs, see \cref{subsec:vqe_fh}) in between. Another way of phrasing this is to say that in \cite{stokes2020quantum} all parametrised gates can be written in the form
\[
e^{i \theta K}
\]
for some Hermitian generator $K$, but as we in general have fixed gates between gates sharing the same angle (which may not be implemented in parallel), this is not always possible. Hence we will only consider 1-dimensional FH systems, as this removes fixed FSWAPs between gates sharing a parameter. The update rule for quantum natural gradient descent is 
\[
\theta_{t+1} = \theta_t - \eta \mathcal{F}^{-1}(\theta_t) \nabla E(\theta_t)
\]
where the {Quantum Fisher Information (QFI) matrix if given by
\[
\mathcal{F}_{ij} = \text{Re} \left [\braket{\partial_i \psi}{\partial_j \psi} - \braket{\partial_i \psi}{\psi}\braket{\psi}{\partial_j \psi} \right ],
\]
and $\ket{\partial_j \psi} := \frac{\partial}{\partial \theta_j} \ket{\psi(\bm{\theta})}$. We will just consider the diagonal terms of this as an approximation, which seems to be well justified in \cite{stokes2020quantum} (see e.g. their Figure 1 \& 2). If we write the ansatz dependent on a single parameter as $\ket{\psi (\theta)} = U e^{i \theta P} \ket{\psi}$ (i.e. absorb previous gates into $\ket{\psi}$ and subsequent gates into $U$), then the derivative is $\partial_\theta \ket{\psi(\theta)} = iU P e^{i \theta P}\ket{\psi} $, and the diagonal term with respect to this parameter is
\begin{align}
\braket{\partial_\theta \psi}{\partial_\theta \psi} - \abs{\braket{\partial_\theta \psi}{\psi}}^2 
& = \bra{\psi} P^2 \ket{\psi} - \abs{\bra{\psi}  P  \ket{\psi}}^2 = \text{Var}_\psi(P)
\end{align}
where we took $P$ to be Hermitian. Now consider having multiple gates sharing the same parameter $\theta$. Assume that the generators $P$ and $Q$ commute, are Hermitian, and the gates are next to each other, so that we have 
\[
\ket{\psi (\theta)} = U e^{i \theta P} e^{i \theta Q} \ket{\psi} = U e^{i \theta (P+Q)} \ket{\psi}
\]

Then the relevant QFI component is
\begin{align}
\braket{\partial_\theta \psi}{\partial_\theta \psi} - \abs{\braket{\partial_\theta \psi}{\psi}}^2 
& = \text{Var}_\psi(P+Q) \\
    &= \text{Var}_\psi(P) + \text{Var}_\psi(Q) + 2\text{Cov}_\psi(P,Q)
\end{align}
where $\text{Cov}$ denotes the covariance, recall that $
\text{Var}(\sum_i X_i) = \sum_{i,j} \text{Cov}(X_i, X_j)$
and $\text{Cov}(X,X) = \text{Var}(X)$. Note that this expression does not actually depend on $\theta$, but on the parameters applied before, so e.g. $\mathcal{F}_{00}$ is independent of all parameters. In our case we have generators $P$ and $Q$ of the form
\begin{align}
    \frac{1}{2} \bigg ( & XX+YY \bigg )~, \hspace{20pt}
     \ketbra{11}
\end{align}
We can approximate the variance and covariance terms from the statistical data collected, as we already measure these terms in the full energy measurement. To summarise our implementation: 
\begin{itemize}
    \item We just look at the diagonal approximation to the quantum Fisher information matrix.
    \item We focus on linear systems to avoid FSWAP networks.
    \item We calculate the variances and covariances numerically.
    \item The number of cost function calls to evaluate the FIM is $\nu$ (number of parameters), where we sequentially set certain parameters to zero.
\end{itemize}

\paragraph{Example: $1 \times 3$ Fermi-Hubbard.}

The six qubits are ordered by spins, then site, i.e. 
\begin{align}
    \text{qubit 1:  } & \text{1st site, spin up,} & \text{qubit 4:  } & \text{1st site, spin down,} \\
    \text{qubit 2:  } & \text{2nd site, spin up,} &  \text{qubit 5:  } & \text{2nd site, spin down,} \\
    \text{qubit 3:  } & \text{3rd site, spin up,} & \text{qubit 6:  } & \text{3rd site, spin down.}
\end{align}

The onsite and hopping terms are then
\begin{align}
    O &= \ketbra{11}_{14} + \ketbra{11}_{25} + \ketbra{11}_{36} \\
    H_1 &= \frac{1}{2} \left ( X_1X_2 + Y_1 Y_2 + X_4X_5 + Y_4 Y_5 \right ) \\
    H_2 &= \frac{1}{2} \left ( X_2X_3 + Y_2 Y_3 + X_5 X_6 + Y_5 Y_6 \right ).
\end{align}

For two layers, parameters
\[
(\theta_1, \theta_2, \theta_3, \theta_4, \theta_5, \theta_6)
\]
would be mapped to the ansatz as

\[
U(\bm{\theta}) \ket{\psi}= e^{i \theta_6 H_2} ~ e^{i \theta_5 H_1} ~ e^{i \theta_4 O} ~ e^{i \theta_3 H_2} ~ e^{i \theta_2 H_1} ~ e^{i \theta_1 O} \ket{\psi}.
\]
For e.g. $\theta_4$:
\begin{align}
    \mathcal{F}_{44} =& \text{Var}_\phi (\ketbra{11}_{14}) + \text{Var}_\phi(\ketbra{11}_{25}) + \text{Var}_\phi(\ketbra{11}_{36} ) \\
    & + 2\text{Cov}_\phi (\ketbra{11}_{14}, \ketbra{11}_{25} )
    + 2\text{Cov}_\phi (\ketbra{11}_{14}, \ketbra{11}_{36} ) \\
    & + 2\text{Cov}_\phi (\ketbra{11}_{25}, \ketbra{11}_{36} ),
\end{align}
where
\[
\ket{\phi}=  e^{i \theta_3 H_2} ~ e^{i \theta_2 H_1} ~ e^{i \theta_1 O} \ket{\psi}.
\]
For e.g. $\theta_2$:
\begin{align}
    \mathcal{F}_{22} =& \text{Var}_\phi \bigg ( \frac{1}{2}  ( X_1X_2 + Y_1 Y_2  ) \bigg ) + \text{Var}_\phi \bigg (\frac{1}{2}  ( X_4 X_5 + Y_4 Y_5  ) \bigg ) \\
    &+ 2\text{Cov}_\phi \bigg ( \frac{1}{2}  ( X_1X_2 + Y_1 Y_2  ) , \frac{1}{2}  ( X_4 X_5 + Y_4 Y_5  )  \bigg ),
\end{align}
where
\[
\ket{\phi} = e^{i \theta_1 O}  \ket{\psi}.
\]
We then calculate the natural gradient as

\[
\nabla_N (\theta_t) =  \mathcal{F}^{-1}(\theta_t) \nabla E = \begin{pmatrix}
\frac{1}{\mathcal{F}_{11}} & \dots & 0 \\
\vdots & \ddots & \vdots \\
0 & \dots & \frac{1}{\mathcal{F}_{66}} 
\end{pmatrix} ~ \nabla E
\]

\subsection{Imaginary Time Evolution}

The update rule for imaginary time evolution (ITE) is \cite{yamamoto2019natural, mcardle2019variational}:

\[
\theta_{t+1} = \theta_t - \mathcal{A}^{-1}(\theta_t) \nabla E(\theta_t)
\]
where
\[
\mathcal{A}_{ij} = \text{Re} \left [\braket{\partial_i \psi}{\partial_j \psi}  \right ]
\]
Hence for $\ket{\psi (\theta)} = U e^{i \theta P} \ket{\psi}$ we can calculate the diagonal terms as
\begin{align}
\braket{\partial_\theta \psi}{\partial_\theta \psi}
& = \bra{\psi} P^2 \ket{\psi}  = \text{Var}(P) + \abs{\bra{\psi}  P  \ket{\psi}}^2 \\
&=  \text{Var}_\psi(P) + \mathbbm{E}_\psi(P)^2
\end{align}

which we can approximate numerically. We then have for $\ket{\psi (\theta)} = U e^{i \theta (P+Q)} \ket{\psi}$
\begin{align}
    \bra{\psi} (P+Q)^2 \ket{\psi}  &= \text{Var}_\psi(P+Q) + \abs{\bra{\psi}  (P+Q)  \ket{\psi}}^2.
\end{align}

 \section{Details of optimisers}
\label{app:optimisers}

\subsection{Hill Climber}
\label{subsec:hill_climber}

\begin{itemize}[-]
    \item \textbf{Overview: } This is perhaps the simplest black-box optimiser, essentially just sampling nearby random points and choosing the best. More specifically, at each iteration, we sample $n$ points according to a multivariate Gaussian centred on our current point, with standard deviation $\sigma$. We then update the current point to the minimum of the new points and current point.
    \item \textbf{Implementation: } in house.
    \item \textbf{Evaluations per iteration: } $n$.
    \item \textbf{Black box: } Yes.
    \item \textbf{gradient-based: } No.
    \item \textbf{References: } \cite{Luke2013Metaheuristics}    
 \item \textbf{Pseudocode:}

\begin{algorithm}
\SetKwFor{RepTimes}{repeat}{times}{end}
\caption{Hill climber}
\KwIn{Initial parameters $x_0$, cost function $f(x)$, termination criteria.}
\Hyperparameters{$n$, $\sigma$.}
$x \gets x_0$ \Comment*[r]{set current parameters to initial parameters}
$v \gets f(x_0)$ \Comment*[r]{set current value to initial value}
\While{not terminate}{
$b \gets x$  \Comment*[r]{set best parameters to current parameters}
\RepTimes{$n$}{
$z \gets x + \mathcal{N}(0, \sigma)$   \Comment*[r]{add normally distributed noise}
$w \gets f(z)$ \Comment*[r]{evaluate the cost function}
\If(\Comment*[f]{if the current value is better}){$w<v$}{ 
$v \gets w$  \Comment*[r]{update the best value found}
$b \gets z$\  \Comment*[r]{update the best parameters this iteration}
}
}
$x \gets b$\  \Comment*[r]{update the parameters}

}
\end{algorithm}

\item \textbf{Hyperparameters: } 2.

\begin{table}[H]
    \centering
    \begin{tabular}{|c|c|c|} \hline
        Hyperparameter & $\sigma$ & $n$  \\\hline
        Description  & standard deviation for the Gaussian & number of points to sample   \\\hline
        Default & 0.1 & 3 \\\hline
        Additional & 0.1 & 14 \\
        & 0.0892 & 5 \\
        & 0.108 & 5 \\
        & 0.1006 & 14 \\\hline
    \end{tabular}
    \caption{Hill climber hyperparameters.}
    
\end{table}

\begin{figure}[H]
    \centering
    \includegraphics[scale=0.3]{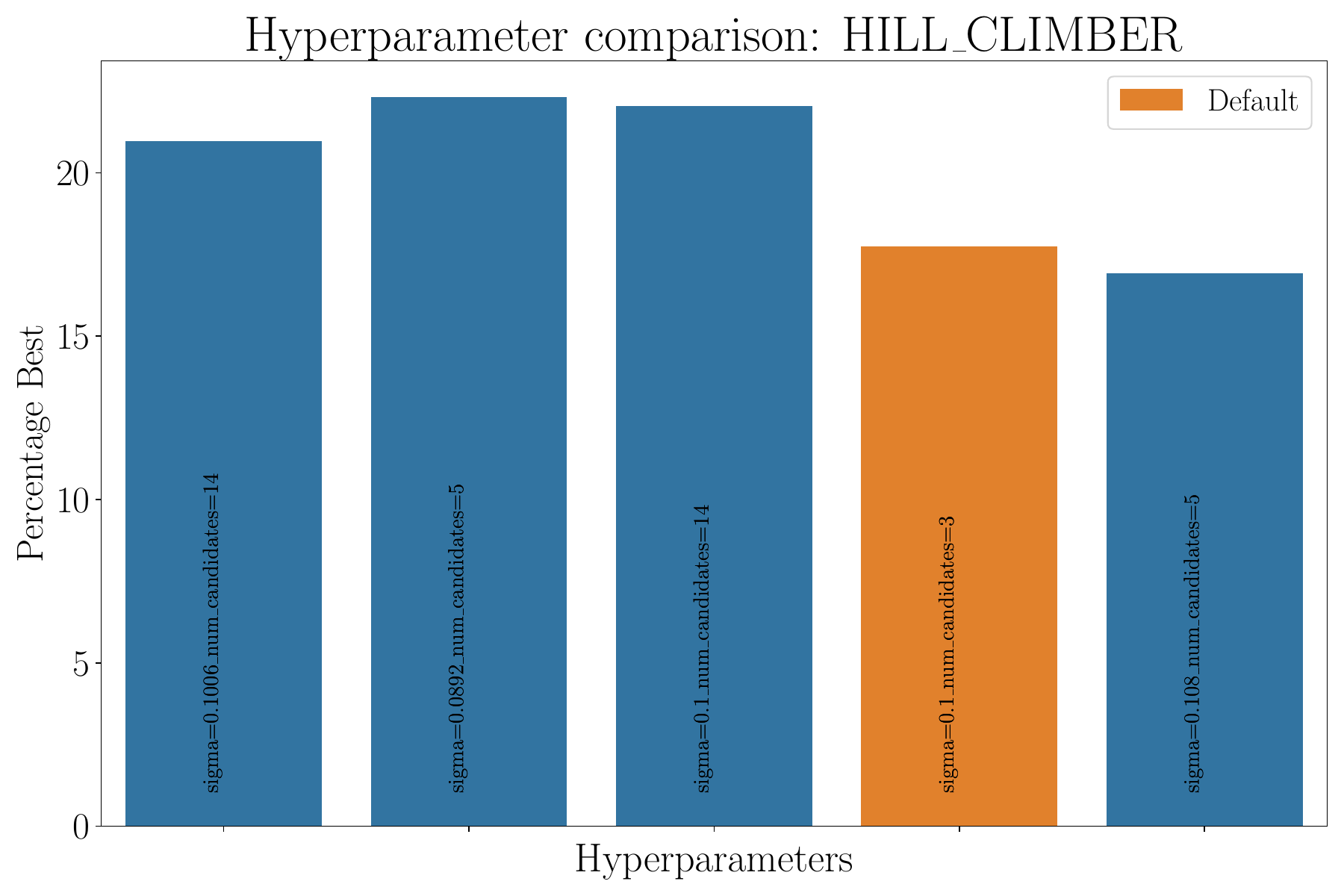}
    \caption{Hill climber hyperparameters.}
    
\end{figure}

\end{itemize}

\subsection{Coordinate Descent}
\label{subsec:coord_descent}

\begin{itemize}[-]
\item \textbf{Overview: } The algorithm proceeds by minimising each parameter in turn, and is discussed in detail in \cite{cade2020strategies}. For each parameter (i.e. keeping all others fixed), a
trigonometric interpolating polynomial is constructed by sampling the cost function at specific points. Once this model
is found (which is exact in theory, although the shot noise introduces an approximation error), then by differentiating
this model and considering the stationary points, we can update each parameter to it’s minimum value along that
parameter slice.

This style of algorithm has also been referred to as `sequential minimal optimization' \cite{nakanishi2020sequential} and `rotosolve' \cite{ostaszewski2021structure}.

We use the algorithm as presented in \cite{cade2020strategies}, except we introduce a hyperparameter \texttt{shuffle} which if set to True randomly shuffles the order at which we loop over the parameters (as opposed to always minimising the parameters
in the same order).
    \item \textbf{Implementation: } in house.
    \item \textbf{Evaluations per iteration: } $(4mn + 1)\nu$
    \item \textbf{Black box: } No, the algorithm is specific to the VQE problem and ansatz considered, and needs the grid size and
number of layers in order to calculate the number of interpolation points required.
    \item \textbf{gradient-based: } No.
    \item \textbf{References: } \cite{cade2020strategies, nakanishi2020sequential, tseng2001convergence, ostaszewski2021structure, parrish2019jacobi}.
    \item \textbf{Pseudocode: } see also \cite{cade2020strategies}.
    \begin{algorithm}
\SetKwFor{RepTimes}{repeat}{times}{end}
\caption{Coordinate Descent}
\KwIn{Initial parameters $x_0$, cost function $f(x)$, termination criteria.}
\Hyperparameters{$n$, $\sigma$.}
$x \gets x_0$ \Comment*[r]{set current parameters to initial parameters}
$v \gets f(x_0)$ \Comment*[r]{set current value to initial value}
\While{not terminate}{
$I \gets [1, \dots, \nu]$\;

\If{\texttt{shuffle}}{
$I \gets \texttt{random\_shuffle}(I)$;\
}

\For{$i$ in I }{
$g \gets \texttt{gradient\_of\_model}(x, f, i)$\;
$S  \gets \texttt{stationary\_points}(g)$\;
$y \gets \text{argmin}_{z \in S} ~ f(z)$\;
$x[i] \gets y$\;
}
}
\end{algorithm}
\item \textbf{Hyperparameters: }
\begin{table}[H]
    \centering
    \begin{tabular}{|c|c|} \hline
        Hyperparameter & \texttt{shuffle} \\\hline
        Description  & Randomises the order of parameters to optimise over \\\hline
        Default & \texttt{False}  \\\hline
        Other & \texttt{True} \\\hline
    \end{tabular}
    \caption{Coordinate descent hyperparameters.}
\end{table}

\begin{figure}[H]
    \centering
    \includegraphics[scale=0.3]{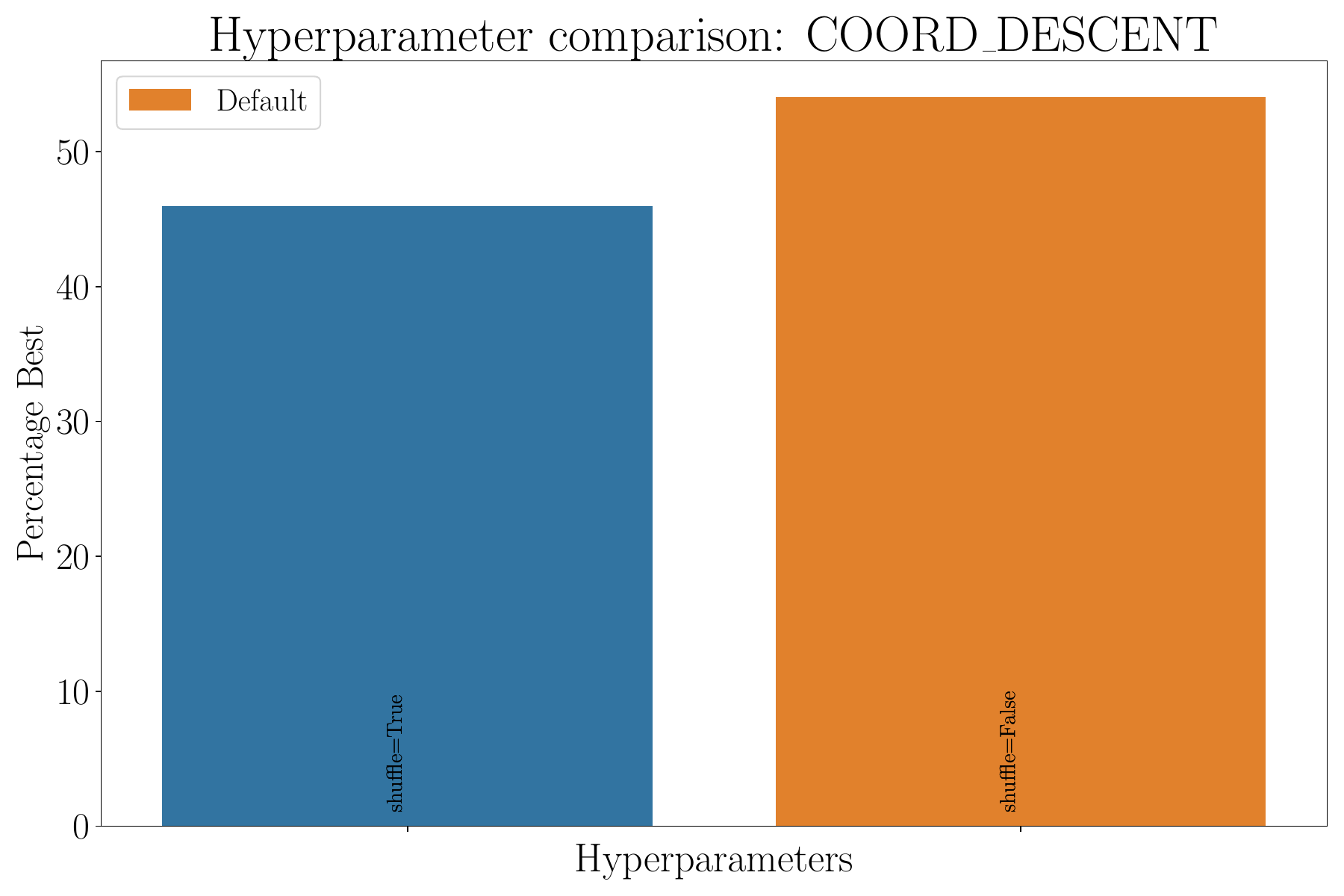}
    \caption{Coordinate descent hyperparameters.}
    
\end{figure}

\end{itemize}

\subsection{BayesMGD}
\label{subsec:bayes_mgd}

\begin{itemize}[-]
    \item \textbf{Overview: } 
Bayes Model Gradient Descent (BayesMGD) was introduced in \cite{stanisic2022observing} as an extension to Model Gradient Descent \cite{sung2020using}.
    \item \textbf{Implementation: } in house.
    \item \textbf{Evaluations per iteration: } $\frac{\eta}{2}(\nu +1)(\nu +2)$
    \item \textbf{Black box: } Yes, but requires a cost function that also returns the uncertainty (or standard error) in the value
returned.
    \item \textbf{gradient-based: } No.
    \item \textbf{References: }    \cite{stanisic2022observing, sung2020using}.
    \item \textbf{Pseudocode: } Detailed pseudocode is provided in \cite{stanisic2022observing}.
    \item \textbf{Hyperparameters: }
    
\begin{table}[H]
    \centering
    \begin{tabular}{|c|c|c|c|c|c|c|c|} \hline
        Hyperparameter & $\alpha$ & $\gamma$ & $A$ & $\delta$ & $\xi$ & $\eta$ & $l_0$ \\\hline
        Description  & \shortstack{learning rate \\ decay}  & \shortstack{learning \\ rate} & \shortstack{stability \\ constant} & \shortstack{sample \\ radius} & \shortstack{radius \\ decay } &  \shortstack{nevals \\ factor }  & \shortstack{length \\ scale}  \\\hline
        Default & 0.602 & 0.3 & 1.0 & 0.6 & 0.101 & 0.6 & 0.2 \\\hline
        Additional & 0.086 & 7.4755 & 0.0485 & 0.1208 & 0.0197 & 0.0185 & 0.0081\\
& 0.0487 & 1.2355 & 15.0934 & 0.234 & 0.0108 & 0.0706 & 0.1125\\
& 0.2337 & 3.5308 & 0.3922 & 0.2262 & 0.0225 & 0.0102 & 0.0924\\
 & 0.5303 & 0.2918 & 0.2582 & 0.3682 & 0.2253 & 0.0071 & 1.4672 \\\hline

    \end{tabular}
    \caption{BayesMGD hyperparameters.}
    \label{fig:bayes_hparams}
\end{table}

\begin{figure}[H]
    \centering
    \includegraphics[scale=0.3]{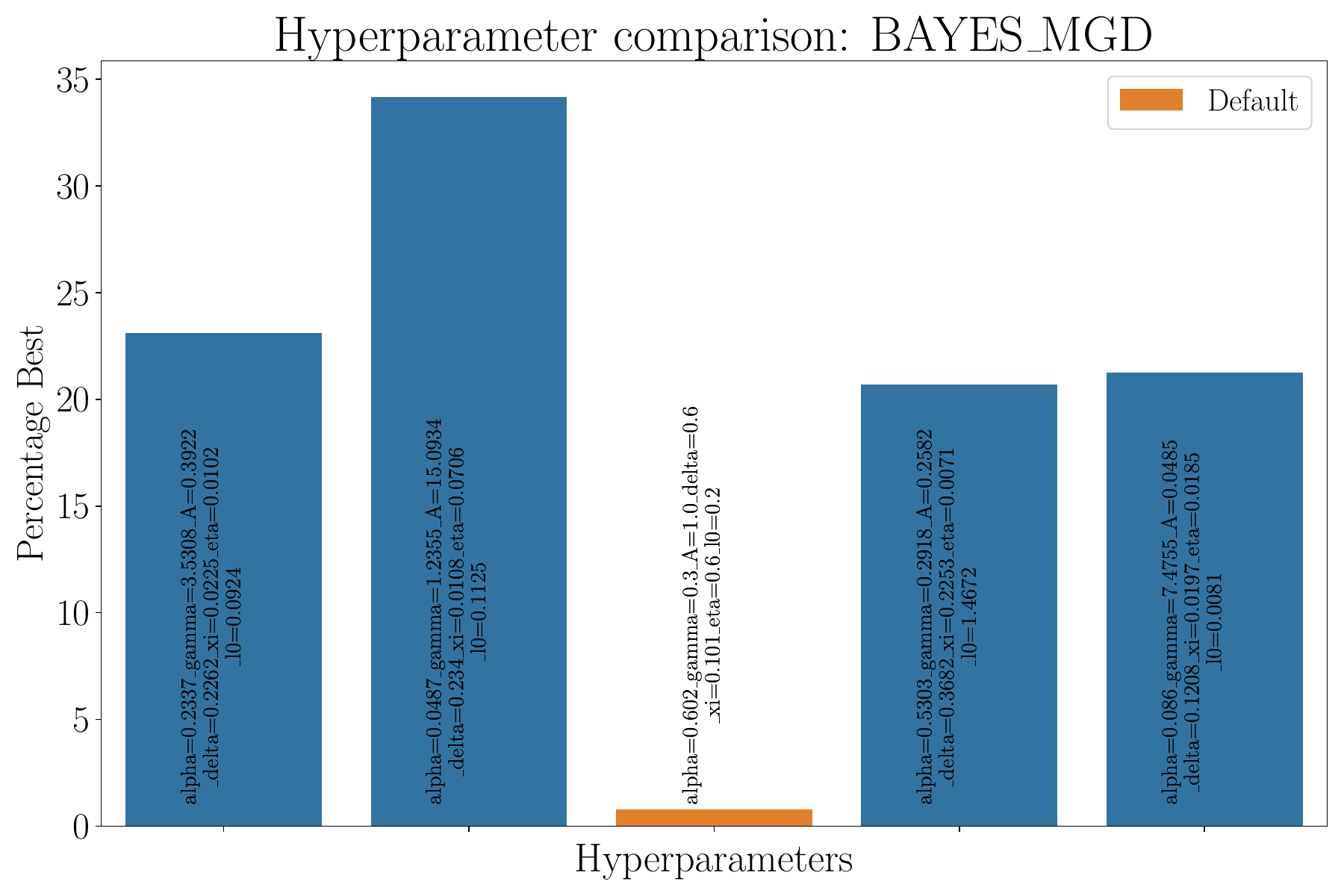}
    \caption{BayesMGD hyperparameters.}
    
\end{figure}

\end{itemize}

\subsection{SPSA}
\label{subsec:spsa}

\begin{itemize}[-]
    \item \textbf{Overview: } The simultaneous perturbation stochastic approximation (SPSA) algorithm follows a standard gradient descent structure, but uses a noisy approximation to the gradient using only two evaluations. In particular, for each parameter one samples randomly from $\{ -1, 1 \}$ to get a list $\Lambda$. For step size $c$, one defines $\theta = c \Lambda$, and then the gradient $g$ is approximated by
    \[
    g_i(x) = \frac{f(x+\theta) - f(x-\theta)}{2\theta_i}
    \]
    which only uses two evaluations of the cost function $f$. Then a gradient descent step follows with learning rate $a$; $x_{t+1} =x_t - a g(x_t)$. Another feature of SPSA is that the step size and learning rate both decay exponentially as the algorithm progresses.
    
    \item \textbf{Implementation: } In house.
    \item \textbf{Evaluations per iteration: } 2,  plus any additional calls to evaluate explicitly at the current parameters (we do this every 20 iterations).
    \item \textbf{Black box: } Yes.
    \item \textbf{gradient-based: } Yes, but approximation of the gradient is built into the algorithm (as opposed to assuming a gradient function is given).
    \item \textbf{References: }   \cite{spall1998overview, spall98implementation}.
    \item \textbf{Pseudocode: } See also  \cite{spall1998overview}.
    \begin{algorithm}
\caption{SPSA}
\KwIn{Initial parameters $x_0$, cost function $f(x)$, termination criteria.}
\Hyperparameters{$\alpha$, $\gamma$, $a$, $c$, $A$.}
$x \gets x_0$ \Comment*[r]{set current parameters to initial parameters}
$v \gets f(x_0)$ \Comment*[r]{set best value to initial value}
$k \gets 1$ \Comment*[r]{iteration tracker}
\While{not terminate}{
$a_k \gets \frac{a}{(k+A)^\alpha}$\;
$c_k \gets \frac{c}{k^\gamma}$\;
$\Delta_i = \texttt{Bernouilli}(-1,1)$\;
$\theta = c_k \Delta$\;
$g_i \gets \frac{f(x+\theta) - f(x-\theta)}{2\theta_i}$\;
$x \gets x - a_k g$\;
$k \gets k+1$\;
}
\end{algorithm}

    \item \textbf{Hyperparameters: } 5.
    \begin{table}[H]
    \centering
    \begin{tabular}{|c|c|c|c|c|c|} \hline
        Hyperparameter & $\alpha$ & $\gamma$ & $a$ & $c$ & $A$  \\\hline
        Description  & \shortstack{learning rate \\ decay}  & \shortstack{step size \\ decay} & \shortstack{learning \\ rate} & \shortstack{step \\ size} &  \shortstack{stability \\ constant}\\\hline
        Default & 0.602 & 0.101 & 0.2 & 0.15 & 1  \\\hline
        Additional &0.0247 & 0.4476 & 0.1208 & 0.4926 & 0.1579\\
& 0.0104 & 0.0074 & 0.045 & 0.0411 & 0.0285\\
& 0.0085 & 0.0381 & 0.0082 & 0.0149 & 0.3002\\
& 0.552 & 0.0383 & 0.1926 & 0.1202 & 61.5955 \\\hline
    \end{tabular}
    \caption{SPSA hyperparameters.}
    \label{fig:spsa_hparams}
\end{table}

\begin{figure}[H]
    \centering
    \includegraphics[scale=0.4]{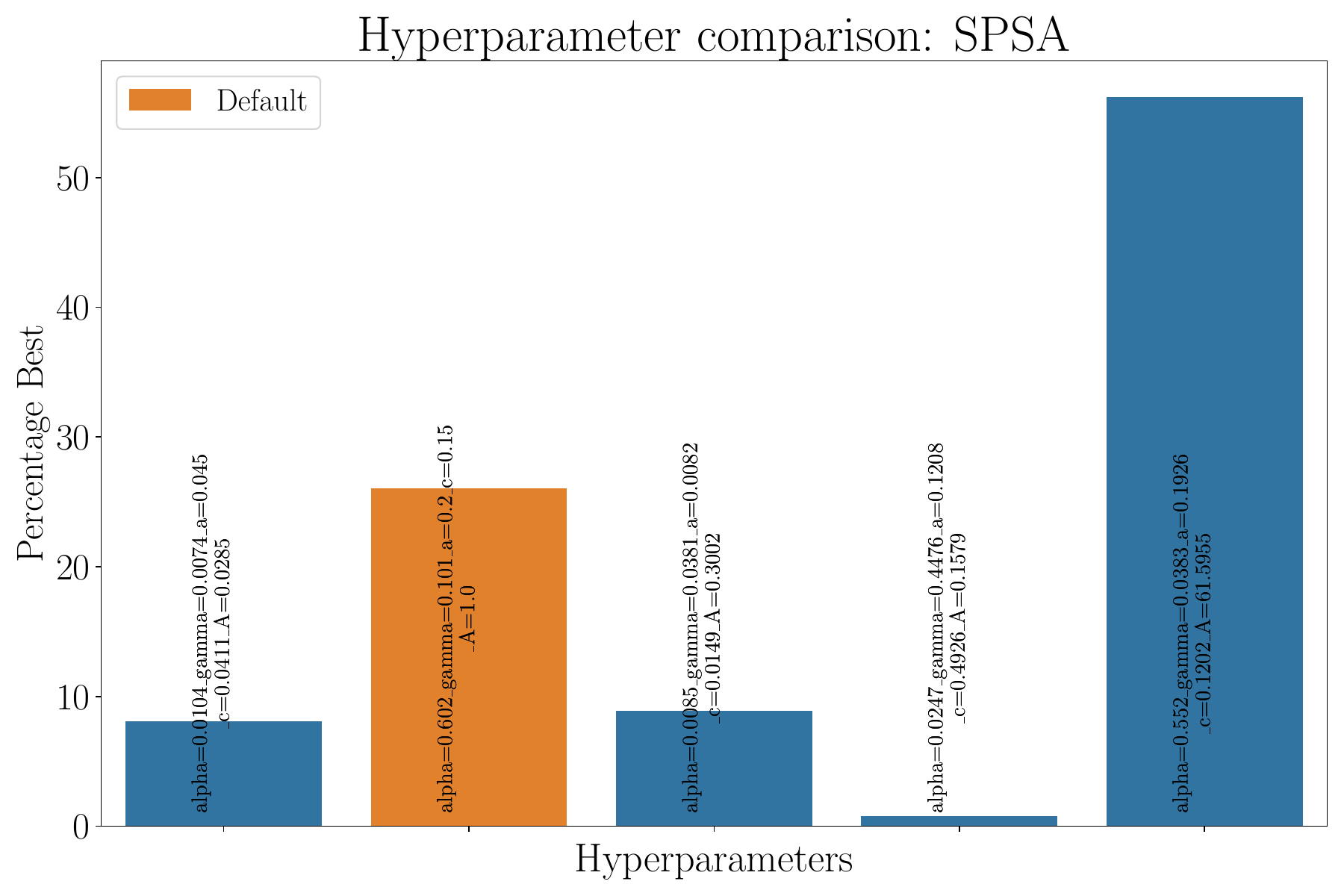}
    \caption{SPSA hyperparameters.}
    
\end{figure}

\end{itemize}

\subsection{Gradient Descent}
\label{subsec:grad_descent}

\begin{itemize}[-]
    \item \textbf{Overview: } Basic (or vanilla) gradient descent simply takes a step in the direction of negative gradient (i.e. downhill) at each iteration. A priori, the algorithm assumes access to some gradient function $\nabla f$, and takes as hyperparameter a learning rate $\eta$, the size of the step to take in the direction of negative gradient. The  parameter update step can be written as

\[
\theta_{t+1} = \theta_t - \eta \nabla f(\theta_t)
\]
    \item \textbf{Implementation: } In house.
    \item \textbf{Evaluations per iteration: } Evaluations required to calculate the gradient function $\nabla f$ , plus any additional calls to evaluate explicitly at the current parameters (we do this every 20 iterations).
    \item \textbf{Black box: } Yes.
    \item \textbf{gradient-based: } Yes.
    \item \textbf{References: } \cite{ruder2016overview}   .
    \item \textbf{Pseudocode: } \cite{ruder2016overview, tilly2022variational},
    \item \textbf{Hyperparameters: } 
    \begin{table}[H]
    \centering
    \begin{tabular}{|c|c|} \hline
    Hyperparameter & $\eta$ \\\hline
    Description  & learning rate \\\hline
    Default & 0.01 \\\hline
    Additional & 0.0157 \\
    & 0.1045 \\
    & 0.0336 \\
    & 0.0063 \\\cline{2-2}
    & 0.0076 \\ 
    & 0.0264 \\ 
    & 0.0095 \\ 
    & 0.005 \\\hline
    \end{tabular}
    \begin{tikzpicture}[remember picture, overlay]
\draw[thick,decorate,decoration={brace,amplitude=10pt,raise=4pt}]
 ($(pic cs:A) + (0, 1.5)$)   --  ($(pic cs:B) + (0, -0.8)$) node [midway, xshift=2cm] {\shortstack{Finite \\ Differences}};
\end{tikzpicture}
    \begin{tikzpicture}[remember picture, overlay]
\draw[thick,decorate,decoration={brace,amplitude=10pt,raise=4pt}]
 ($(pic cs:A) + (-0.1, -1.0)$)   --  ($(pic cs:B) + (-0.1, -3.2)$) node [midway,xshift=2.1cm] {\shortstack{Simultaneous \\ Perturbation}};
\end{tikzpicture}
    \caption{Gradient descent hyperparameters.}
    
\end{table}

\begin{figure}[H]
     \begin{subfigure}[b]{0.45\textwidth}
         \centering
         \includegraphics[width=\textwidth]{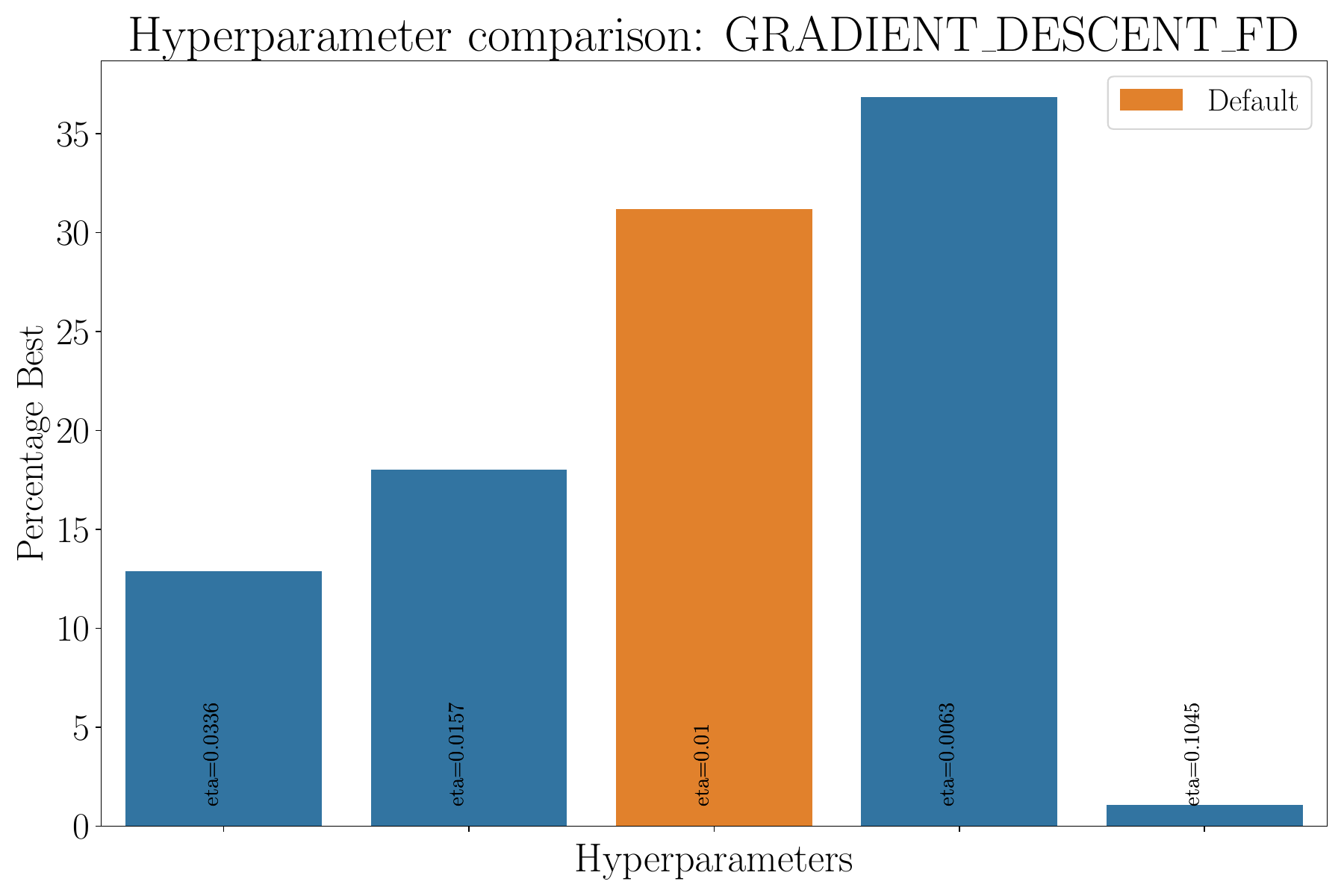}
         \caption{}
     \end{subfigure}
      \begin{subfigure}[b]{0.45\textwidth}
         \centering
         \includegraphics[width=\textwidth]{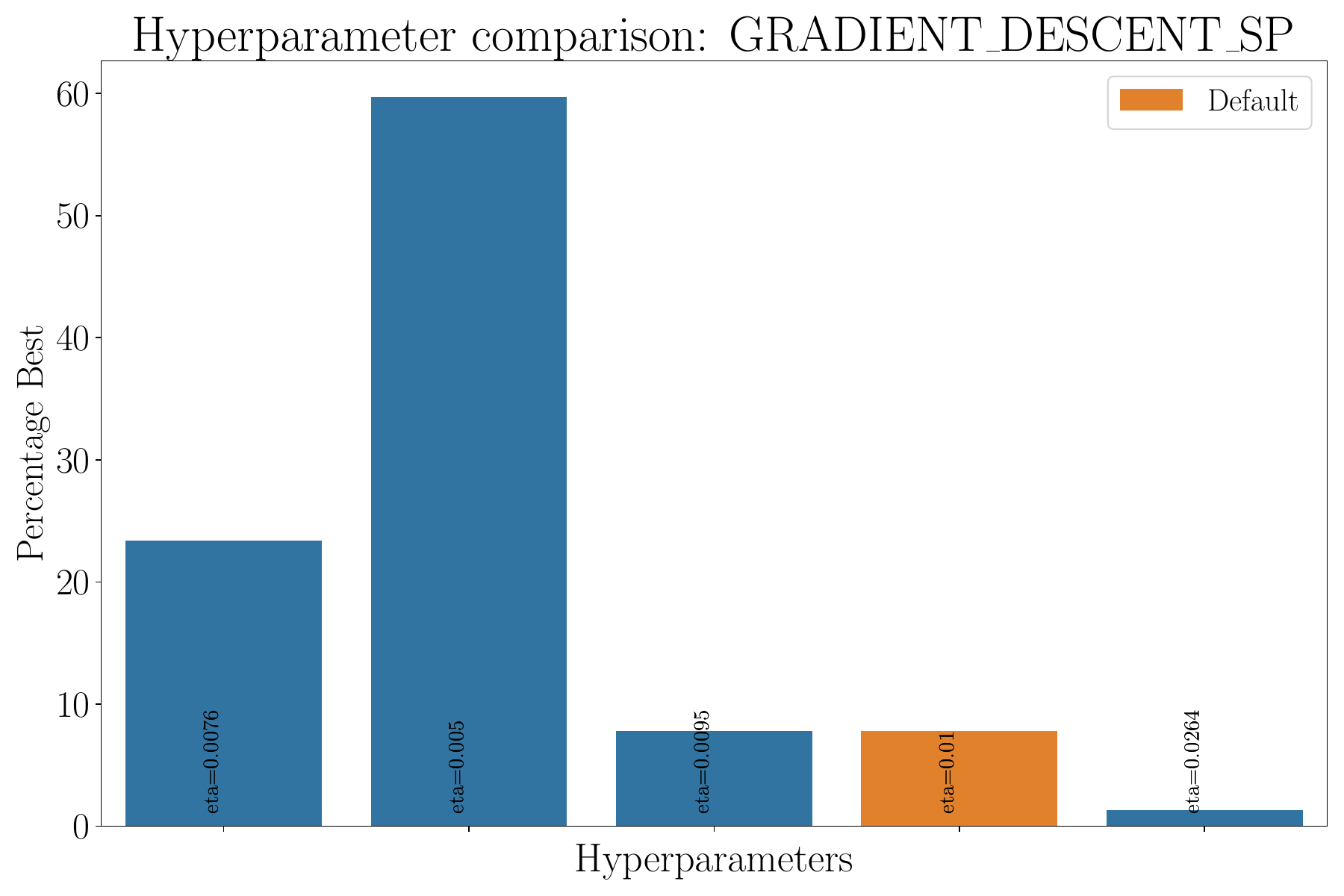}
         \caption{}
     \end{subfigure}
    \caption{Gradient descent hyperparameters.}
    
\end{figure}\end{itemize}

\subsection{Momentum}
\label{subsec:Momentum}

\begin{itemize}[-]
    \item \textbf{Overview: } One can extend vanilla gradient descent by including a velocity type term.
    \begin{align}
    v_{t+1} &= \gamma  v_t + \eta \nabla f(\theta_t) \\
\theta_{t+1} &= \theta_t - v_t 
    \end{align}

    A variant of this is known as nesterov, in which we instead evaluate the gradient at the approximate next parameters:
    \[
    v_{t+1} =\gamma  v_t + \eta \nabla f(\theta_t - \gamma v_t)
    \]
    We include this possibility as a boolean hyperparameter \texttt{nesterov}.
    
    \item \textbf{Implementation: } In house.
    \item \textbf{Evaluations per iteration: } Evaluations required to calculate the gradient function $\nabla f$ , plus any additional calls to evaluate explicitly at the current parameters (we do this every 20 iterations).
    \item \textbf{Black box: } Yes.
    \item \textbf{gradient-based: } Yes.
    \item \textbf{References: } \cite{ruder2016overview, qian1999momentum}   .
    \item \textbf{Pseudocode: }    \cite{ruder2016overview}.
    \item \textbf{Hyperparameters: } 3.
     
\begin{table}[H]
    \centering
    \begin{tabular}{|c|c|c|c|} \hline
        Hyperparameter & $\eta$ & $\gamma$ & \texttt{nesterov} \\\hline
        Description  & learning rate & momentum term & nesterov term  \\\hline
        Default & 0.01 & 0.9 & \texttt{False}  \\\hline
        Additional   & 0.7113 & 0.7238 & \texttt{False} \\
        & 0.3476 & 0.462  & \texttt{False} \\
        & 0.0657 & 0.7631  & \texttt{False} \\
        & 0.0739 & 0.6769 & \texttt{True}\\\cline{2-4}
        & 0.0001 & 0.3808  & \texttt{False} \\
        & 0.0182& 0.8154  & \texttt{True} \\
        & 0.7013 & 0.4951  & \texttt{True} \\
        & 0.0455 & 0.3182  & \texttt{False} \\\hline
    \end{tabular}
    \begin{tikzpicture}[remember picture, overlay]
\draw[thick,decorate,decoration={brace,amplitude=10pt,raise=4pt}]
 ($(pic cs:A) + (0, 1.5)$)   --  ($(pic cs:B) + (0, -0.8)$) node [midway, xshift=2cm] {\shortstack{Finite \\ Differences}};
\end{tikzpicture}
    \begin{tikzpicture}[remember picture, overlay]
\draw[thick,decorate,decoration={brace,amplitude=10pt,raise=4pt}]
 ($(pic cs:A) + (-0.1, -1.0)$)   --  ($(pic cs:B) + (-0.1, -3.2)$) node [midway,xshift=2.1cm] {\shortstack{Simultaneous \\ Perturbation}};
\end{tikzpicture}
    \caption{Momentum hyperparameters.}
    
\end{table}

\begin{figure}[H]
     \begin{subfigure}[b]{0.45\textwidth}
         \centering
         \includegraphics[width=\textwidth]{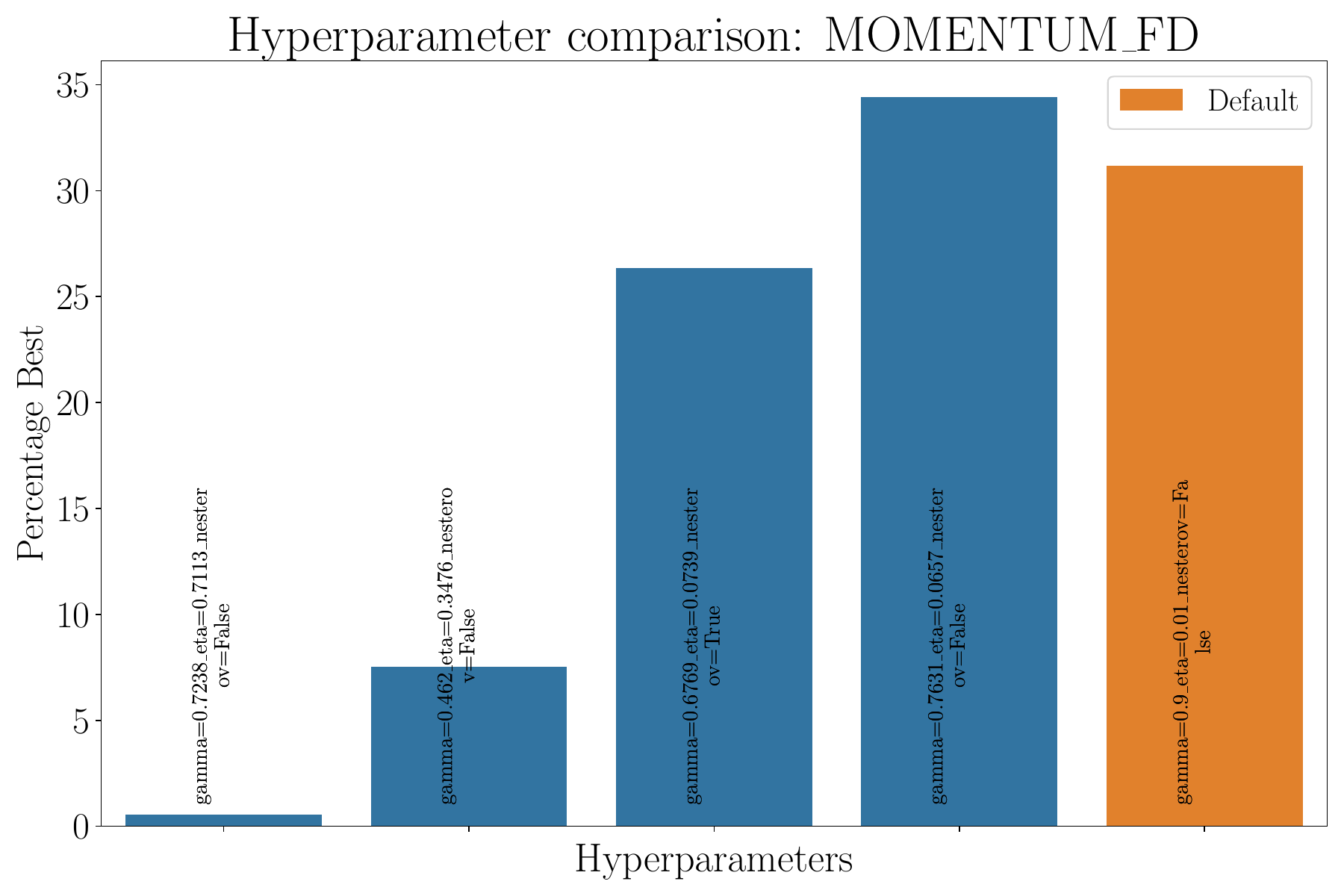}
         \caption{}
     \end{subfigure}
      \begin{subfigure}[b]{0.45\textwidth}
         \centering
         \includegraphics[width=\textwidth]{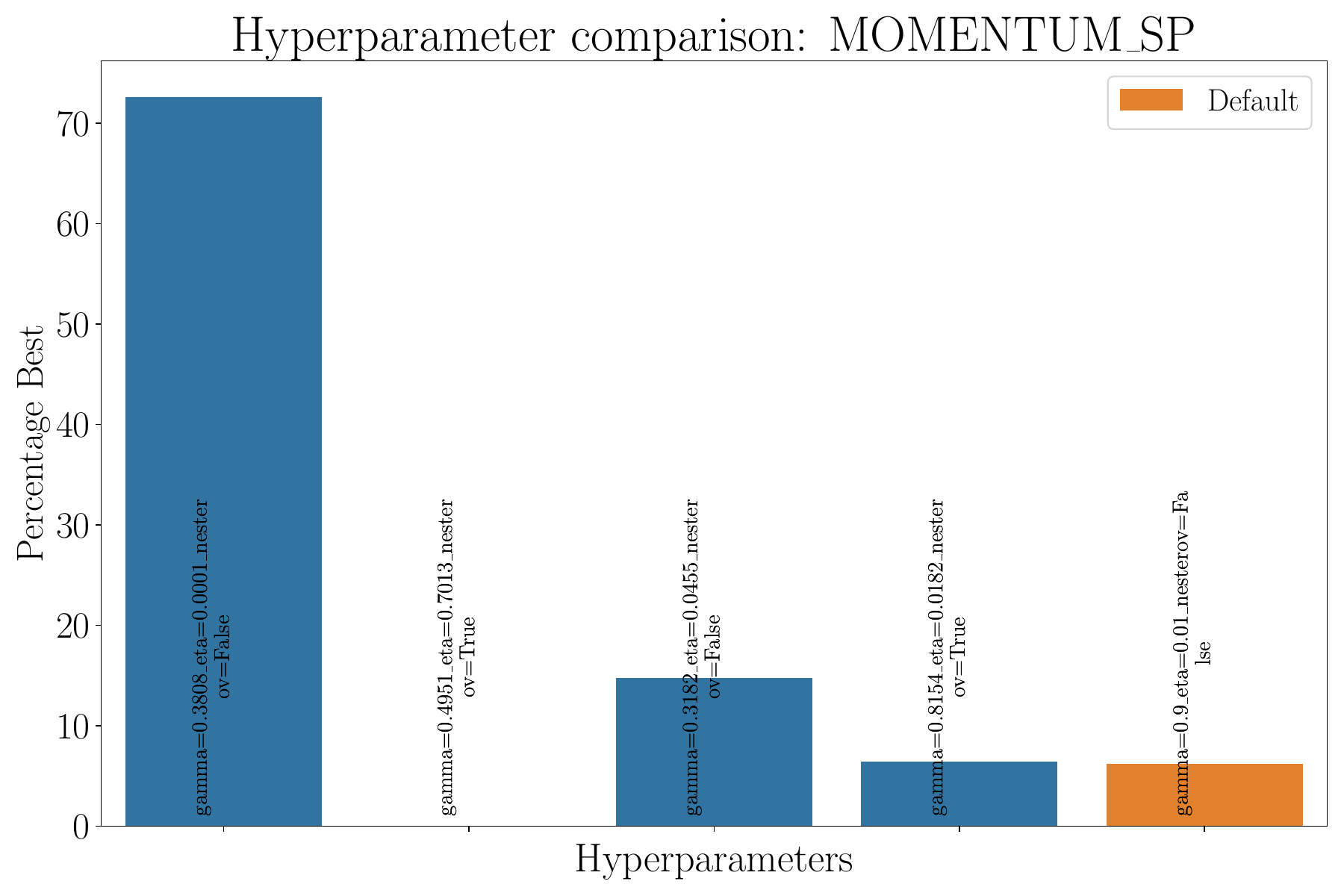}
         \caption{}
     \end{subfigure}
    \caption{Momentum hyperparameters.}
    
\end{figure}\end{itemize}

\subsection{AdaDelta}
\label{subsec:adadelta}

\begin{itemize}[-]
    \item \textbf{Overview: } AdaDelta \cite{zeiler2012adadelta} is an extension to Momentum (more specifically, to another method known as AdaGrad \cite{duchi2011adaptive}). Some of the motivations for this optimiser include removing the need to manually set a learning rate, using previous gradient information in a fixed window (and not for all time), and enforcing that the update step is consistent with the units of the parameters.
    \item \textbf{Implementation: } In house.
    \item \textbf{Evaluations per iteration: } Evaluations required to calculate the gradient function $\nabla f$ , plus any additional calls to evaluate explicitly at the current parameters (we do this every 20 iterations).
    \item \textbf{Black box: } Yes.
    \item \textbf{gradient-based: } Yes.
    \item \textbf{References: } \cite{zeiler2012adadelta, ruder2016overview}   .
    \item \textbf{Pseudocode: } \cite{ruder2016overview}    \item \textbf{Hyperparameters: } 
    
\begin{table}[H]
    \centering
    \begin{tabular}{|c|c|} \hline
        Hyperparameter & $\gamma$  \\\hline
        Description  & momentum term \\\hline
        Default & 0.9  \\\hline
        Additional   & 0.3318 \\
        & 0.8995 \\ 
        & 0.6603 \\
        & 0.99 \\\hline
        & 0.99 \\
        & 0.9921 \\ 
        & 0.0005 \\\hline
    \end{tabular}
    \begin{tikzpicture}[remember picture, overlay]
\draw[thick,decorate,decoration={brace,amplitude=10pt,raise=4pt}]
 ($(pic cs:A) + (0, 1.3)$)   --  ($(pic cs:B) + (0, -1.0)$) node [midway, xshift=2cm] {\shortstack{Finite \\ Differences}};
\end{tikzpicture}
    \begin{tikzpicture}[remember picture, overlay]
\draw[thick,decorate,decoration={brace,amplitude=10pt,raise=4pt}]
 ($(pic cs:A) + (-0.1, -1.3)$)   --  ($(pic cs:B) + (-0.1, -3.1)$) node [midway,xshift=2.1cm] {\shortstack{Simultaneous \\ Perturbation}};
\end{tikzpicture}
    \caption{AdaDelta hyperparameters.}
    
\end{table}

\begin{figure}[H]
     \begin{subfigure}[b]{0.45\textwidth}
         \centering
         \includegraphics[width=\textwidth]{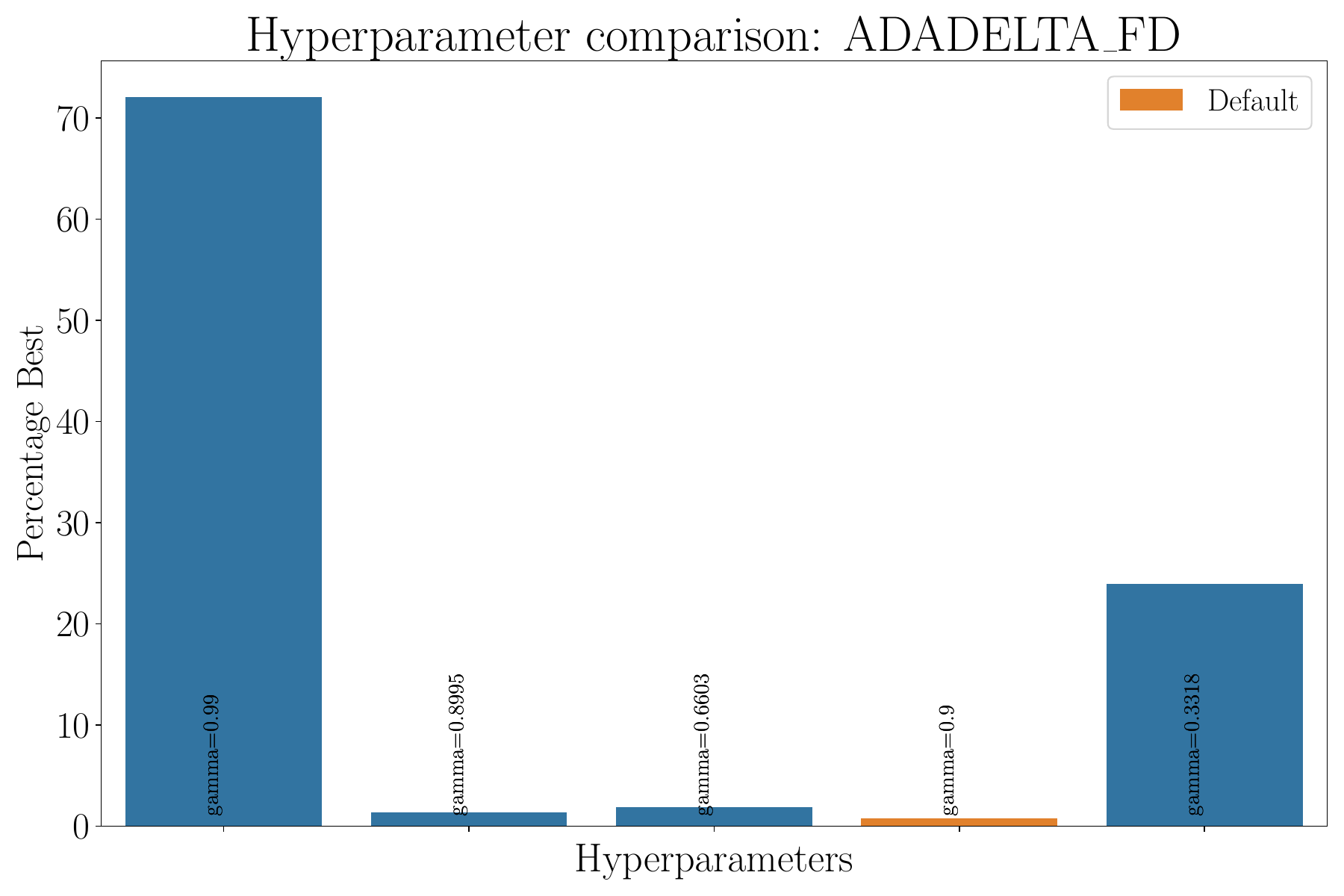}
         \caption{}
         
     \end{subfigure}
      \begin{subfigure}[b]{0.45\textwidth}
         \centering
         \includegraphics[width=\textwidth]{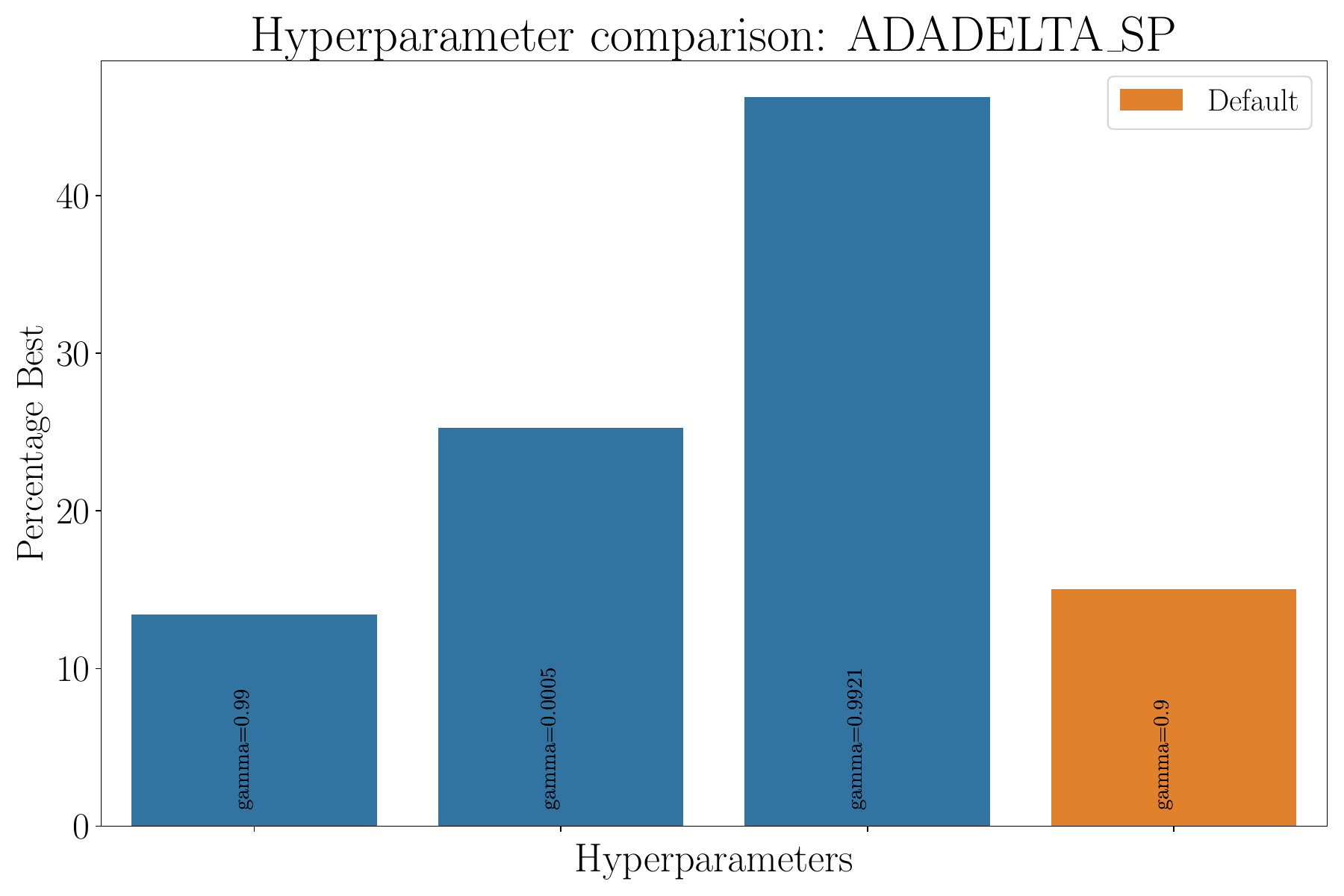}
         \caption{}
         
     \end{subfigure}
    \caption{AdaDelta hyperparameters.}
    
\end{figure}

   \end{itemize}

\subsection{Adam}
\label{subsec:adam}

\begin{itemize}[-]
    \item \textbf{Overview: } Adaptive Moment Estimation (Adam) \cite{kingma2014adam} has become a popular method in machine learning, storing exponentially decaying averages of the first and second moments of the gradient, and then uses a similar update step to that of AdaDelta. 
    \item \textbf{Implementation: } In house.
    \item \textbf{Evaluations per iteration: } Evaluations required to calculate the gradient function $\nabla f$ , plus any additional calls to evaluate explicitly at the current parameters (we do this every 20 iterations).
    \item \textbf{Black box: } Yes.
    \item \textbf{gradient-based: } Yes.
    \item \textbf{References: } \cite{ruder2016overview, kingma2014adam}. 
    \item \textbf{Pseudocode: } \cite{kingma2014adam, ruder2016overview, tilly2022variational}. 
    \item \textbf{Hyperparameters: }

\begin{table}[H]
    \centering
    \begin{tabular}{|c|c|c|c|c|} \hline
        Hyperparameter & $\alpha$ & $\beta_1$ & $\beta_2$ &  \texttt{nadam} \\\hline
        Description  & step size & first moment & second moment & nadam term  \\\hline
        Default \cite{kingma2014adam} & 0.001 & 0.9 & 0.999 & \texttt{False}  \\\hline
        Additional &0.0701 & 0.6798 & 0.0928& \texttt{True} \\
        &0.0992 & 0.5233 & 0.0652 & \texttt{True} \\
        &0.0 & 0.6223 & 0.8977 & \texttt{False}\\
        &0.0727 & 0.0168 & 0.479 & \texttt{False} \\\cline{2-5}
        & 0.0692 & 0.0996 & 0.4881& \texttt{False} \\& 0.0575 & 0.0904 & 0.3212& \texttt{False} \\& 0.0424 & 0.5542 & 0.2278& \texttt{False} \\& 0.0256 & 0.7416 & 0.3236& \texttt{False} \\\hline
    \end{tabular}
    \begin{tikzpicture}[remember picture, overlay]
\draw[thick,decorate,decoration={brace,amplitude=10pt,raise=4pt}]
 ($(pic cs:A) + (0, 1.5)$)   --  ($(pic cs:B) + (0, -0.8)$) node [midway, xshift=2cm] {\shortstack{Finite \\ Differences}};
\end{tikzpicture}
    \begin{tikzpicture}[remember picture, overlay]
\draw[thick,decorate,decoration={brace,amplitude=10pt,raise=4pt}]
 ($(pic cs:A) + (-0.1, -1.0)$)   --  ($(pic cs:B) + (-0.1, -3.2)$) node [midway,xshift=2.1cm] {\shortstack{Simultaneous \\ Perturbation}};
\end{tikzpicture}
    \caption{Adam hyperparameters.}
    
\end{table}

\begin{figure}[H]
     \begin{subfigure}[b]{0.45\textwidth}
         \centering
         \includegraphics[width=\textwidth]{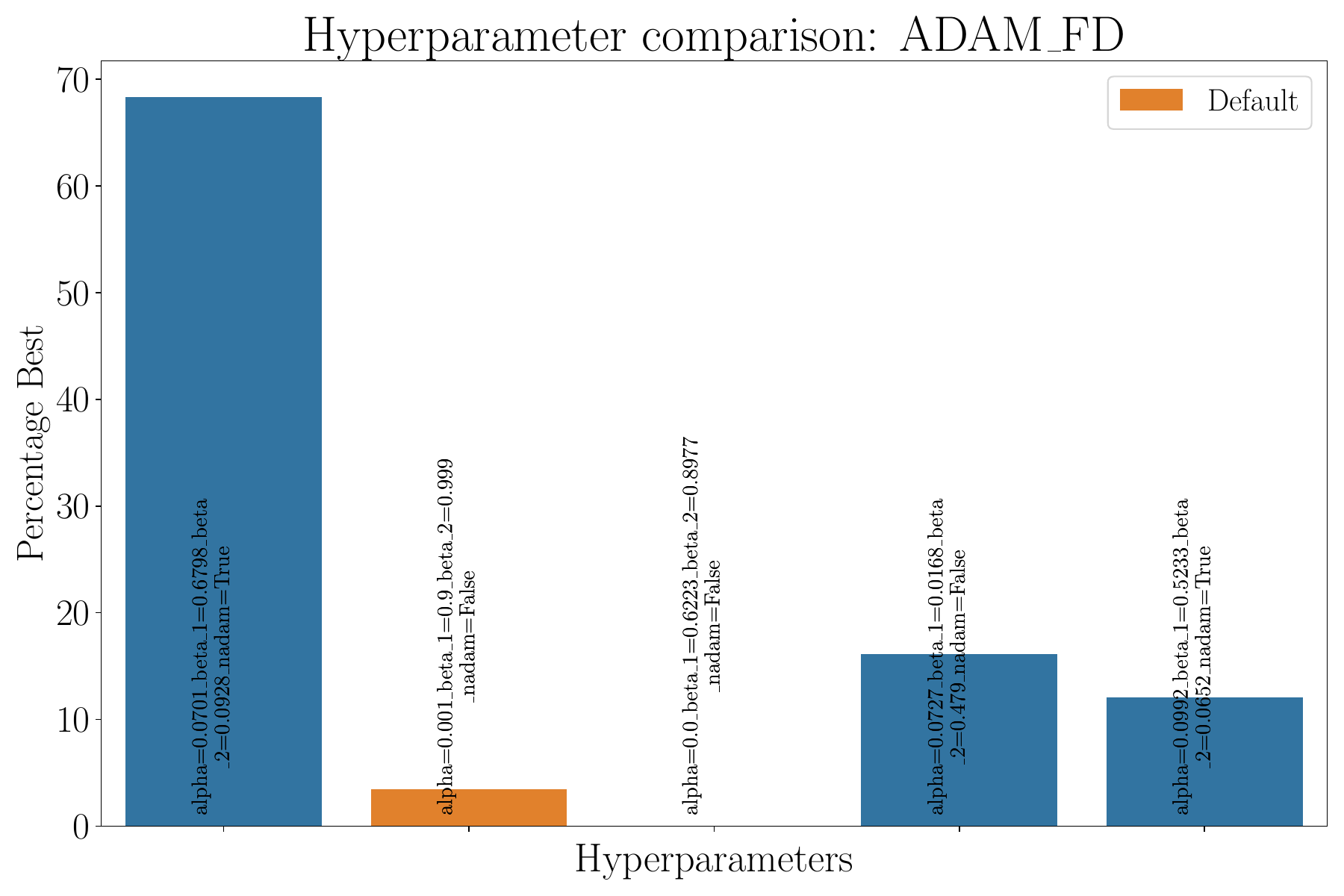}
         \caption{}
         
     \end{subfigure}
      \begin{subfigure}[b]{0.45\textwidth}
         \centering
         \includegraphics[width=\textwidth]{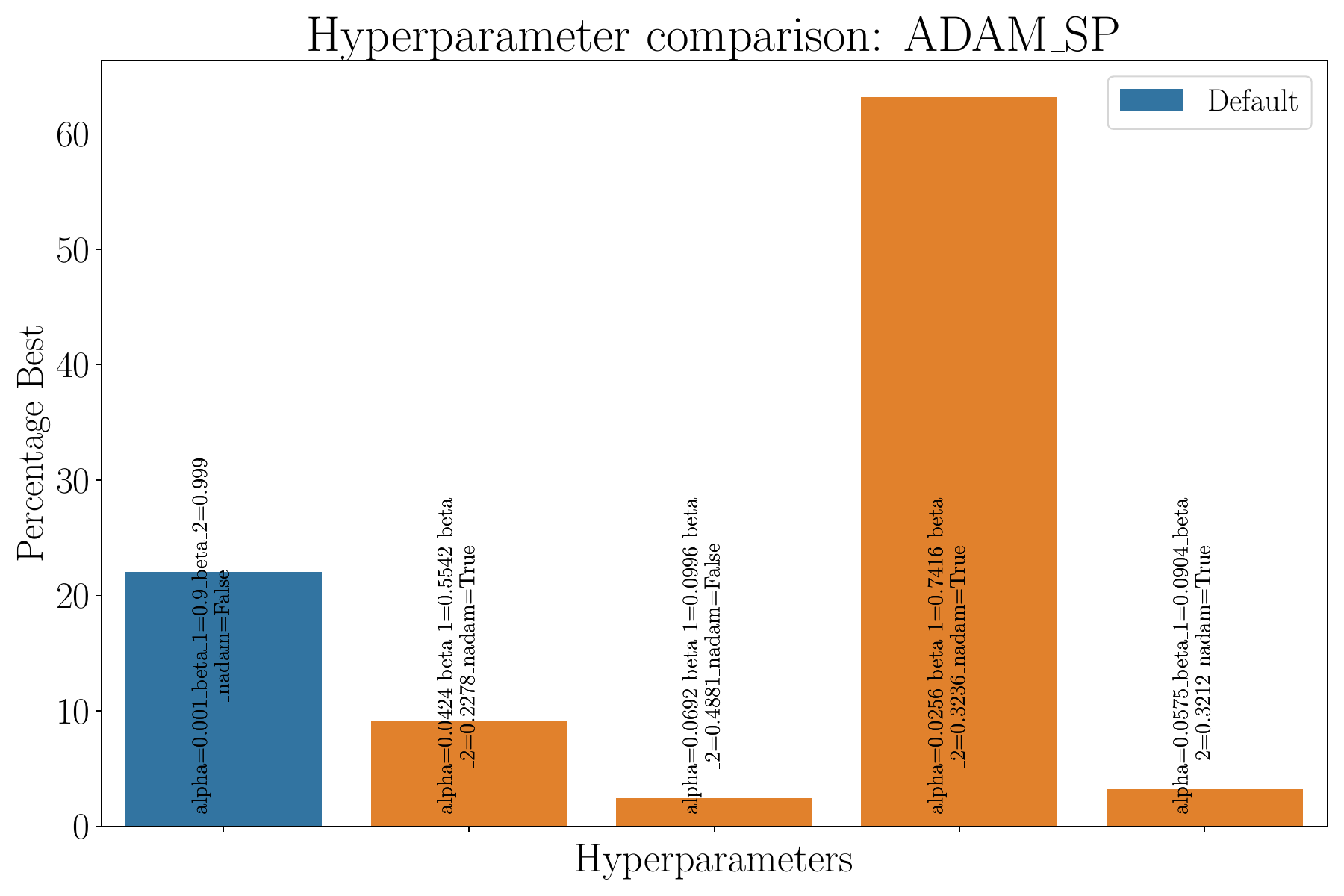}
         \caption{}
         
     \end{subfigure}
    \caption{Adam hyperparameters.}
    
\end{figure}
    
    \end{itemize}

\subsection{$\lambda + \mu$}
\label{subsec:lam_mu}

\begin{itemize}[-]
    \item \textbf{Overview: } 
In a basic evolutionary strategy, at each iteration (generation), $\mu$ `parents' are selected to produce $\lambda$ `offspring'. Two functions must be specified: how to combine two solutions to produce a child solution (crossover, mating), and a mutation function that acts on an individual solution. 

Once crossover and mutation have occurred, we then select the best $\mu$ individuals to continue the process. If the original parents are included in this consideration, then the algorithm is termed $\mu + \lambda$ (``mu plus lambda''), an alternative only select the next $\mu$ parents from the $\lambda$ children produced, this is known as a ($\mu$, $\lambda)$ (``mu comma lambda'') evolutionary strategy.
    \item \textbf{Implementation: } DEAP \cite{deap}.
    \item \textbf{Evaluations per iteration: } $\lambda$.
    \item \textbf{Black box: } Yes.
    \item \textbf{gradient-based: } No.
    \item \textbf{References: } \cite{deap, Luke2013Metaheuristics}.    
    \item \textbf{Pseudocode: }  Page 34 of \cite{Luke2013Metaheuristics}   \item \textbf{Hyperparameters: } 11.
    
    \begin{table}[H]
    \centering
    \begin{tabular}{|c|c|c|c|c|c|} \hline
         & \texttt{min\_strat} & \texttt{max\_strat} & $\mu$ & $\lambda$-\texttt{factor} & $\alpha$ \\\hline
         Description & \shortstack{lower bound on \\ standard deviation} & \shortstack{upper bound on \\ standard deviation} & \shortstack{population \\ size} & \shortstack{ number of children \\ $= \mu \times \lambda$-factor}  & \shortstack{crossover \\ blend} \\\hline
        Default &0.01 & 5.0 & 2 & 5 & 0.1  \\\hline
        Additional & 0.4857 & 2.0073 & 10 & 6 & 0.1091     \\
                    & 0.1164 & 3.5681 & 7 & 1 & 8.1756 \\
                    & 0.7165 & 3.4974 & 5 & 8 & 3.9816 \\
                    & 0.0555 & 4.8834 & 12 & 6 & 0.0209\\\hline
    \end{tabular}

    \begin{tabular}{|c|c|c|c|c|c|c|} \hline
         & $\sigma$ & $c$ & \texttt{indpb} & \texttt{pb\_sum} & \texttt{pb\_cut} & \texttt{tournsize} \\\hline
         Description & \shortstack{normal std of \\ initial population \\ about initial vector} & \shortstack{strategy \\ learning \\parameter} & \shortstack{mutation \\ probability} & \shortstack{sum of \\ mutpb\\ and cxpb}  &  \shortstack{split of mutpb \\ and cxpb over\\ pb\_sum} &  \shortstack{number of \\ individuals in \\ each tournament.} \\\hline
        Default & 0.1 & 1.0 & 0.03 & 0.9 & 0.333 & 3 \\\hline
        Additional  & 0.1853 & 0.7551 & 0.3484 & 0.305 & 0.0337 & 6 \\
         & 0.0161 & 0.3634 & 0.108 & 0.2048 & 0.7677 & 5 \\
         & 0.0431 & 0.2467 & 0.5829 & 0.9431 & 0.0374 & 10 \\
         & 0.1759 & 0.0479 & 0.4767 & 0.2732 & 0.3227 & 5 \\\hline
    \end{tabular}
    \caption{$\lambda + \mu$ hyperparameters.}
    
\end{table}

\begin{figure}[H]
    \centering
    \includegraphics[scale=0.3]{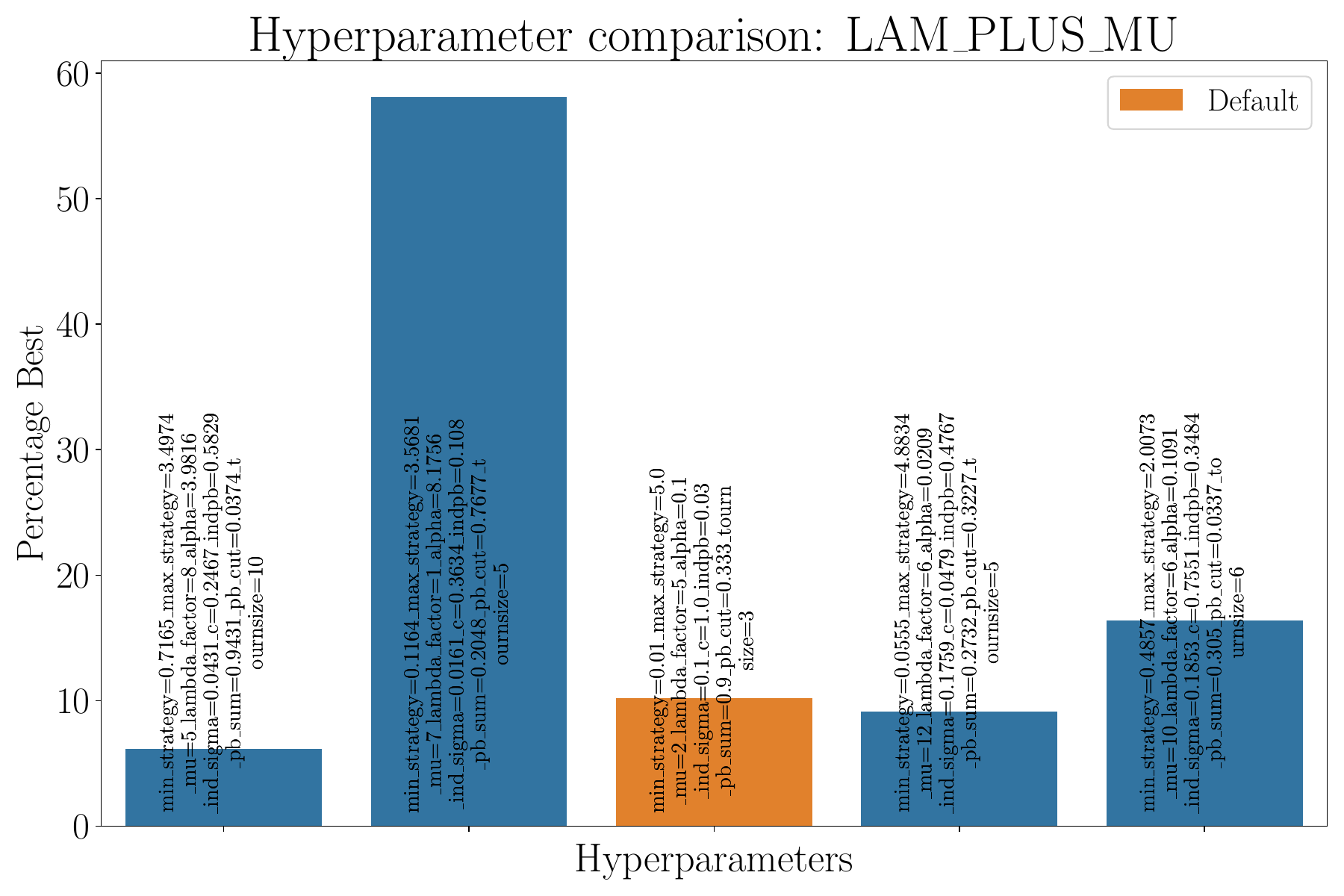}
    \caption{$\lambda + \mu$ hyperparameters.}
\end{figure}

\end{itemize}

\subsection{Particle Swarm Optimisation}
\label{subsec:pso}

\begin{itemize}[-]
\item \textbf{Overview: } Particle Swarm Optimisation (PSO) is an evolutionary algorithm consisting of a set of candidate solutions (particles) exploring the search space. At each iteration, the location and velocity of each particle is updated according to the individual particle fitness, as well as the global best fitness found.
    \item \textbf{Implementation: } DEAP \cite{deap}.
    \item \textbf{Evaluations per iteration: } The size of the population (default is 5).
    \item \textbf{Black box: } Yes.
    \item \textbf{gradient-based: } No.
    \item \textbf{References: } \cite{deap, Luke2013Metaheuristics}.    
    \item \textbf{Pseudocode: }  Page 57 of \cite{Luke2013Metaheuristics}.
    \item \textbf{Hyperparameters: }

\begin{table}[H]
    \centering
    \begin{tabular}{|c|c|c|c|c|c|c|} \hline
         & \texttt{pop\_size} & \texttt{ind\_sigma} & \texttt{smin} & \texttt{smax} & $\phi_1$ & $\phi_2$ \\\hline
         Description & \shortstack{Number of \\ particles} & \shortstack{normal std \\ of initial \\ population \\ about initial \\ vector} & \shortstack{lower speed \\ limit} & \shortstack{upper speed \\ limit} & \shortstack{max speed \\ change towards \\ particle best}  & \shortstack{max speed \\ change towards \\ overall best} \\\hline
         Default & 5 & 0.1 & -3.0 & 3.0 & 2.0 & 2.0 \\\hline
        Other    & 14 & 0.072 & -9.4335 & 0.3746 & 3.0398 & 2.5012 \\
            & 17 & 0.0327 & -8.8692 & 0.267 & 2.9392 & 6.8784 \\
            & 2 & 0.0869 & -4.1556 & 0.1707 & 7.5111 & 1.6902 \\
            & 10 & 0.143 & -8.1941 & 0.2252 & 5.0192 & 3.398 \\\hline
    \end{tabular}
    \caption{PSO hyperparameters.}
    
\end{table}

\begin{figure}[H]
    \centering
    \includegraphics[scale=0.3]{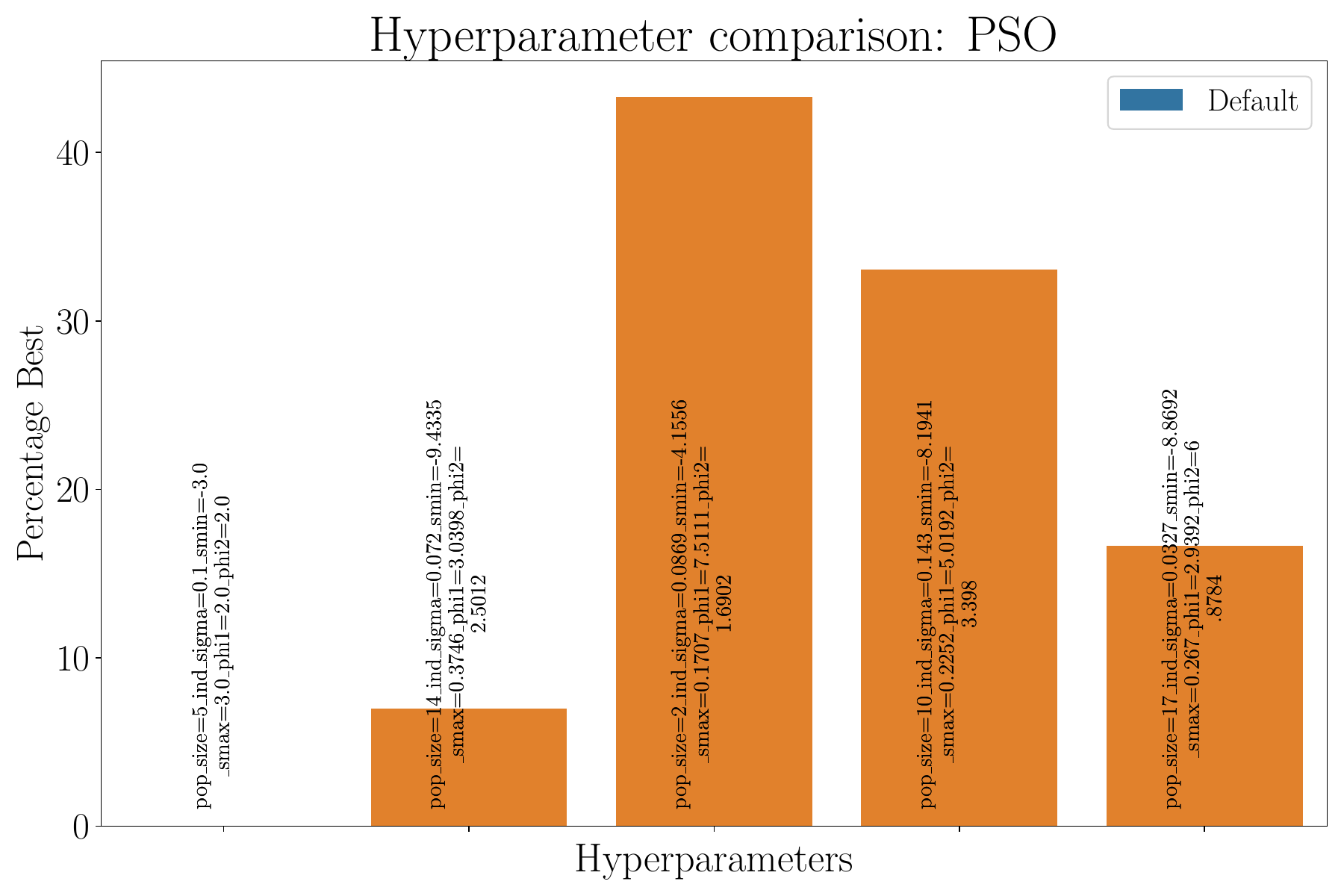}
    \caption{PSO hyperparameters.}
    
\end{figure}

    \end{itemize}

\subsection{CMAES}
\label{subsec:cmaes}

\begin{itemize}[-]
    \item \textbf{Overview: } The Covariance Matrix Adaption Evolution Strategy (CMAES) is a black-box numerical optimiser. At each generation, new candidate solutions are sampled from a multivariate normal distribution subject to a covariance matrix (which encodes dependencies between the individual variables). The technical aspect of the algorithm stems from the update rule of the covariance matrix.
    \item \textbf{Implementation: } DEAP \cite{deap}.
    \item \textbf{Evaluations per iteration: } $\lambda$.
    \item \textbf{Black box: } Yes.
    \item \textbf{gradient-based: } No.
    \item \textbf{References: } \cite{deap, Luke2013Metaheuristics, hansen2001completely, hansen2016cma}.    
    \item \textbf{Pseudocode: } \cite{hansen2016cma}     \item \textbf{Hyperparameters: } 

\begin{table}[H]
    \centering
    \begin{tabular}{|c|c|} \hline
        Hyperparameter & $\sigma$  \\\hline
        Description  & initial standard deviation\\\hline
        Default & 0.1  \\\hline
        Other & 0.1104  \\
        & 0.085  \\
        & 0.1726  \\
        & 0.1332  \\\hline
    \end{tabular}
    \caption{CMAES hyperparameters.}
    
\end{table}

\begin{figure}[H]
    \centering
    \includegraphics[scale=0.3]{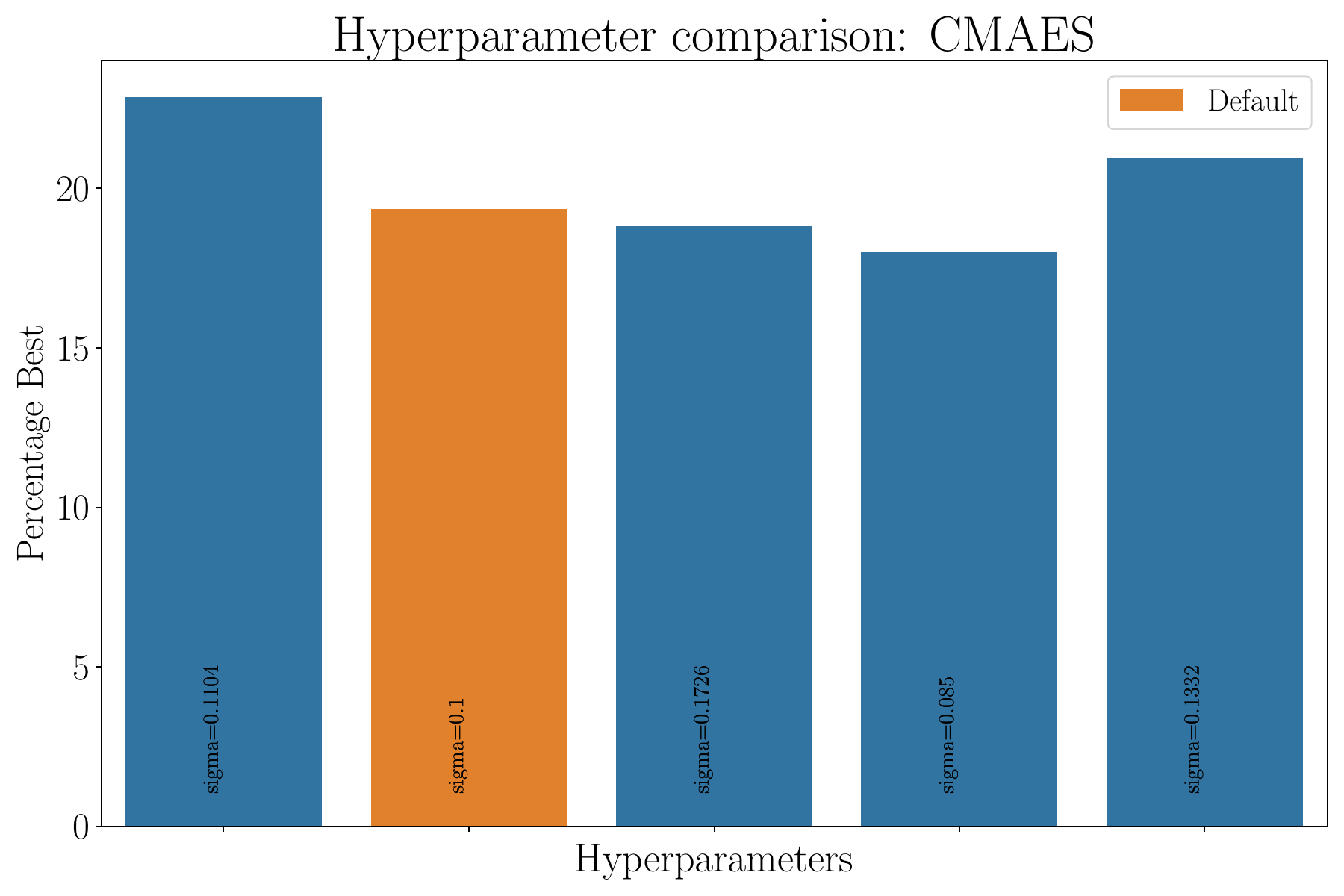}
    \caption{CMAES hyperparameters.}
    \label{fig:cmaes_hparams}
    
\end{figure}

    \end{itemize}

\subsection{Nelder-Mead}
\label{subsec:nelder_mead}

\begin{itemize}[-]
    \item \textbf{Implementation: } \texttt{scipy} \cite{scipy}.
    \item \textbf{Evaluations per iteration: } $ \nu +1$ (in the worst case).
    \item \textbf{Black box: } Yes.
    \item \textbf{gradient-based: } No.
    \item \textbf{References: }   \cite{scipy, gao2012implementing, nelder1965simplex}. 
    \item \textbf{Pseudocode: } \cite{nocedal1999numerical}.
    \item \textbf{Hyperparameters: } 
    \begin{table}[H]
    \centering
    \begin{tabular}{|c|c|c|} \hline
        Hyperparameter & \texttt{adaptive} & \texttt{bounds}  \\\hline
        Description \cite{scipy}  & \shortstack{adapt parameters \\ to dimension} & \shortstack{bound \\parameters} \\\hline
        Default & \texttt{True} & \texttt{True} \\\hline
        Other & \texttt{True}&  \texttt{False}\\
          & \texttt{False}&  \texttt{True}\\
           & \texttt{False}&  \texttt{False}\\\hline
    \end{tabular}
    \caption{Coordinate descent hyperparameters.}
\end{table}
    \begin{figure}[H]
    \centering
    \includegraphics[scale=0.3]{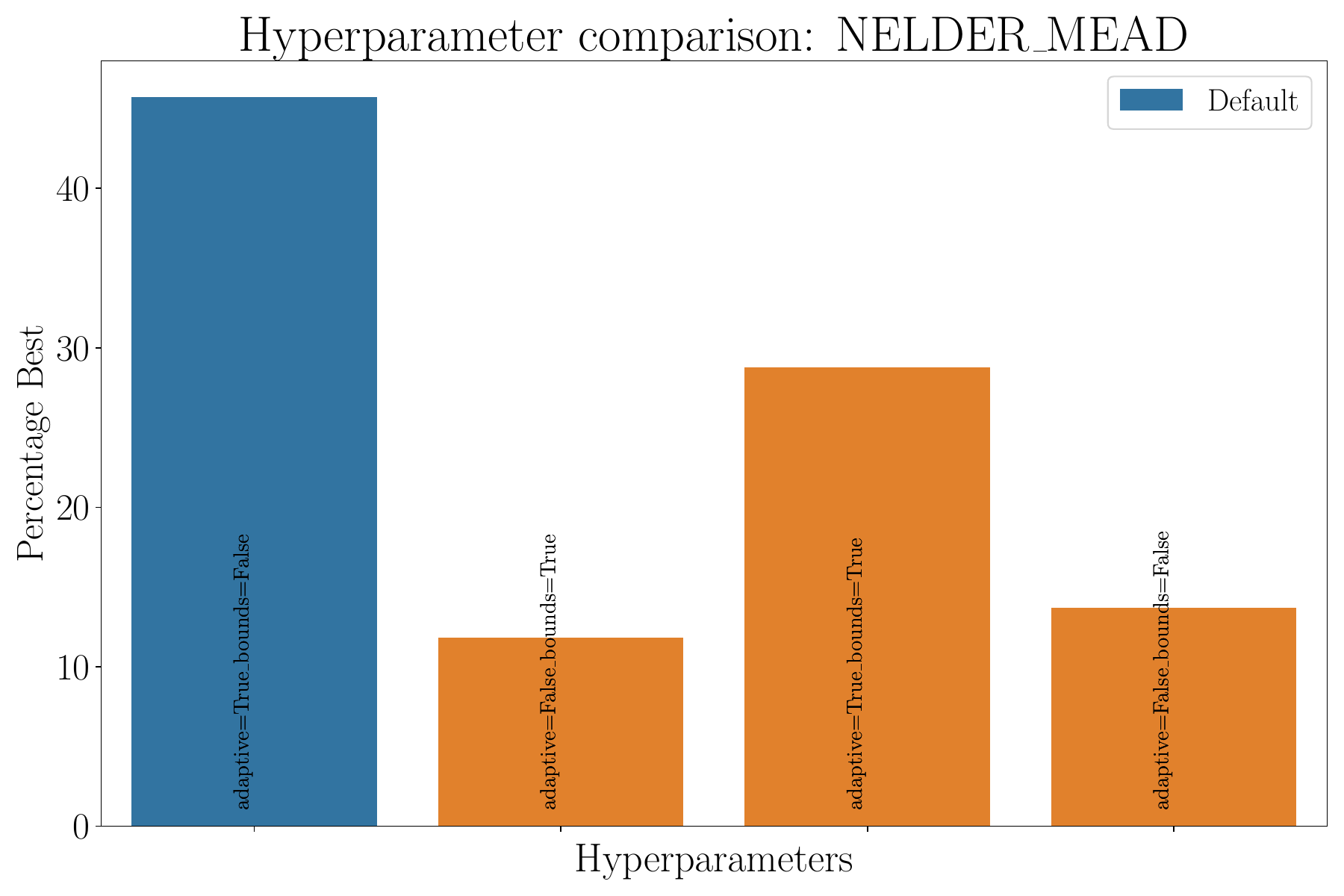}
    \caption{Nelder-Mead hyperparameters.}
    
\end{figure}\end{itemize}

\subsection{Other optimisers}

\subsection*{COBYLA}
\label{subsec:cobyla}

\begin{itemize}[-]
    \item \textbf{Implementation: } \texttt{scipy} \cite{scipy}.
    \item \textbf{Black box: } Yes.
    \item \textbf{gradient-based: } No.
    \item \textbf{References: }   \cite{powell1994direct}. 
    \end{itemize}

\subsection*{BFGS}
\label{subsec:bfgs}

\begin{itemize}[-]
    \item \textbf{Implementation: } \texttt{scipy} \cite{scipy}.
    \item \textbf{Black box: } Yes.
    \item \textbf{gradient-based: } Yes.
    \item \textbf{References: }   \cite{scipy, kochenderfer2019algorithms}. 
    \end{itemize}

\subsection*{L-BFGS-B}
\label{subsec:l_bfgs_b}

\begin{itemize}[-]
    \item \textbf{Implementation: } \texttt{scipy} \cite{scipy}.
    \item \textbf{Black box: } Yes.
    \item \textbf{gradient-based: } Yes.
    \item \textbf{References: }   \cite{scipy, kochenderfer2019algorithms}. 
    \end{itemize}

\subsection*{Powell}
\label{subsec:powell}

\begin{itemize}[-]
    \item \textbf{Implementation: } \texttt{scipy} \cite{scipy}.
    \item \textbf{Black box: } Yes.
    \item \textbf{gradient-based: } No.
    \item \textbf{References: }   \cite{scipy, powell1994direct, powell2007view, kochenderfer2019algorithms} . 
    \end{itemize}

\subsection*{SLSQP}
\label{subsec:slsqp}

\begin{itemize}[-]
    \item \textbf{Implementation: } \texttt{scipy} \cite{scipy}.
    \item \textbf{Black box: } Yes.
    \item \textbf{gradient-based: } Yes.
    \item \textbf{References: }   \cite{scipy, nocedal1999numerical}. 
    \end{itemize}

\subsection*{TNC}
\label{subsec:tnc}

\begin{itemize}[-]
    \item \textbf{Implementation: } \texttt{scipy} \cite{scipy}.
    \item \textbf{Black box: } Yes.
    \item \textbf{gradient-based: } Yes.
    \item \textbf{References: }   \cite{scipy, nocedal1999numerical, kochenderfer2019algorithms}. 
    \end{itemize}

\subsection*{CG}
\label{subsec:cg}

\begin{itemize}[-]
    \item \textbf{Implementation: } \texttt{scipy} \cite{scipy}.
    \item \textbf{Black box: } Yes.
    \item \textbf{gradient-based: } Yes.
    \item \textbf{References: }   \cite{scipy, nocedal1999numerical, kochenderfer2019algorithms}. 
    \end{itemize}

\subsection*{Newton-CG}
\label{subsec:newton_cg}

\begin{itemize}[-]
    \item \textbf{Implementation: } \texttt{scipy} \cite{scipy}.
    \item \textbf{Black box: } Yes.
    \item \textbf{gradient-based: } Yes.
    \item \textbf{References: }   \cite{scipy, nocedal1999numerical, kochenderfer2019algorithms}. 
    \end{itemize}

\end{document}